\documentclass[12pt]{article}

\usepackage[margin=1.25in]{geometry}

\usepackage{setspace}
\usepackage[utf8]{inputenc}
\usepackage[utf8]{inputenc}
\usepackage{amsmath}
\usepackage{caption}   %% HUGE PACKAGE FOR TABLE  CAPTION!!
\usepackage{mathrsfs}

\usepackage{hyperref}
\usepackage{afterpage}
\usepackage{mathtools}
\usepackage{eurosym}
\usepackage{multirow}
\usepackage{pdfpages}
\usepackage[normalem]{ulem}
\usepackage{latexsym}
\usepackage{amssymb}
\usepackage{lipsum}
\usepackage{csquotes}
\usepackage{setspace}
\usepackage{float}
\usepackage[english]{babel}
\usepackage{fancyhdr, afterpage}
\usepackage{latexsym}
\usepackage{bbm}
\usepackage{amsmath}
\usepackage{amsfonts}
\usepackage{amssymb}
\usepackage[utf8]{inputenc}
\usepackage{pgfplots}
\usepackage{caption}
\usepackage[flushleft]{threeparttable}
\usepackage{bm}
\usepackage{pdflscape} 
\usepackage{tikz}
\usepackage[subnum]{cases}
\usepackage{mathtools}% Loads amsmath
\DeclareMathOperator*{\E}{\mathbb{E}}
\usepackage{booktabs}
\usepackage{hyperref} 
\usepackage{subcaption}
\usepackage{lscape}
\usepackage{graphicx}  % remove 'demo' option for your real document
\hypersetup{
colorlinks=true,
linkcolor=blue,
filecolor=black, 
urlcolor=blue,
citecolor=blue
}
\usepackage[english]{babel}

\usepackage[
  backend=bibtex,
  style=authoryear,
  natbib=true,
  maxbibnames=99,
  maxcitenames=1,%modified
  citetracker=true%added
  ,doi=false,isbn=false,url=false,eprint=false
]{biblatex}

\hypersetup{
  pdfborderstyle={/S/U/W 1}, % thanks, http://tex.stackexchange.com/a/26085/22939
}

\DefineBibliographyStrings{english}{%
  bibliography = {References},
}

\defbibheading{bibintocindent}[\refname]{%
  \section*{#1}%
  \addcontentsline{toc}{section}{\protect\numberline{}#1}%
  \markboth{\MakeUppercase{#1}}{\MakeUppercase{#1}}}

\usepackage{xpatch}
\xpatchbibmacro{cite}{\printnames{labelname}}%
{\ifciteseen{\printnames{labelname}}{\printnames[][-\value{listtotal}]{labelname}}}
{}{}

\xpatchbibmacro{textcite}{\printnames{labelname}}%
{\ifciteseen{\printnames{labelname}}{\printnames[][-\value{listtotal}]{labelname}}}
{}{}

\renewcommand{\footnoterule}{%
  \kern -3pt
  \hrule width \textwidth height 1pt
  \kern 2pt
}
 \pgfplotsset{compat=1.18}

\title{Theory coherent shrinkage of Time-Varying Parameters in VARs {\footnote{{I would like to thank Frank Schorfheide and Andrea Carriero for invaluable guidance and support. I also thank participants at the NBER-NSF SBIES Conference at the Federal Reserve of Philadelphia, the Workshop in Empirical Macroeconomics at King's College London, the  RCEA 2024 Bayesian Workshop, the Barcelona Summer Forum 2024 and the JRC/ECFIN seminar series. I thank Agostino Consolo, Luca Fanelli, Florian Huber, Massimiliano Marcellino, Michele Piffer, Tommaso Tornese, Joshua C. Chan for helpful comments.}}}}
\author{Andrea Renzetti \\ \small Bocconi University}
\begin{document}
\maketitle
\begin{abstract}
 %This paper introduces a Theory Coherent TVP-VAR model that exploits the cross equations restrictions implied by an arbitrary theory about the variables in the system as a prior. 
%This paper introduces a shrinkage prior for TVP-VARs which encodes the cross equations restrictions implied by an arbitrary theory about the variables in the system. 
This paper introduces a novel theory-coherent shrinkage prior for Time-Varying Parameter VARs (TVP-VARs). The prior centers the time-varying parameters on a path implied a priori by an underlying economic theory, chosen to describe the dynamics of the macroeconomic variables in the system. Leveraging information from conventional economic theory using this prior significantly improves inference precision and forecast accuracy compared to the standard TVP-VAR. In an application, I use this prior to incorporate information from a New Keynesian model that includes both the Zero Lower Bound (ZLB) and forward guidance into a medium-scale TVP-VAR model. This approach leads to more precise estimates of the impulse response functions, revealing a distinct propagation of risk premium shocks inside and outside the ZLB in US data. 

% The idea is to leverage on economic theory to obtain sharper finite sample inference, achieve identification of macroeconomic shocks and obtain more reliable predictions.

% The approach is a direct extension to TVP-VARs of the general equilibrium prior proposed by \citet{DNS2004} for nt parameters VARs. The idea is to consider an even more flexible model for the data, able to capture evolving non-linearities in the relationship among the variables of interest and 

%Below I show an application in which I use a small scale New-Keynesian model to form a prior for the parameters of a trivariate TVP-VAR for GDP-growth, inflation and interest rate on US data. This application is preliminary and meant to understand how the properties of the proposed prior translate in the posterior estimates of the time varying coefficients. 
\end{abstract}
\vspace{0.2cm}

%\emph{J.E.L Classification Code: C32, C53   } 

%\emph{Keywords:} \small Time Varying Parameters VARs, Bayesian Econometrics, DSGE-VARs\\

\clearpage
\section{Introduction}

%Importantly, the proposed prior can be derived from theories that imply both constant and time varying moments for the data. As well, the model can be use to evaluate and compare competing theories. 
%%%%%%%%%%%%%%%%%%%%%%%%%%%%%%%%%%%%%%%%%%%%%%%%%
%%%%%%%%%%% Background and motivation  %%%%%%%%%%
%%%%%%%%%%%%%%%%%%%%%%%%%%%%%%%%%%%%%%%%%%%%%%%%%
\normalsize
Over the past four decades vector autoregressive models have become the leading tool for description, forecasting, structural inference and policy analysis of macroeconomic data \citep{sims1980,SW2012}. A natural  progression in the literature was to allow for time-varying parameters to capture changes in the complex dynamic interrelationship among the variables in the system \citep{CS2002,primiceri2005, COGLEY2005262}.  On the one side, this class of models known as Time Varying Parameters VARs (TVP-VARs) can be flexible enough to fit many different forms of structural instabilities and evolving nonlinear relationships among the macroeconomic variables.  On the other side, due to the growing number of parameters, the model can become overly parameterized, which may negatively affect the precision of inference for typical objects of interest, such as impulse response functions, and reduce the reliability of the forecasts. 

%%%%%%%%%%%%%%%%%%%%%%%%%%%%%%%%%%
%%%%%%%%%%% Contribution %%%%%%%%%
%%%%%%%%%%%%%%%%%%%%%%%%%%%%%%%%%%

In this paper, I propose using economic theory to sharpen inference in TVP-VARs. The approach consists in leveraging on prior information coming from an underlying economic theory regarding the macroeconomic variables in the system. This prior information is incorporated into a shrinkage prior for the time-varying parameters, guiding the time-varying coefficients towards a predetermined path implied by the chosen economic theory. The resulting model, equipped with this prior, which I label Theory Coherent TVP-VAR (TC-TVP-VAR), remains a flexible statistical model for the data that exploits economic theory to enhance inference about the time varying parameters. 
The model features two crucial hyper-parameters governing the behavior of the time varying coefficients: the first one determines their intrinsic time variation, while the other one determines their degree of theory coherence stemming from the amount of shrinkage towards the path for those coefficients dictated by the economic theory. Both the optimal degree of time variation and of theory coherence of the time varying parameters can be optimally tuned by maximizing the marginal data density of the model which is available in closed form. The TC-TVP-VAR can leverage prior information from economic theories that imply both constant and time varying paths for the time varying parameters, thereby extending the DSGE-VAR framework developed in \citet{DNS2004} to a time-varying framework. The TC-TVP-VAR can also serve as a tool for estimating the deep parameters from the underlying economic theory. These parameters are as another set of hyper-parameters in the model and are indirectly estimated by mapping the TVP-VAR estimates onto the theoretical restrictions imposed by the structural model.%Thus, in this approach, learning about the deep parameters from the economic theory happens indirectly through learning about the TVP-VAR parameters. 

%%%%%%%%%%%%%%%%%%%%%%%%%%%%%%%%%%
%%%%%%%%%%% Results   &%%%%%%%%%%%
%%%%%%%%%%%%%%%%%%%%%%%%%%%%%%%%%%

In the paper I show that incorporating information from conventional economic theory into a prior for the coefficients of TVP-VARs can be beneficial to improve forecast accuracy and to obtain more precise estimates of typical objects of interest such as the impulse response functions. In particular, I find that exploiting a basic three equations New-Keynesian model to form a prior for the parameters of a trivariate TVP-VAR for output growth, inflation rate and the interest rate improves both point and density forecast accuracy of both output growth and the inflation rate at all the horizons considered (one quarter ahead, two quarters ahead and one year ahead). Then, I exploit the TC-TVP-VAR to investigate changes in the propagation of risk premium shocks inside and outside the Zero lower Bound (ZLB) period in the US economy. According to a standard NK, the economy is expected to exhibit a different response to demand and supply shocks when the ZLB constraint is in effect. However, and more importantly, the short length of the ZLB period in the US makes the standard TVP-VAR unfit to detect the change in the responses predicted by the NK model \citep{lubikbenati}. In other words, whether or not there was a change in the propagation of risk premium shocks during the ZLB as predicted by a standard NK model, cannot be directly inferred by using a standard TVP-VAR. Based on a simulation study, I show that the TC-TVP-VAR can be used to solve this inferential problem. In particular, I exploit the time varying restriction functions implied by a medium scale NK model that accounts for forward guidance and the ZLB period to parametrize my prior in the TC-TVP-VAR. I show that this approach allows to estimate more precisely the response of the economy to macroeconomic shocks inside and outside the ZLB period, solving the inferential problems of the standard TVP-VAR. Estimating the model on US data, I find that there are convincing evidences supporting a different propagation of risk premium shocks inside the ZLB period similar to the one predicted by a standard NK model. This finding has clearly important policy implications for the conduct of fiscal and macroprudential policies at the ZLB. \\

%limits the extent to which inference can be made using a standard TVP-VAR \citep{lubikbenati}.

%%%%%%%%%%%%%%%%%%%%%%%%%%%%%%%%%%%%%%%%%%%%%%%%%%%%
%%%%%%%%%%%%%%%%%%%%%%%%%%%%%%%%%%%%%%%%%%%%%%%%%%%%
%%%%%%%%%   Related literature            %%%%%%%%%%
%%%%%%%%%%%%%%%%%%%%%%%%%%%%%%%%%%%%%%%%%%%%%%%%%%%%
%%%%%%%%%%%%%%%%%%%%%%%%%%%%%%%%%%%%%%%%%%%%%%%%%%%%

\textbf{Related Literature} \hspace{0.2cm} 
This paper shows how to exploit prior information grounded on the basis of an economic theory to impose parsimony on the coefficients of TVP-VARs. In this aspect, the contribution conceptually borrows from ideas from the seminal work of \citet{WHITEMAN} and operationally from the insights in \citet{DNS2004} which show how to exploit the non-linear cross equation restrictions implied by a DSGE to form a prior for the parameters of a constant parameters VAR model. Extending the framework of  \citet{DNS2004} to TVP-VARs is important at least for two reasons. First, because in macroeconomic applications the assumption of constant coefficients is often restrictive. Indeed, instabilities in the autoregressive coefficients of VARs used to model the dynamics of key macroeconomic indicators such as output and inflation have been widely documented in the literature \citep{CS2002,primiceri2005,giannonegambetti}. In this setting, as the model becomes more flexible, additional shrinkage can be particularly beneficial to reduce overfitting. Second, because economic theories themselves might imply time-varying paths for the coefficients. For example, macroeconomic theories assuming rational expectations extended so as to allow some parameters to vary according to Markov process with given transition probabilities, lead to state space representation with time varying coefficients \citep{FARMER20091849}. At the same time, solutions for linear stochastic rational expectations models in the face of a finite sequence of anticipated structural changes lead to state space representation with time varying coefficients \citep{kulishcagliarini}.\footnote{Another notable case within the rational expectations framework are models log-linearized around time varying trend for inflation \citep{cogleysbordone,sbordoneascari,ascari2023long}.} Likewise, outside the rational expectation framework, macroeconomic theories that assume learning can lead to state space representation with time varying coefficients \citep{MILANI20072065}. In all those cases, the proposed TC-TVP-VAR can be used both to incorporate the implied time varying restriction functions into a prior for the time varying coefficients of the VAR and to estimate the deep parameters of the underlying economic model. In this sense, the paper is also related to the strand of literature that exploits an auxiliary flexible statistical model for the data to make indirect inference on the deep parameters of a structural model from the economic theory (see for example \citet{gallant} \citet{kasyfessler2019}).\footnote{In the TC-TVP-VAR the structural parameters from the underlying economic theory are estimated by implicitly minimizing the weighted discrepancy between the unrestricted TVP-VAR estimates and the restriction functions. This approach can be thought as a Bayesian version of \citet{smith1993} as in \citet{DNS2004}.}

The paper is also related to the increasing number of studies that has recently focused on the issue of mitigating complexity and over-parametrization in TVP-VARs. One strand of literature has focused on identifying fixed versus time varying coefficients, by concentrating on the variance selection problem in the generic state equations of each of the TVP-VARs’ coefficients. \citep{FRUHWIRTHSCHNATTER201085,belmontekoopkorobilis,kalli,BITTO201975,onorante}. This literature has produced shrinkage priors for variances aimed at ``automatically'' reducing time-varying coefficients to static ones if the model is overfitting.\footnote{Similarly, from a frequentist perspective, \citet{coulombe2021timevarying} showed that time varying parameters can be framed as ridge regressions problems and used cross validation to tune the optimal amount of time variation in each of the state equations of the coefficients of the TVP-VAR.} While treating the coefficients of the model as independent stochastic processes and just focusing on the problem of tuning the optimal amount of time variation of the single coefficients, this strand of literature typically entirely neglects co-movement and correlation among the coefficients. However, in macroeconomic applications, the  high degree of co-movement in the parameters is an empirical regularity.  This fact was already found and stressed by \citet{COGLEY2005262} in one of the papers that introduced TVP-VARs in the field. In the same paper, the authors envisaged that the reduced-form parameters should move in a highly structured way because of the cross-equation restrictions suggesting that \textit{``a formal treatment of cross-equation restrictions with parameter drift is a priority for future work}'' (p. 274). More in line with these considerations, a smaller number of studies \citep{gambettidewind,stevanovic,CHAN2020105} proposed to use a factor structure to model the time variation of the parameters. Despite being compatible with the idea of the coefficients varying in a highly structured way because of the cross-equation restrictions associated with macroeconomic equations,  this approach is purely statistical and abstracts from any macroeconomic theory disciplining the behavior of the coefficients.\footnote{An exception to this literature is the TVP-VAR in \citet{leon} where the identified structural innovations are allowed to influence the dynamics of the coefficients.} This paper fills this gap in the literature by showing how to exploit prior information grounded on the basis of an economic theory to state a priori a plausible correlation structure among the time varying parameters of the model.  While focusing on economic theory as a source of potentially useful information on the time varying parameters, the paper methodologically contributes to literature on priors for TVP-VARs, by specifying a joint shrinkage prior for the entire path of the time varying parameters. This approach allows to model the shape of the path the time varying coefficients are expected to follow. Operationally this is done by writing the TVP-VAR in static compact form and exploiting band matrices to specify a joint prior for the time varying parameters.

Finally, the paper is more broadly related to the econometric literature showing that moment conditions from economic theory can successfully be exploited for forecasting macroeconomic and financial variables (\citet{GIACOMINI2014145,ccm} most notably).\\

\textbf{Outline} \hspace{0.2cm}  The paper is organized as follows. In Section \ref{sec: tctvpvar} I introduce the TC-TVP-VAR, presenting an analytical derivation of the proposed theory coherent prior and a simulation from the prior to showcase its main properties. In the section, I discuss the fit-complexity trade-off linked to the calibration of the hyper-parameters determining the intrinsic time variation of the time varying-coefficients and the degree of shrinkage towards the restriction functions from the economic theory. I conclude section \ref{sec: tctvpvar} by presenting an MCMC sampler used for estimation of the TC-TVP-VAR. In section \ref{sec:tvolatility} I also discuss how to exploit the proposed prior in TVP-VARs featuring stochastic volatility. Then, in section \ref{oos} I exploit the well known 3-equations New-Keynesian (NK) block to form a prior for the parameters of a TVP-VAR for GDP growth, inflation and the interest rate for the US economy. I compare the forecasts from a TC-TVP-VAR that encodes the restrictions from the NK model as a prior for the time varying coefficients to the forecasts from a standard TVP-VAR, showing that the former outperforms the latter both in terms of point and density forecast accuracy. Then, in section \ref{sec:application} I focus on impulse response analysis at the ZLB. I conduct a simulation study and show that a standard TVP-VAR struggles to detect the change in the response of the economy to macroeconomic shocks during the ZLB period as predicted by a standard NK model. I show that the TC-TVP-VAR can  be used to solve this inferential problem. Finally, I use the TC-TVP-VAR to investigate the propagation of risk premium shocks inside and outside the ZLB period. Section \ref{conclòus} concludes.

\section{Theory coherent TVP-VAR}\label{sec: tctvpvar}

%The construction of a theory coherent prior builds on the idea that an economic theory implies restrictions on the parameters of the TVP-VAR. Based on this idea, in this section I derive analytically a prior for a TVP-VAR which is theory coherent, in the sense that it centers the time varying coefficients on a path implied by an underlying economic theory about the variables in the system. I consider the case in which the population moments implied by the model from the economic theory are available in closed form as a function of the deep structural parameters. De facto, this allows to avoid stochastic simulation of the artificial observations from the economic theory and write the prior for the time varying parameters of the VAR directly as a function of the deep structural parameters from the economic theory. \\

%What follows clarifies this point, by analytically showing  how to construct a prior for based on these steps. 
%In the paper, I will consider the situation in which the theory provides the population moments for the data $\boldsymbol{Y}$ as a function of the deep parameters $\boldsymbol{\theta}$ from the theory. This assumption, together with the features of the parsimonious TVP-VAR considered in what follows, will allow at the same moment to make posterior inference both on the parameters of the structural model from the theory and the reduced form parameters of the TVP-VAR.  
%Good, flexible statistical model for the data versus tight restrictions from a DSGE model. 

\textbf{Notation} \hspace{0.25cm} Before moving on, I introduce some notations conventions used in the paper. Scalars are in lowercase and normal weight. Vectors are in lowercase and in bold. Matrices are in uppercase and bold.
\subsection{Construction of the prior} \label{sec:constr_prior}
A TVP-VAR is given by: 
\begin{equation}   \label{tvpvar}
\underbrace{\boldsymbol{y}'_t}_{\text{$1 \times N$}} = \underbrace{\boldsymbol{x'}_t}_{\text{$1 \times k$}} \underbrace{\boldsymbol{\Phi}_t}_{\text{$k \times N$}} + \underbrace{\boldsymbol{u_t}'}_{\text{$1 \times N$}} \hspace{3cm}  \boldsymbol{u_t} \sim \mathcal{N}(\boldsymbol{0}_{N \times 1}, \boldsymbol{\Sigma_u})
\end{equation}
where $\boldsymbol{y_t}$ is an $N$ dimensional random vector observed for $t = 1, \ldots,T$ periods, $\boldsymbol{x_t} = [1,\boldsymbol{y}_{t-1}', \ldots, \boldsymbol{y}_{t-p}']'$, $p$ is the lag order and $k = 1+Np$ is the number of coefficients in each equation of the VAR. For convenience, we can rewrite the TVP-VAR in (\ref{tvpvar}) in ``static'' compact form as
\begin{equation} \label{static}   \underbrace{\boldsymbol{Y}}_{\text{$T \times N$}} = \underbrace{\boldsymbol{X}}_{\text{$T \times Tk$}} \underbrace{\boldsymbol{\Phi}}_{\text{$Tk \times N$}} + \underbrace{\boldsymbol{U}}_{\text{$T \times N$}} \hspace{3cm} \boldsymbol{U} \sim MVN(\boldsymbol{0},\boldsymbol{\Sigma_u},\boldsymbol{I_T}) 
\end{equation}
where
\begin{equation}
\boldsymbol{Y} = \begin{bmatrix}
\boldsymbol{y}_1'   \\
\vdots   \\
\boldsymbol{y}_T' \\
\end{bmatrix}\hspace{1cm} \boldsymbol{X} = \begin{bmatrix}
\boldsymbol{x}_1' &  \boldsymbol{0} & \boldsymbol{0}  \\
\boldsymbol{0}  & \ddots & \boldsymbol{0}  \\
\boldsymbol{0}  & \boldsymbol{0} & \boldsymbol{x}_T' \\
\end{bmatrix}
\end{equation}
For now it is assumed a constant error covariance matrix $\boldsymbol{\Sigma_u}$, while in section \ref{sec:tvolatility}  the model is extended to the case of a time-varying error covariance. The likelihood implied by (\ref{static}) is
\begin{equation}
\begin{aligned}
p(\boldsymbol{Y}| \boldsymbol{\Phi}, \boldsymbol{\Sigma_u}) & =  (2\pi)^{-\frac{T N}{2}}|\boldsymbol{ \Sigma_u}|^{-\frac{T}{2}} exp \left[-\frac{1}{2}tr\left(( \boldsymbol{\Sigma_u}^{-1})(\boldsymbol{Y'Y} - \boldsymbol{\Phi'X'Y} - \boldsymbol{Y'X\Phi} + \boldsymbol{\Phi'X'X\Phi} )\right)  \right]  \\
\end{aligned}
\end{equation}
Note that in this representation, the matrix $\boldsymbol{\Phi}$ stores the matrices with the time varying coefficients one on the top of the other, for all the time periods $t=1, \ldots, T$ (see \ref{compact_states_rep}). This paper introduces a prior for $(\boldsymbol{\Phi},\boldsymbol{\Sigma_u})$, with its final form presented in equations (\ref{NIW_PRIOR}). This prior centers the time-varying coefficients in $\boldsymbol{\Phi}$ on a path predefined by an underlying economic theory. The aim of this section is to guide the reader through the construction of this prior, which begins by considering a simple dynamic model for the time-varying coefficients, namely 
\begin{equation}
    vec(\boldsymbol{\Phi_t}) = vec(\boldsymbol{\Phi_{t-1}}) + vec(\boldsymbol{\Xi}_t) \hspace{2cm} vec(\boldsymbol{\Xi}_t) \sim \mathcal{N}(\boldsymbol{0},\boldsymbol{\Omega})
\end{equation} 
This dynamics can be rewritten in compact form for all the coefficients as 
\begin{equation}\label{compact_states}
    \boldsymbol{H}_{Tk} \boldsymbol{\Phi} = \boldsymbol{\Phi_{00}} + \boldsymbol{\Xi}
\end{equation} 
%
\iffalse
\[ 
\underbrace{
\begin{bmatrix}
\boldsymbol{I}_k & \boldsymbol{0}  &\ldots& \boldsymbol{0}  \\
-\boldsymbol{I}_k & \boldsymbol{I}_k& \ldots & \boldsymbol{0}  \\
\boldsymbol{0} & -\boldsymbol{I}_k& \ddots & \boldsymbol{0}  \\
\vdots & \vdots&\vdots & \vdots\\
\boldsymbol{0}  & \boldsymbol{0} & -\boldsymbol{I}_k & \boldsymbol{I}_k \\
\end{bmatrix}}_{\text{$\boldsymbol{H_{Tk}}$}}
\underbrace{
\begin{bmatrix}
    \boldsymbol{\Phi_1} \\
    \boldsymbol{\Phi_2} \\ 
    \boldsymbol{\Phi_3}  \\
    \vdots              \\
    \boldsymbol{\Phi_T} \\
\end{bmatrix}}_{\text{$\boldsymbol{\Phi}$}} = 
\underbrace{
\begin{bmatrix}
    \boldsymbol{\Phi_0} \\
    \boldsymbol{0} \\ 
    \boldsymbol{0} \\
    \vdots         \\
    \boldsymbol{0} \\
\end{bmatrix}}_{\text{$\boldsymbol{\Phi_{00}}$}}  + 
\underbrace{
\begin{bmatrix}
 \boldsymbol{\Lambda_k^{-1}} & \boldsymbol{0}  &\ldots& \boldsymbol{0}  \\
0 & \boldsymbol{\Lambda_k^{-1}}& \ldots & \boldsymbol{0}  \\
\boldsymbol{0} & 0& \ldots & \boldsymbol{0}  \\
\vdots & \vdots&\vdots & \vdots\\
\boldsymbol{0}  & \boldsymbol{0} & 0& \boldsymbol{\Lambda_k^{-1}} \\
\end{bmatrix}
\begin{bmatrix}
    \boldsymbol{u}^*_1 \\
    \boldsymbol{u}^*_2  \\ 
    \boldsymbol{u}^*_3  \\
    \vdots         \\
    \boldsymbol{u}^*_T  \\
\end{bmatrix}}_{\text{$\boldsymbol{\eta}$}} 
\]
\fi
which graphically is
\begin{equation}\label{compact_states_rep}
  \underbrace{
\begin{bmatrix}
\boldsymbol{I}_k & \boldsymbol{0}  &\ldots& \boldsymbol{0}  \\
-\boldsymbol{I}_k & \boldsymbol{I}_k& \ldots & \boldsymbol{0}  \\
\boldsymbol{0} & -\boldsymbol{I}_k& \ddots & \boldsymbol{0}  \\
\vdots & \vdots&\vdots & \vdots\\
\boldsymbol{0}  & \boldsymbol{0} & -\boldsymbol{I}_k & \boldsymbol{I}_k \\
\end{bmatrix}}_{\text{$\boldsymbol{H_{Tk}}$}}
\underbrace{
\begin{bmatrix}
    \boldsymbol{\Phi_1} \\
    \boldsymbol{\Phi_2} \\ 
    \boldsymbol{\Phi_3}  \\
    \vdots              \\
    \boldsymbol{\Phi_T} \\
\end{bmatrix}}_{\text{$\boldsymbol{\Phi}$}} = 
\underbrace{
\begin{bmatrix}
    \boldsymbol{\Phi_0} \\
    \boldsymbol{0} \\ 
    \boldsymbol{0} \\
    \vdots         \\
    \boldsymbol{0} \\
\end{bmatrix}}_{\text{$\boldsymbol{\Phi_{00}}$}}  + 
\underbrace{
\begin{bmatrix}
    \boldsymbol{\Xi}_1 \\
    \boldsymbol{\Xi}_2  \\ 
    \boldsymbol{\Xi}_3  \\
    \vdots         \\
    \boldsymbol{\Xi}_T \\
\end{bmatrix}}_{\text{$\boldsymbol{\Xi}$}}   
\end{equation}
%This representation just rewrites the state equations of the time varying coefficients in compact form for all the time periods and is conditional on the initial state $\boldsymbol{\Phi_0}$.\footnote{In the estimation process I will consider $\boldsymbol{\Phi_0}$ as fixed.} 
%with the prior for the initial state centered on the OLS estimates from a constant parameters VAR model, that is $vec(\Phi_1) \sim \mathcal{N} \left( vec(\hat{\Phi}_0), \boldsymbol{\Sigma} \otimes \boldsymbol{V(\hat{\Phi}_0^{OLS}} \right)$ with a diffuse variance. 
For now, I think of just an hyperparameter $\lambda$ affecting the variance of the innovations in the random walk dynamics of all the elements in $\boldsymbol{\Phi_1}, \ldots, \boldsymbol{\Phi_T}$.  In particular, I assume that the variance covariance matrix $\boldsymbol{\Omega}$ has the Kronecker structure $\boldsymbol{\Omega} = \boldsymbol{\Sigma_u} \otimes ( \lambda^2 \boldsymbol{I_k})^{-1}$. This assumption implies that the variance in the state equation of the coefficients is proportional to the variance of the innovations of the equation to which the coefficients appertain $\sigma^2_{ii}$. In practice this means that the generic state equation of the coefficient attached to the $j^{th}$ regressor in the $i^{th}$ equation reads as follows
%of in the state equation of the time varying coefficients proportionally to t,  namely that that the generic state equation attached to the $j^th$ regressor in the $i^{th}$ equation reads as follows 
\begin{equation}
\phi_{jt}^{(i)} = \phi_{jt-1}^{(i)} + \xi_{jt}^{(i)} \hspace{2cm} \xi_{jt}^{(i)} \sim \mathcal{N}\left(0, \frac{\sigma^2_{ii}}{\lambda^2}\right)     
\end{equation}
%, namely that $\boldsymbol{\Omega}$ has the Kronecker structure $\boldsymbol{\Omega} = \boldsymbol{\Sigma_u} \otimes \lambda\boldsymbol{( I_k)}^{-1}$, where $\boldsymbol{\Lambda_k}$ is a diagonal matrices with elements $\lambda_j$ with $j=1,\ldots,k$. : 
%with $j = 1, \ldots, k$ and $i=1, \ldots, N$. Proportionally to $\Sigma_{ii}$ i.e the variance of the innovations in each equation of the VAR the time variation of the coefficients of the TVP-VAR is determined by the $\lambda_j$'s parameters . 
with $i=1, \ldots, N$ and $j=1, \ldots, k$.  As $\lambda \rightarrow 0$ the variance of the \textit{Normal} prior centering a coefficient at time $t$ on the realization at time $t-1$ increases, meaning that this prior becomes more and more diffuse and the coefficients of two consecutive time periods are not forced a priori to be close to each other. Conversely as $\lambda \rightarrow \infty$ the prior centering the coefficient at time $t$ on the realization at time $t-1$ becomes more and more tight. Clearly the calibration of $\lambda$ involves a fit complexity trade off since by letting $\lambda \rightarrow \infty$ we allow the coefficient at time $t$ to be arbitrary distant from the coefficient at $t-1$ up to the point that we can almost fit the data perfectly in sample.\footnote{This point will be covered more in detail in Section \ref{fitcomplex}, through the lenses of the marginal likelihood of the model.} The assumption of a unique hyperparameter $\lambda$ determining the amount of time variation of the time varying coefficients can be easily replaced by assuming $\lambda_j$'s hyper-parameters with $j = 1, \ldots, k$ as it will be clarified in section \ref{distinct_l}.
%where $vec(\boldsymbol{\eta}) \sim \mathcal{N}(0, \boldsymbol{\Sigma} \otimes (\boldsymbol{\Lambda'\Lambda})^{-1})$
%and  $\boldsymbol{\eta_1} \sim \mathcal{N} \left(\boldsymbol{0},5\boldsymbol{V(\hat{\Phi}_0^{OLS})}\right)$ with $\boldsymbol{\Phi_0} = \boldsymbol{\hat{\Phi}_0}^{OLS}$ based on OLS estimates of a constant parameter VAR on a pre-sample of observations and while $t = 2, \ldots, T$ we have $\boldsymbol{\eta_t} \sim \mathcal{N} \left(\boldsymbol{0}, \boldsymbol{\Sigma} \otimes \frac{1}{\lambda^2} \boldsymbol{I_k}\right)$.
% 
%
%
Exploiting representation (\ref{compact_states}) and the Kronecker structure of $\boldsymbol{\Omega}$, we can conclude that equation (\ref{compact_states}) implies a joint normal distribution for $\boldsymbol{\Phi}$, that is
\begin{equation}\label{jointprior}
vec(\boldsymbol{\Phi})| \boldsymbol{\Sigma}_{u},\lambda \sim \mathcal{N}(vec\underbrace{(\boldsymbol{H_{Tk}}^{-1}\boldsymbol{\Phi_{00}}}_{\text{$\underline{\Phi_0}$}}), \boldsymbol{\Sigma}_{u} \otimes  (\underbrace{\lambda^2\boldsymbol{H}_{Tk}'\boldsymbol{H}_{Tk})^{-1}}_{\text{$\underline{\Psi}(\lambda)$}}) 
\end{equation}
since $|\boldsymbol{H}_{Tk}| = 1$, $\boldsymbol{H}_{Tk}$ is always invertible and the determinant of Jacobian matrix of the linear transformation cancels out.\footnote{Notice that this is true also if we consider an AR(1) dynamics for the time varying coefficients. In that case the elements off the main diagonal of the band matrix $\boldsymbol{H_{Tk}}$ would store the AR coefficients of the state equations.} Combining this conditional distribution for the time varying parameters with an \textit{Inverse-Wishart} distribution for $\boldsymbol{\Sigma}_u$ 
\begin{equation}
    p(\boldsymbol{\Sigma_u}) \sim \mathcal{IW}(\underline{\textbf{S}}, \underline{\nu})
\end{equation}
leads to 
\begin{equation}\label{niw}
    p(\boldsymbol{\Phi, \Sigma_u}| \lambda ) \sim \mathcal{NIW}
\end{equation}
Equation (\ref{niw}) states a prior for the time varying parameters and the variance covariance matrix of the innovations of the TVP-VAR which is conditional on $\lambda$, the crucial hyper-parameter determining the amount of time variation of the coefficients.\footnote{To be precise this prior is also conditional on $\boldsymbol{\Phi_0}$, $\boldsymbol{\underline{S}}$ and $\underline{\nu}$. However, for the purpose of the paper I will consider $\boldsymbol{\Phi_0}$, $\boldsymbol{\underline{S}}$ and $\underline{\nu}$ as fixed.} Importantly, this prior is conjugate to the Gaussian likelihood. We exploit the conjugacy to update equation (\ref{niw}) with  the likelihood of some artificial observations from a structural model coming from an underlying economic theory. As a resulting distribution, we obtain a \textit{Normal-Inverse-Wishart} distribution which is now theory coherent, meaning that it is centered on the moment restrictions from the economic theory. \footnote{Note that also assuming $p(\boldsymbol{\Sigma_u}) \propto |\boldsymbol{\Sigma_u}|^{-\frac{N +1}{2}}$ would lead to a \textit{Normal-Inverse Wishart} distribution.} In the specific, updating (\ref{niw}) with observations from the theory we get

%In our framework we exploit a particular kronecker structure for the variance covariance matrix in the state equations of the time varying parameters to obtain 
%We exploit this conjugacy to form a prior for $\boldsymbol{\Phi}$ 

%I combine dummy observations from the theory to form a prior for the time varying parameters as follows: % In particular, we can combine the \textit{Normal-Inverse-Wishart} prior (\ref{niw}) and the dummy observations,  to construct a prior that reflects both the intrinsic persistence of the coefficients of the VAR and the correlation structure among the coefficients implied by the DSGE. In the specific,  we have: 

\begin{equation}\label{combination}
    p(\boldsymbol{\Phi, \Sigma_u}|\lambda,\boldsymbol{\theta},\gamma) = c(\lambda,\boldsymbol{\theta},\gamma)^{-1} p(\boldsymbol{\Phi, \Sigma_u}|\lambda)p(\boldsymbol{Y(\theta)}| \gamma, \boldsymbol{\Phi}, \boldsymbol{\Sigma_u})
\end{equation}
where $p(\boldsymbol{Y(\theta)}| \boldsymbol{\Phi}, \boldsymbol{\Sigma_u}, \gamma) $ is the likelihood of $\gamma$ simulated samples of $\boldsymbol{Y(\theta)}$ from the theory and $\boldsymbol{\theta}$ are the deep parameters from the economic theory while  
\begin{equation}
 c(\lambda,\boldsymbol{\theta},\gamma) = \int_{-\infty}^{\infty}  p(\boldsymbol{\Phi, \Sigma_u}|\lambda)p(\boldsymbol{Y(\theta)}| \gamma, \boldsymbol{\Phi}, \boldsymbol{\Sigma_u}) d\boldsymbol{\Phi} d\boldsymbol{\Sigma_u}    
\end{equation}
is an integrating constant which ensures that the prior density is proper and integrates to one.\footnote{In the appendix \ref{ic_prior} I report the details on the integrating constant.} Equation (\ref{combination}) makes clear that the theory-coherent prior distribution for $(\boldsymbol{\Phi}, \boldsymbol{\Sigma_u})$ is just the posterior distribution of $(\boldsymbol{\Phi}, \boldsymbol{\Sigma_u})$ obtained by updating the hierarchical prior $p(\boldsymbol{\Phi, \Sigma_u}|\lambda)$ with the likelihood of $\gamma$ artificial samples of observations generated by the model from the theory. As in \citet{DNS2004}, in $ p(\boldsymbol{Y(\theta)}| \boldsymbol{\Phi}, \boldsymbol{\Sigma_u})$  I replace the sample moments $\boldsymbol{Y(\theta)'Y(\theta)}$, $\boldsymbol{Y(\theta)'X(\theta)}$, and $\boldsymbol{X(\theta)'X(\theta)}$ by their expected values  $\boldsymbol{\Gamma_{yy}(\theta)} \equiv \E[\boldsymbol{Y'Y}|\boldsymbol{\theta}]$ , $\boldsymbol{\Gamma_{xy}(\theta)} \equiv \E[\boldsymbol{X'Y}|\boldsymbol{\theta}]$ , $\boldsymbol{\Gamma_{xx}(\theta)} \equiv \E[\boldsymbol{X'X}|\boldsymbol{\theta}]$. Note that  since they relate to $\boldsymbol{Y}$ and $\boldsymbol{X}$ in the static representation of the TVP-VAR in (\ref{static}), these moments matrices are storing the moments $\E[\boldsymbol{y}_t\boldsymbol{y}_t'|\boldsymbol{\theta}]$, $\E[\boldsymbol{y}_t\boldsymbol{x}_t'|\boldsymbol{\theta}]$, $\E[\boldsymbol{x}_t\boldsymbol{x}_t'|\boldsymbol{\theta}]$,  for $t=1,\ldots,T$. \footnote{The structure of these matrices is provided in the Appendix \ref{popmoments}.} These theory implied population moments are assumed to be available in closed form as a function of the deep parameters from the economic theory $\boldsymbol{\theta}$ and in most of the cases can be derived from the state space representation of the structural model from the economic theory. This assumption let us write the likelihood of the artificial observations directly as a function of the vector of deep parameters from the economic theory $\boldsymbol{\theta}$. To sum up, the \textit{Normal-Inverse-Wishart} distribution in equation (\ref{combination}) is de-facto obtained as the posterior distribution that combines a \textit{Normal-Inverse Wishart} prior for $(\boldsymbol{\Phi},\boldsymbol{\Sigma_u})$ with a \textit{Gaussian} likelihood for the artificial data from the theory. Therefore the theory coherent prior in equation (\ref{combination}) takes the form 

\begin{equation}
\begin{aligned}  \label{NIW_PRIOR}
    p(\boldsymbol{\Sigma}_u) &\sim \mathcal{IW} \left({\underline{\underline{\boldsymbol{S}}}}, \underline{\underline{\nu}}\right) \\
    p(vec(\boldsymbol{\Phi})|\boldsymbol{\Sigma}_u) & \sim \mathcal{N}(vec(\boldsymbol{\underline{\underline{\Phi}}}),\boldsymbol{\Sigma}_u \otimes \boldsymbol{\underline{\underline{\Psi}}})     
    \end{aligned}
\end{equation}
\begin{equation}\label{prior_S}
\underline{\underline{\boldsymbol{S}}} = \underline{\boldsymbol{S}} + \gamma \boldsymbol{\Gamma_{yy}(\theta)} +   \boldsymbol{\underline{\Phi_{0}}}'(\lambda^2\boldsymbol{H_{Tk}}' \boldsymbol{H_{Tk}}) \boldsymbol{\underline{\Phi_{0}}} - \boldsymbol{\underline{\underline{\Phi}}}'\boldsymbol{\underline{\underline{\Psi}}}^{-1}\boldsymbol{\underline{\underline{\Phi}}}   
\end{equation}
\begin{equation}\label{prior_nu}
\underline{\underline{\nu}} = \underline{\nu} + T\gamma
\end{equation}
\begin{equation}
vec(\boldsymbol{\underline{\underline{\Phi}}}) = vec( \left(\gamma\boldsymbol{\Gamma_{xx}(\theta)} + \lambda^2\boldsymbol{H_{Tk}'H_{Tk}}\right)^{-1}( \gamma\boldsymbol{\Gamma_{xy}(\theta)} + \lambda^2 \boldsymbol{H}_{Tk}'\boldsymbol{H}_{Tk}\boldsymbol{\underline{\Phi_{0}}}) ) 
\end{equation} 
\begin{equation}\label{prior_PSI}
\boldsymbol{\underline{\underline{\Psi}}} = \left( \gamma\boldsymbol{\Gamma_{xx}(\theta)}  +  \lambda^2\boldsymbol{H_{Tk}'H_{Tk}}\right)^{-1} 
\end{equation}
This prior for $\boldsymbol{\Phi}$ encompasses two different pieces of information about the time varying coefficients. The first piece of information is their intrinsic time variation determined by the shrinkage hyperparameter $\lambda$. The second, is instead the degree of theory coherence implied by the shrinkage hyperparameter $\gamma$ which defines the tightness of the \textit{Normal} prior around the restriction function defined by the economic theory
\begin{equation}\label{restriction_function}
\underbrace{\boldsymbol{\Phi(\theta)^*}}_{Tk\times N} = \underbrace{\boldsymbol{\Gamma_{xx}(\theta)^{-1}}}_{Tk \times Tk}\underbrace{\boldsymbol{\Gamma_{xy}(\theta)}}_{Tk \times N}
\end{equation}
This restriction function implies that at each time period $t$ the matrix of coefficients  is centered on 
\begin{equation}\label{approxt}
\boldsymbol{\Phi_t (\theta)}^*= \boldsymbol{\Gamma_{xx,t}(\theta)}^{-1}\boldsymbol{\Gamma_{xy,t}(\theta)}     
\end{equation}
where $\boldsymbol{\Gamma_{xx,t}(\theta)} \equiv \E [\boldsymbol{x_tx_t'}|\boldsymbol{\theta}]$  and $\boldsymbol{\Gamma_{xy,t}(\theta)} \equiv \E [\boldsymbol{x_ty_t'}|\boldsymbol{\theta}]$ are respectively of dimension $k \times k$ and $k \times N$. Equation (\ref{approxt}) shows that the prior is centering the time varying coefficients on  $\boldsymbol{\Phi_t(\theta)} ^*$, which is a local OLS formula evaluated at the theory implied population moments. In general, these population moments  are assumed to be available in closed form as function of the deep parameters from the economic theory $\boldsymbol{\theta}$. However, in the case they are not directly available in closed form from the chosen economic theory, they can easily be obtained  by stochastic simulation as in \citet{LORIA2022105}. Importantly, as the moments stored in the matrices $\boldsymbol{\Gamma_{xx}}$, $\boldsymbol{\Gamma_{xy}}$, $\boldsymbol{\Gamma_{yy}}$ can be either constant or time varying conditionally on the deep parameters from the theory, (\ref{restriction_function}) can define both constant and time varying paths for the time varying coefficients. Finally, it is worth to mention that this structure of the prior can also be used more in general to model any a-priori beliefs about the evolution of the time-varying coefficients over time, even in the case in which these beliefs are not coming from an explicit economic theory.

%As a last note, the joint prior (\ref{NIW_PRIOR}) implies that the conditional prior $p(vec(\boldsymbol{\Phi_t}|vec(\boldsymbol{\Phi_{t-1}}))$ is still \textit{Normal} and the prior for $vec(\boldsymbol{\Phi_t})$ is still a Markov process. \footnote{The full derivation of the conditional form of the prior as a function of the elements of the precision matrix $\boldsymbol{\underline{\underline{\Psi}}}^{-1}$ and the matrix of means $\boldsymbol{\underline{\underline{\Phi}}}$ can be found in the appendix \ref{sec:conditional_prior}.}

%\begin{equation} p(vec(\boldsymbol{\Phi_t})|vec(\boldsymbol{\Phi_{t-1}}),vec(\boldsymbol{\Phi_{t-2}}), \ldots, vec(\boldsymbol{\Phi_{1}})) =    p(vec(\boldsymbol{\Phi_t})|vec(\boldsymbol{\Phi_{t-1}}))
%\end{equation}

\subsection{Simulation of the prior}\label{sec:simulation} 
\begin{figure}[H]
\centering
\begin{subfigure}[b]{\textwidth}
   \makebox[\textwidth][c]{\includegraphics[scale=0.24]{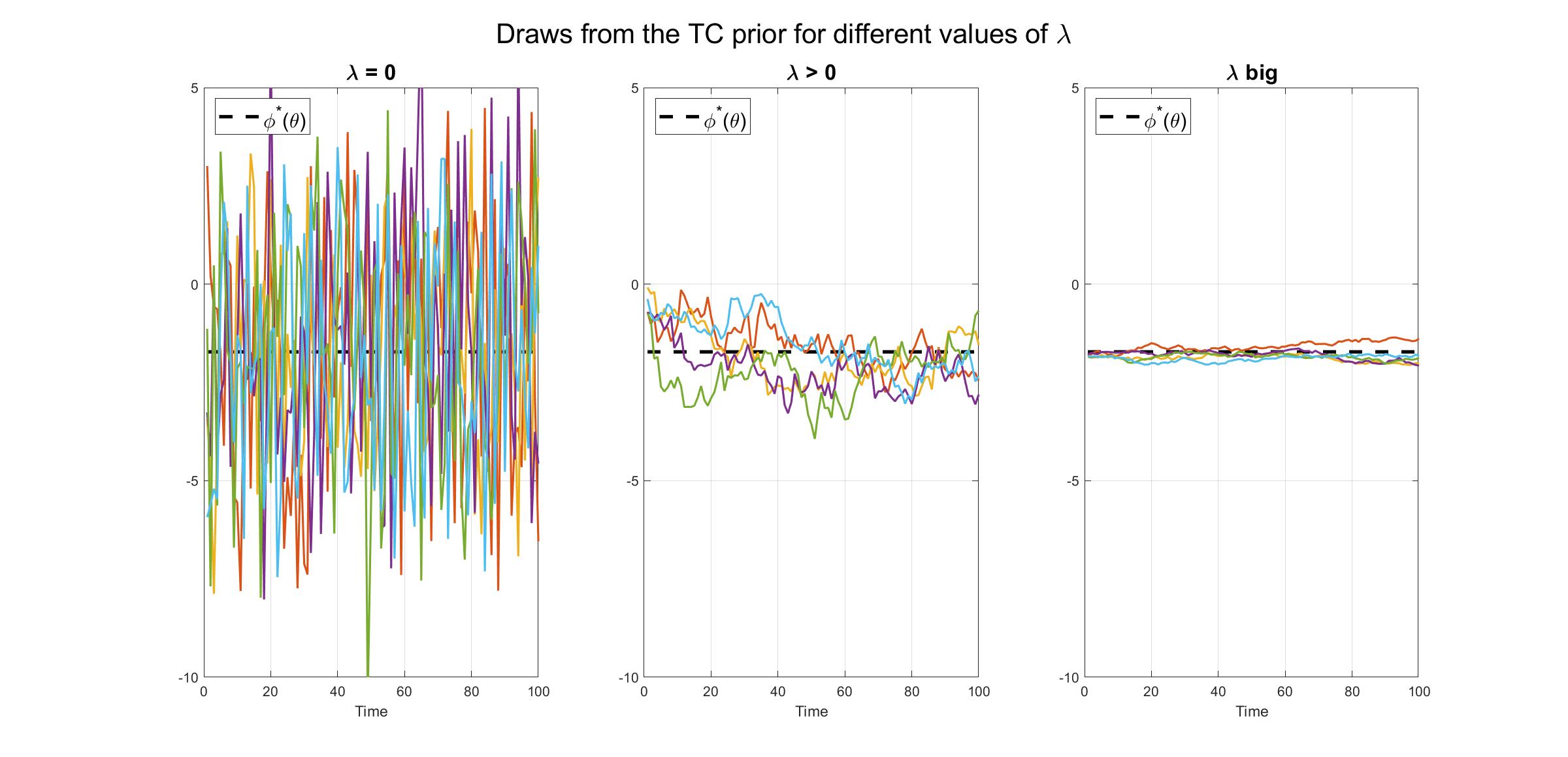}}
   %{prior_flambda_2.jpg}}
\end{subfigure}
\begin{subfigure}[b]{\textwidth}
   \makebox[\textwidth][c]{\includegraphics[scale=0.24]{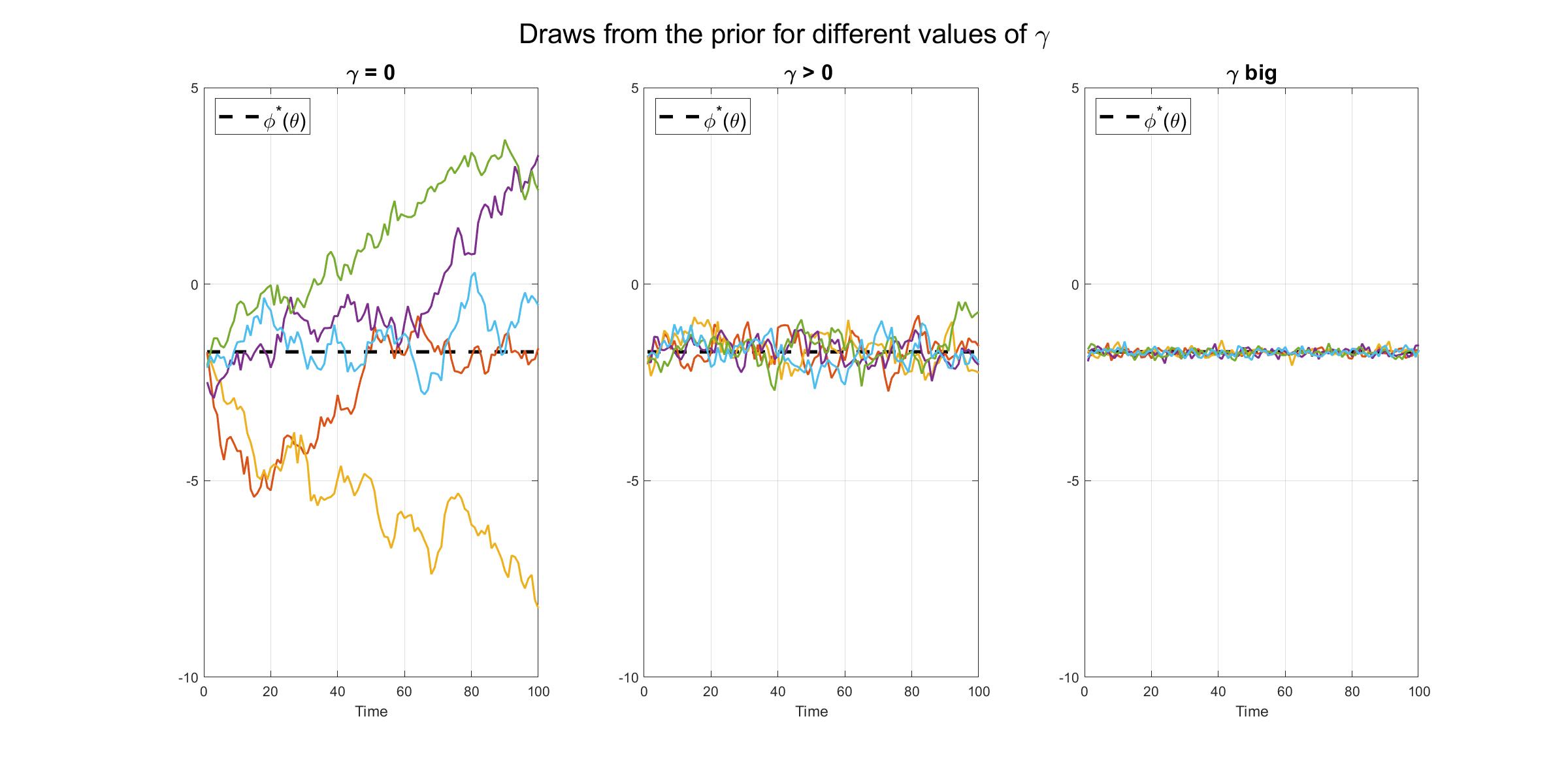}}%{prior_fgamma_2.jpg}}
\end{subfigure}
\caption{ \small Each subfigure presents five draws from the theory coherent prior for a generic coefficient of the TVP-VAR for different values of the hyperparameters $\lambda$ (shrinkage towards constant) and $\gamma$ (theory coherent shrinkage). The first row, presents draws for $\lambda=0$, $\lambda>0$ and a big $\lambda$, conditionally on $\gamma>0$. The second row presents draws for $\gamma=0$, $\gamma>0$ and a big $\gamma$, conditionally on $\lambda>0$. The dashed black line represents the restriction functions coming  from the theory. }\label{simulation}
\end{figure}

\begin{figure}[htp!] 
   \centering    \includegraphics[scale= 0.85]{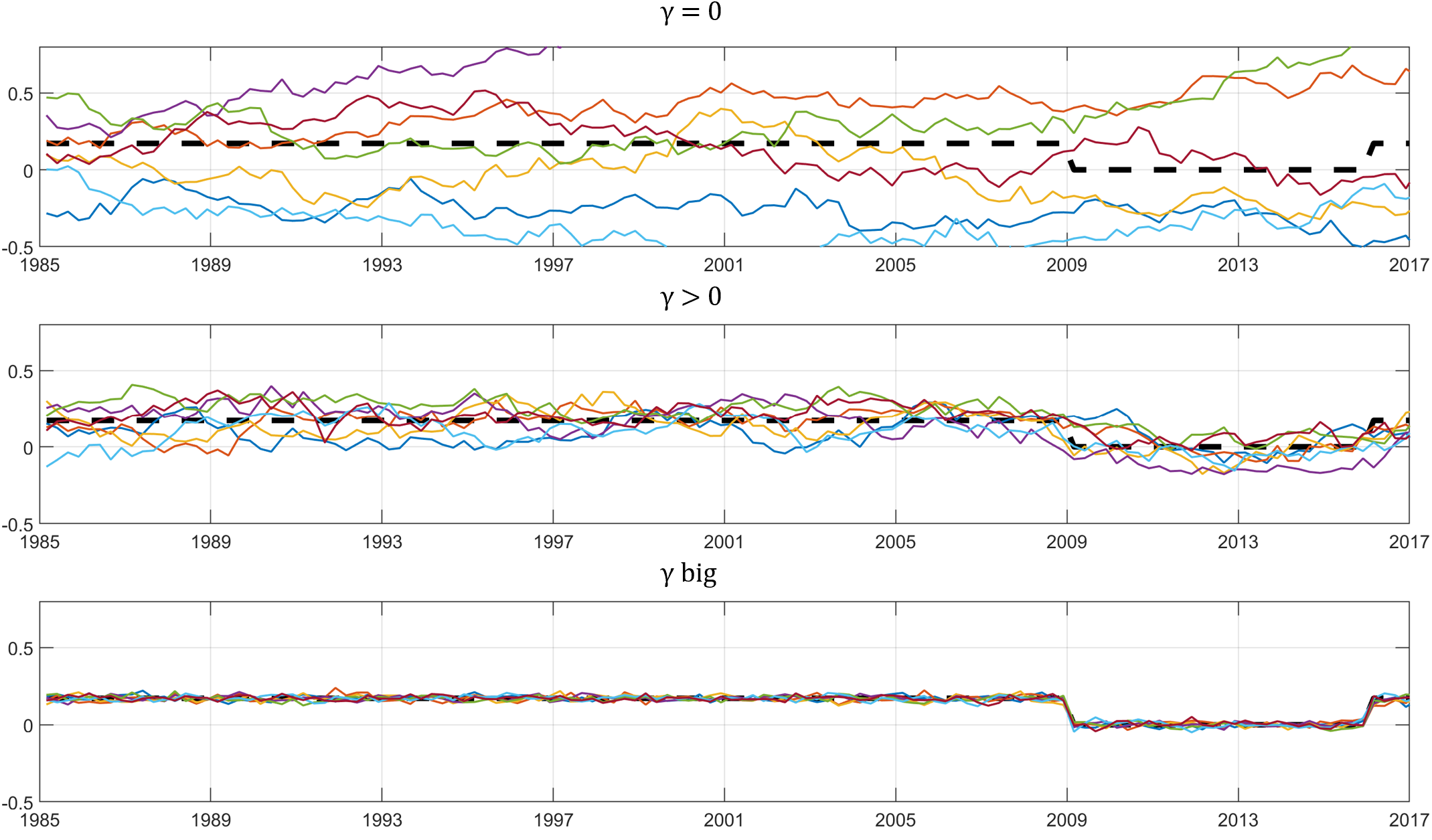}
    \caption{The figure shows draws from the prior for the coefficient on the first lag of the inflation rate in the equation of the Fed Fund rate for different values of the hyperparameter $\gamma$. The model is a seven variable TVP-VAR for the US economy. The dashed black line represents the restriction function implied by the medium scale NK model.}
    \label{fig:tvres}
\end{figure}

To showcase the properties of the proposed theory coherent prior, figure \ref{simulation} shows random draws for the time series of a generic $\phi_{jt}^{(i)}$ coefficient of the TVP-VAR for $t=1, \ldots, 250$ from the prior. The dashed black line represents the path i.e. the restriction function $(\ref{restriction_function})$ defined by the economic theory for that coefficient, that I label $\phi_{jt}^{(i)}(\boldsymbol{\theta})^*$. Just for now, for illustrative purposes  I consider the case of a constant restriction function, however having time varying moments stored in the matrices $\boldsymbol{\Gamma_{xx}}$ and $\boldsymbol{\Gamma_{xy}}$, would directly imply a time varying restriction function through (\ref{restriction_function}). The first row shows five draws from the theory coherent prior for different values of $\lambda$ conditioning on a positive $\gamma$, meaning that for a given degree of theory coherence I change the value of hyper-parameter which determines the intrinsic time variation of the coefficient.  When $\lambda = 0$, the draws for the coefficient $\phi_{jt}^{(i)}$  are independent for $t = 1, \ldots, T$ and centered on the restriction function coming from the theory. Hence, nothing prevents a coefficient at time $t$ to be arbitrary distant from a coefficient at time $t-1$ except the fact both coefficients are centered, with a precision determined by $\gamma$, on the restriction function defined by the theory. As $\lambda$ increases the time series of the coefficient becomes less and less time-varying up to the point that when $\lambda$ is very big, the coefficient becomes almost constant over time. Instead, in the second row, I let the degree of theory coherence change conditionally on a given intrinsic time variation of the time varying coefficients. In other words, I show random draws for $\phi_{jt}^{(i)}$  for different values of $\gamma$ conditionally on $\lambda > 0$. When $\gamma = 0$, the coefficients are just random walks with variance equal to $\frac{\sigma_{ii}}{\lambda^2}$, and they are completely unrelated to the restriction function implied by the economic theory $\phi_{jt}^{(i)}(\boldsymbol{\theta})^*$.\footnote{Note that the draws fore the time varying coefficients were initiated near the restriction functions just for visualization purposes.} As $\gamma$ increases the draws for the coefficient are centered on the restriction function defined by the theory with an increasing precision and on the limit, with $\gamma\rightarrow \infty$, they go to $\phi_{jt}^{(i)}(\boldsymbol{\theta})^*$.\footnote{The coefficients would have a degenerate distribution, with point mass on the restriction function.} Importantly, the hyperparameter $\gamma$ is contemporaneously shrinking all the time varying coefficients  $\phi_{t,j}^{(i)}$ in the equations of the VAR towards the restrictions implied by the economic theory. This reflects the idea that it is economic theory that a priori postulates a plausible correlation structure among the coefficients of VAR model through the restriction function in  (\ref{restriction_function}).  
It is important to remark that this framework can accommodate both constant and time-varying restrictions for the coefficients through (\ref{restriction_function}). This allows to extend the DSGE-VAR framework developed in \citet{DNS2004}
to a broader set of economic theories, which might imply time varying restriction functions for the parameters of the VAR. To show this property, I exploit the medium scale New-Keynesian model in \citet{delnegroschorfheide2005} to parametrize a prior for a medium scale TVP-VAR for output growth, consumption growth, investment growth, real wage growth, hours worked, inflation and the Fed Fund rate. The NK model accounts for the ZLB and forward guidance and it is solved using the method proposed by \citet{kulishcagliarini} for linear rational equation expectation systems in the face of anticipated structural changes. The solution of the model implies a state space representation that exhibits time varying coefficients. Once the population moments implied by the state space representation are used to parameterize (\ref{restriction_function}) we obtain time varying restrictions functions for the time varying coefficients.\footnote{The details on the derivation of the moments are available in the section \ref{sec:application}.} Figure \ref{fig:tvres}
makes this point by showing the implied path for the coefficient of the first lag of the inflation rate in the equation of the Fed Fund Rate. Together with the time varying restriction function, the figure shows draws from the prior for different values of the hyper-parameter $\gamma$ conditioning on a given value of $\lambda$ and a set of structural parameters of the NK model. More in the specific,  while outside the zero lower bound the Fed Fund rate is expected to increase when the lagged value of inflation increases, inside the zero lower bound the interest rate is not expected to respond to the lagged value of the inflation rate and therefore the prior mean for this coefficients becomes centered around zero.  As expected, as $\gamma$ increases the draws for the time varying coefficient are centered with increasing precision on the path defined by the economic theory.  

\clearpage
\subsection{Conditional posterior of \texorpdfstring{$\Phi$}{TEXT} and \texorpdfstring{$\Sigma_u$}{TEXT} }
It is straightforward to derive the posterior distribution of $(\boldsymbol{\Phi},\boldsymbol{\Sigma}_u)$ conditional on the hyper-parameters $\lambda$, $\gamma$ and on the deep parameters from the theory $\boldsymbol{\theta}$ since this is just a \textit{Normal-Inverse-Wishart} distribution obtained by updating a Gaussian likelihood with the \textit{Normal-Inverse-Wishart} prior (\ref{NIW_PRIOR}), namely

\begin{equation}
\begin{aligned}\label{condposterior}
  p(vec(\boldsymbol{\Phi})|\boldsymbol{\Sigma}_u, \gamma, \lambda, \boldsymbol{\theta}, \boldsymbol{Y}) & \sim \mathcal{N}(vec(\boldsymbol{\tilde{\Phi}}),\boldsymbol{\Sigma}_u \otimes \boldsymbol{\tilde{\Psi}})  \\
  p(\boldsymbol{\Sigma}_u|\gamma, \lambda, \boldsymbol{\theta}, \boldsymbol{Y}) & \sim \mathcal{IW} \left({\boldsymbol{\tilde{S}}}, \tilde{\nu}\right)
\end{aligned}
\end{equation}
\begin{equation}\label{postphi}
  vec(\boldsymbol{\tilde{\Phi}}) = vec(\left( \boldsymbol{X'X} 
+ \gamma\boldsymbol{\Gamma_{xx}(\theta)} + \lambda^2\boldsymbol{H_{Tk}'H_{Tk}}\right)^{-1}( \boldsymbol{X'Y} +  \gamma\boldsymbol{\Gamma_{xy}(\theta)} + \lambda^2\boldsymbol{H}_{Tk}'\boldsymbol{H}_{Tk}\boldsymbol{\underline{\Phi_{0}}}))   
\end{equation}
\begin{equation}
     \boldsymbol{\tilde{\Psi}} =  \left(\boldsymbol{X'X} + \gamma\boldsymbol{\Gamma_{xx}(\theta)}  +  \lambda^2\boldsymbol{H_{Tk}'H_{Tk}}\right)^{-1}
\end{equation}
\begin{equation}
 \tilde{\boldsymbol{S}} = \boldsymbol{Y'Y} + \underline{\underline{\boldsymbol{S}}} + \boldsymbol{\underline{\underline{\Phi}}}'\boldsymbol{\underline{\underline{\Psi}}}^{-1}\boldsymbol{\underline{\underline{\Phi}}} - \boldsymbol{\tilde{\Phi}}'\boldsymbol{\tilde{\Psi}}^{-1}\boldsymbol{{\tilde{\Phi}}}   
\end{equation}
\begin{equation}
\tilde{\nu} = \underline{\underline{\nu}} + T 
\end{equation}
Equation (\ref{postphi}) makes it clear that thanks to the conjugacy of the prior the formula for the mean of the conditional posterior of $\boldsymbol{\Phi}$ translates in a standard OLS regression formula based on a sample augmented with a set of dummy observations that determines the degree of time variation of the time varying coefficients and another set of dummy observations that centers the coefficients on the restriction function implied by the theory. While the first set of dummy observations shapes the correlation of the time varying coefficients over time by imposing a set of linear restrictions as a function of the hyper-parameter $\lambda$, the second set of dummy observations induces at each time period a correlation structure implied by the non-linear cross equation restrictions coming from the theory, as a function of the deep parameters from the theory $\boldsymbol{\theta}$ and of the shrinking parameter $\gamma$.\footnote{In the appendix \ref{sec:dummy_tvp} I show that thanks to the Kronecker structure of the prior (\ref{jointprior}) the time variation of the coefficients can be modelled by dummy observations.} It is worth to remark that from a computational perspective equation (\ref{postphi}) makes the estimation of the time varying coefficients very efficient since it allows to draw all the history of the time varying coefficients in all the equations of the VAR in a single step, avoiding forward filtering and backward smoothing algorithms à la \citet{kc1994}. \footnote{Indeed, as in the precision sampler by \citet{chan2009efficient}, we can draw all the latent states from $t=1, \ldots, T$ in a single step and thanks to the Kronecker structure of the posterior, we can do it for all the $N$ equations of the TVP-VAR jointly. More in general, the Kronecker structure (\ref{kron_transition}) coupled with the precision sampler by \citet{chan2009efficient} can be exploited to estimate medium to large scale TVP-VARs. }

\subsection{\texorpdfstring{$\lambda, \gamma$}{TEXT} and fit complexity trade off}\label{fitcomplex}

\begin{figure}[H]
    \centering
    \includegraphics[scale=0.42]{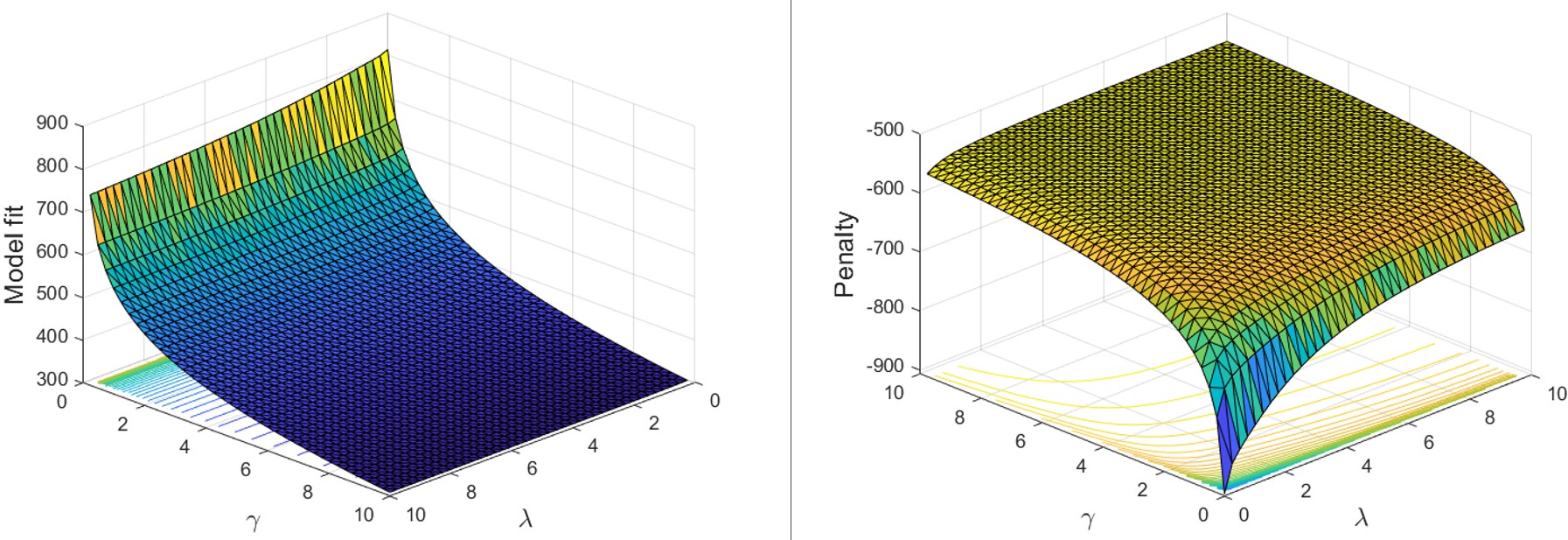}
    \caption{ \small The figure shows the model fit term   (on the left) and the penalty 
    term (on the right) of the marginal likelihood (\ref{ML}) as a function of the hyper-parameters $\lambda$ (intrinsic time variation) and $\gamma$ (theory coherence).}\label{fig:mdd}
\end{figure}

Tuning of the optimal degree of theory coherence and the intrinsic amount of time variation of the coefficients is a delicate matter, since it clearly involves a fit-complexity trade off. As a matter of fact, very low values of both $\gamma$ and $\lambda$ imply a priori that the coefficients in two consecutive time periods can potentially be very distant from each others and from the restriction functions defined by the theory.\footnote{When both $\gamma=0$ and $\lambda=0$ the model is left totally unrestricted, with the prior variance covariance of the coefficients being equal to infinity. Clearly, in this case there are more parameters than you can feasibly estimate with a flat prior meaning that the conditional posterior of $\boldsymbol{\Phi}$ cannot be computed due to the non-invertibility of $\boldsymbol{X'X}$  (this can be seen from equation (\ref{condposterior})).}   Intuitively, this model will fit the data very well in sample but will perform badly for forecasting out-of-sample. Indeed, decreasing $\gamma$ and $\lambda$ will in general increase in-sample fit of the model at the expense of out-of-sample accuracy. Based on this argument, I recommend to base the optimal choice of both the hyper-parameters on the maximization of the marginal likelihood of the model or equivalently on the maximization of the posterior of the hyper-parameters $\lambda$ and $\gamma$ under a flat prior for these hyper-parameters. This translates into maximizing the one-step-ahead out-of-sample forecasting ability of the model. Indeed, the log-marginal likelihood (or Bayesian evidence) can be interpreted as the sum of the one step-ahead predictive scores, since it is equivalent to the scoring rule of the form
\begin{equation}
  S(\boldsymbol{Y}) = \sum_{t=1}^{T}s(\boldsymbol{y}_t|\boldsymbol{y}_{t-1})=  \sum_{t=1}^{T}log(p(\boldsymbol{y}_t|\boldsymbol{y}_{t-1}))
\end{equation}
An attractive feature of the TC-TVP-VAR is that the marginal likelihood $p(\boldsymbol{Y}|\gamma, \lambda, \boldsymbol{\theta})$ obtained by integrating out $\boldsymbol{\Phi},\boldsymbol{\Sigma}_u $ from the conditional posterior $p(\boldsymbol{\Phi, \Sigma_u} |\lambda,\boldsymbol{\theta}, \gamma,\boldsymbol{Y}) $ is available in closed form and it is equal to
\begin{equation}\label{ML}
p(\boldsymbol{Y}|\lambda,\boldsymbol{\theta},\gamma) = \left(\pi\right)^{-\frac{T N}{2}}\frac{\Gamma_N\left(\frac{\tilde{{\nu}} }{2} \right) |\tilde{{S}}|^{-\frac{\tilde{\nu}}{2}} |\tilde{{\boldsymbol{\Psi}}}|^{\frac{N}{2}}}{\Gamma_N\left(\frac{\underline{\nu}}{2} \right) |\underline{\underline{S}}|^{-\frac{\underline{\underline{\nu}}}{2}}|\boldsymbol{\underline{\underline{\Psi}}}|^{\frac{N}{2}}} 
\end{equation}
As a consequence, calibrating $\gamma$ and $\lambda$ to maximize (\ref{ML}) corresponds to finding $\gamma$ and $\lambda$ maximizing the one-step-ahead out-of-sample forecasting ability of the model. This strategy of estimating hyper-parameters by maximizing the marginal likelihood is an empirical Bayes method  which has a clear frequentist interpretation. In what follows, and in particular in the estimation algorithm detailed in the next section \ref{sec:estimation} I will regard $\gamma$ and $\lambda$ as random variables and perform full posterior inference on the hyper-parameters, but analogously maximizing the posterior of the hyper-parameters will correspond to maximizing the one-step-ahead out of sample forecasting ability of the model. Following the same steps as in \citet{giannonelenzaprimicieri}, we can rewrite equation (\ref{ML}) as
\begin{equation} \label{fitcomplexity}
p(\boldsymbol{Y}|\lambda,\boldsymbol{\theta},\gamma)  \propto | \boldsymbol{(V_{\varepsilon}^{post})^{-1}V^{prior}_{\varepsilon}}|^{\frac{T + \underline{\underline{\nu}}}{2}} \prod_{t=1}^{T}|\boldsymbol{V_{t|t-1}}|^{-\frac{1}{2}}
\end{equation}
where $\boldsymbol{V_{\varepsilon}^{post}}$ and $\boldsymbol{V^{prior}_{\varepsilon}}$ are the posterior and prior means (or modes) of the residual variance, while $\boldsymbol{V_{t|t-1}}$ is equal to the variance (conditional on $\boldsymbol{\Sigma_u}$) of the
one-step-ahead forecast of $\boldsymbol{y}$, averaged across all possible a-priori realizations of $\boldsymbol{\Sigma_u}$. Complete formulas of these objects are reported in the appendix \ref{mdd_anal} and are the analog of the formulas reported in \citet{giannonelenzaprimicieri}. Equation (\ref{fitcomplexity}) makes clear that the marginal likelihood involves two terms:  a reward for model fit, $| \boldsymbol{(V_{\varepsilon}^{post})^{-1}V^{prior}_{\varepsilon}}|^{\frac{T + \underline{\underline{\nu}}}{2}}$ and a penalty term for model complexity $\prod_{t=1}^{T}|\boldsymbol{V_{t|t-1}}|^{-\frac{1}{2}}$.  Figure \ref{fig:mdd} plots the model fit term and the penalty term of the marginal likelihood as a function of the two hyper-parameters $\gamma$ and $\lambda$ conditionally on a set of deep parameters from the theory $\boldsymbol{\theta}$. As $\lambda$ decreases the tightness of the prior centering each coefficient on its realization in the previous period eases and in sample model fit improves, meaning that $\boldsymbol{V}^{post}$ decreases. However, at the same time, as  $\lambda$ decreases $\boldsymbol{V_{t|t-1}}$ increases, since the variance (conditional on $\boldsymbol{\Sigma}$) of the one-step-ahead forecast of $\boldsymbol{y}$ increases. Analogously, as $\gamma$ decreases the restriction functions from the theory become less and less binding, meaning that the model will fit better in sample, namely $\boldsymbol{V}^{post}$ decreases, but the variance of the one step ahead forecast error $\boldsymbol{V_{t|t-1}}$ will increase. 
\subsection{Estimation}\label{sec:estimation}
 In this section I describe a posterior simulation sampler to make inference on the TVP-VAR parameters $\boldsymbol{\Phi}$ and $ \boldsymbol{\Sigma}_u$ and on the shrinkage hyper-parameters  $\lambda$ and $\gamma$ together with the deep parameters from the theory $\boldsymbol{\theta}$ .\footnote{For the purpose of the paper we fix $\boldsymbol{\Phi}_0 =  \boldsymbol{\hat{\Phi}_0}$ and $\boldsymbol{\underline{S}} = \boldsymbol{\hat{S}} = diag(\hat{s}_1^2, \ldots, \hat{s}_N^2)$ and  $\nu = N+2$. Alternatively inference on these hyper-parameters can be made by extending the Random Walk Metropolis step in the MCMC sampler described below.} We can simulate draws from the posterior $p(\boldsymbol{\Phi, \Sigma_u}, \lambda, \boldsymbol{\theta}, \gamma|\boldsymbol{Y})$ exploiting the factorization: 
\begin{equation}
p(\boldsymbol{\Phi, \Sigma_u}, \lambda, \boldsymbol{\theta}, \gamma|\boldsymbol{Y})=p(\boldsymbol{\Phi,\Sigma_u}|\lambda, \boldsymbol{\theta}, \gamma, \boldsymbol{Y})p(\lambda, \boldsymbol{\theta}, \gamma|\boldsymbol{Y})
\end{equation}
 This factorization makes clear that we can sample from the posterior of $\boldsymbol{\Phi}, \boldsymbol{\Sigma_u}, \boldsymbol{\theta}, \gamma, \lambda$  building a MCMC algorithm that iterates the following two steps:        
    \begin{enumerate}
    \item Draw from $p(\lambda, \boldsymbol{\theta}, \gamma|\boldsymbol{Y}) \propto p(\boldsymbol{Y}|\lambda,\boldsymbol{\theta}, \gamma)p(\lambda, \boldsymbol{\theta}, \gamma)$.
    \item Draw from $p(\boldsymbol{\Phi, \Sigma_u} | \lambda,\boldsymbol{\theta}, \gamma,\boldsymbol{Y})$ %\sim \mathcal{NIW} $
    \end{enumerate} 
In the first step, estimation of both the shrinkage parameters $\gamma$ and $\lambda$ and of deep parameters from the theory $\boldsymbol{\theta}$ happens by posterior evaluation of 
\begin{equation}
 p(\lambda, \boldsymbol{\theta},\gamma| Y) \propto p(\boldsymbol{Y}|\lambda,\boldsymbol{\theta},\gamma)p(\boldsymbol{\theta})p(\gamma)p(\lambda)  
\end{equation}
where $p(\gamma)$ and $p(\lambda)$ are the priors for the shrinkage hyper-parameters \footnote{For both $\gamma$ and $\lambda$ I consider the Uniform prior $\gamma \sim \mathcal{U} (0, b_{\gamma}) $ and $\lambda \sim \mathcal{U} (0, b_{\lambda})$ where $b_{\gamma} = 10^{10}$ and $b_{\lambda} = 10^{10}$} while $p(\boldsymbol{\theta})$ is the prior of the deep parameters from the theory. Hence, learning about the structural parameters happens implicitly by projecting the VAR estimates onto the restrictions implied by the model from the theory. More precisely, the estimates of the deep parameters minimizes the weighted discrepancy between the TVP-VAR unrestricted estimates and the restriction function (\ref{restriction_function}). This approach  can be thought as a Bayesian version of \citet{smith1993} and was pioneered by \citet{DNS2004}.  In particular, the TVP-VAR is used to summarize the statistical properties of both the observed data and the theory-simulated data and an estimate of the deep parameters from the theory is obtained by matching as close as possible TVP-VAR parameters from observed data and from the simulated data.\footnote{See Proposition 1 and Proposition 2 in \citet{DNS2004} for further details on this. } To sample from $p(\lambda, \boldsymbol{\theta}, \gamma|\boldsymbol{Y})$ I consider a two blocks random walk metropolis algorithm. This is basically a Gibbs Sampler where in the first block, I draw the hyper-parameters $\gamma$ and $\lambda$ conditionally on $\boldsymbol{\theta}$ while in the second block I draw $\boldsymbol{\theta}$ conditionally on the shrinking hyper-parameters $\gamma$ and $\lambda$. Step 2, instead, consists just on Monte Carlo draws from the posterior of $(\boldsymbol{\Sigma_u}, \boldsymbol{\Phi})$ conditional on $\lambda,\gamma$ and $ \boldsymbol{\theta}$ which is the \textit{Normal-Inverse-Wishart} distribution in (\ref{condposterior})  that is:

            \begin{enumerate}
                \item [2.1)] $p(\boldsymbol{\Sigma_u} |\boldsymbol{\Phi}, \lambda,\boldsymbol{\theta}, \gamma,\boldsymbol{Y}) \sim \mathcal{IW} \left({\boldsymbol{\tilde{S}}}, \tilde{\nu}\right) $
                \item [2.2)] $p(vec(\boldsymbol{\Phi})|\boldsymbol{\Sigma_u}, \lambda,\boldsymbol{\theta}, \gamma,\boldsymbol{Y}) \sim \mathcal{N} (\boldsymbol{\tilde{\Phi}},\boldsymbol{\Sigma}_u \otimes \boldsymbol{\tilde{\Psi}})  $
            \end{enumerate}
In this approach, the time varying parameters and the variance covariance matrix are drawn in a single step from their \textit{Normal-Inverse-Wishart} conditional posterior distribution. As a last note, to let the estimated path of time varying coefficients $\boldsymbol{\Phi_1}, \ldots,\boldsymbol{\Phi_T}$ and the estimate of the hyper-parameter $\lambda$ less affected by the choice of the initial condition $\boldsymbol{\Phi_0}$ I assume that in equation (\ref{compact_states}) $\boldsymbol{\xi_1} \sim \mathcal{N}(\boldsymbol{0}, \boldsymbol{\Sigma_u} \otimes 5 (\sum_{t=1}^{T_{pre}}\boldsymbol{x}_t\boldsymbol{x}_t')^{-1} ) $ such that $\boldsymbol{\Phi_1} \sim \mathcal{N}(\boldsymbol{\hat{\Phi}_0}, \boldsymbol{\Sigma_u} \otimes 5 (\sum_{t=1}^{T_{pre}}\boldsymbol{x}_t\boldsymbol{x}_t')^{-1} ) $ where $T_{pre}$ is the size of a pre-sample of observations while $\boldsymbol{\hat{\Phi}_0}=0$.

\normalsize

\subsection{Allowing for distinct $\lambda_j$}\label{distinct_l}
For simplicity of exposition in the previous section I assumed a unique hyper-parameter i.e.  $\lambda$ determining the intrinsic time variation of the coefficients. This assumption can easily be replaced by assuming $\lambda_j$'s hyper-parameters with $j = 1, \ldots, k$ such that $\boldsymbol{\Omega}$ can be factorized as 
\begin{equation}\label{kron_transition}
 \boldsymbol{\Omega} = \boldsymbol{\Sigma} \otimes (\boldsymbol{\Lambda_k}'\boldsymbol{\Lambda_k})^{-1}    
\end{equation}
where $\boldsymbol{\Lambda_k} = diag(\lambda_1, \ldots, \lambda_k) $ and the Kronecker structure of the \textit{Normal-Inverse-Wishart} prior (\ref{jointprior}) is preserved. In terms of estimation, in the case we allow for regressor specific shrinkage hyper-parameters $\lambda_j$ for $j = 1, \ldots, k$ nothing changes conceptually, since the algorithm should just be adapted to draw all the $\lambda_j$'s hyper-parameters together with $\gamma$ in the second block of the random walk metropolis in Step 1. Formulas for the marginal likelihood and the conditional posterior distribution of $\boldsymbol{\Phi}$ and $\boldsymbol{\Sigma}_u$ can be found in the appendix \ref{sec:distinctlambda}. In general, keeping the assumption of a Kronecker structure for $\boldsymbol{\Omega}$ allows to have a closed form expression for the likelihood of $\gamma, \lambda$ and $\boldsymbol{\theta}$ marginally of the time varying parameters $\boldsymbol{\Phi_1,}, \ldots, \boldsymbol{\Phi_T}$ and of $\boldsymbol{\Sigma_u}$. In practice this allows to make inference on $\boldsymbol{\Phi}, \boldsymbol{\Sigma}_u$, the shrinking hyper-parameters $\gamma$ and $\lambda$ and also on the deep parameters from the theory $\boldsymbol{\theta}$ jointly as in \citet{DNS2004}.  Considering a more general structure for $\boldsymbol{\Omega}$, would in general lead to a non-conjugate structure of the prior. This would complicate inference on the parameters of the model, slowing down the convergence of the algorithm and produce more correlated draws, especially when drawing the deep parameters from the economic theory $\boldsymbol{\theta}$ jointly with the parameters of the TVP-VAR.  

\subsection{Time-varying error covariance}\label{sec:tvolatility}

As remarked by \citet{sims2002} commenting on \citet{CS2002}, a model with only time variation in parameters could mistakenly result in a substantial amount of time variation even
though the true data generating process only features stochastic volatility. In this section I show how to extend the theory coherent prior introduced in the previous section to a TVP-VAR featuring heteroskedasticity. More in the specific, to allow for time-varying error covariance, I replace the assumption $vec(\boldsymbol{U}) \sim \mathcal{N}( \boldsymbol{0}, \boldsymbol{\Sigma}_u \otimes \boldsymbol{I}_T)$ with 
\begin{equation}
vec(\boldsymbol{U}) \sim \mathcal{N}( \boldsymbol{0}, \boldsymbol{\Sigma}_u \otimes \boldsymbol{D} )
\end{equation}
where $\boldsymbol{D}$ is a $T \times T$ diagonal matrix storing the time-varying volatilities, namely $\boldsymbol{D} = diag(d_{1}, \ldots, d_{T})$. This class of VARs with flexible Kronecker error covariance structure is considered in \citet{jcc_kron}.  The likelihood of the model becomes
\begin{equation}\scriptsize
\begin{aligned}
p(\boldsymbol{Y}| \boldsymbol{\Phi}, \boldsymbol{\Sigma},\boldsymbol{D}) & =  (2\pi)^{-\frac{T N}{2}}|\boldsymbol{ \Sigma_u \otimes D}|^{-\frac{1}{2}} exp \left[-\frac{1}{2}tr\left(( \boldsymbol{\Sigma_u}^{-1})(\boldsymbol{Y'D^{-1}Y} - \boldsymbol{\Phi'X'D^{-1}Y} - \boldsymbol{Y'D^{-1}X\Phi} + \boldsymbol{\Phi'X'D^{-1}X\Phi} )\right)  \right]  \\
% & \propto (2\pi)^{-\frac{T N}{2}}|\boldsymbol{ \Sigma_u \otimes D}|^{-\frac{1}{2}} exp \left[-\frac{1}{2}tr\left(( \boldsymbol{\Sigma_u}^{-1})(\boldsymbol{Y'D^{-1}Y} - \boldsymbol{\Phi'X'D^{-1}Y} - \boldsymbol{Y'D^{-1}X\Phi} + \boldsymbol{\Phi'X'D^{-1}X\Phi} )\right)  \right]  \\
\end{aligned}
\end{equation}
This common specification for the volatility process was introduced in \citet{ccm2016}. For the elements in $\boldsymbol{D}$, i.e. the time varying volatilities, I follow \citet{jcc_kron} and consider the following\textit{ log-normal} prior
\begin{equation}
    log(d_t) =  \rho log(d_{t-1}) + \eta_t  \hspace{1cm} \eta_t \sim \mathcal{N} (0, \sigma^2_{\eta})
\end{equation}
The choice of the priors on $\rho$ and $\sigma^2_d$  also follows \citet{jcc_kron} with $\rho \sim \mathcal{N}(\rho_0,V_{\rho})\mathbbm{1}(|\rho|<1)$ and $\sigma^2_{\eta} \sim \mathcal{IG}(\nu_{\eta}, S_{\eta})$.\footnote{Following \citet{jcc_kron} I set $\rho_0 = 0.9$, $V_{\rho}= 0.2^2$ $\nu_{\eta} = 5$ and $S_{\eta} = 0.04$.} The theory coherent prior for $(\boldsymbol{\Phi},\boldsymbol{\Sigma}_u)$ is still the \textit{Normal-Inverse-Wishart} prior in equation (\ref{NIW_PRIOR}).  In order to draw from the posterior distribution of $(\boldsymbol{\Phi,\Sigma_u},\lambda,\gamma,\boldsymbol{\theta}, \boldsymbol{D}, \sigma^2_d, \rho )$ I consider an MCMC sampler that iterates the following steps:

    \begin{enumerate}
    \item Draw from $p(\lambda, \boldsymbol{\theta}, \gamma|\boldsymbol{Y}, \boldsymbol{D}) \propto p(\boldsymbol{Y}|\boldsymbol{D},\lambda,\boldsymbol{\theta},\gamma)p(\lambda, \boldsymbol{\theta}, \gamma)$. 
    \item Draw from $p(\boldsymbol{\Phi, \Sigma_u} | \lambda,\boldsymbol{\theta}, \gamma,\boldsymbol{D},\boldsymbol{Y})$ %\sim \mathcal{NIW} $
    \item Draw from $p(\boldsymbol{D},\rho, \sigma^2_{\eta}| \boldsymbol{\Phi}, \boldsymbol{\Sigma}_u,\boldsymbol{Y})$ \end{enumerate} 
This sampler generalizes the algorithm in \citet{ccm} to a VAR that features both time varying volatility and time varying parameters.  As in the algorithm used for the estimation of the homoskedastic model, in step 1  I resort to the random walk Metropolis to draw $\lambda, \boldsymbol{\theta}$ and $\gamma$. \footnote{Note that the conditional posterior distribution of $\lambda, \boldsymbol{\theta}$ and $\gamma$ is now combining the prior distribution $p(\lambda,\boldsymbol{\theta},\gamma)$ with $p(\boldsymbol{Y}|\boldsymbol{D},\lambda,\boldsymbol{\theta},\gamma)$. The formula $p(\boldsymbol{Y}|\boldsymbol{D},\lambda,\boldsymbol{\theta},\gamma)$ is available in the appendix \ref{sec:tctvp_var_sv} .}  In step 2, the formulas of the conditional posteriors of $(\boldsymbol{\Phi},\boldsymbol{\Sigma_u})$ modify as follows

\begin{equation}
\begin{aligned}\label{condposterior_SV}
  p(vec(\boldsymbol{\Phi})|\boldsymbol{\Sigma}_u, \boldsymbol{D}, \gamma, \lambda, \boldsymbol{\theta}, \boldsymbol{Y}) & \sim \mathcal{N}(vec(\boldsymbol{\tilde{\Phi}}),\boldsymbol{\Sigma}_u \otimes \boldsymbol{\tilde{\Psi}})  \\
  p(\boldsymbol{\Sigma}_u|\boldsymbol{D}, \gamma, \lambda, \boldsymbol{\theta}, \boldsymbol{Y}) & \sim \mathcal{IW} \left({\boldsymbol{\tilde{S}}}, \tilde{\nu}\right)
\end{aligned}
\end{equation}
\begin{equation}\label{postphi}
  vec(\boldsymbol{\tilde{\Phi}}) = vec(\left( \boldsymbol{X'D^{-1}X} 
+ \gamma\boldsymbol{\Gamma_{xx}(\theta)} + \lambda^2\boldsymbol{H_{Tk}'H_{Tk}}\right)^{-1}( \boldsymbol{X'D^{-1}Y} +  \gamma\boldsymbol{\Gamma_{xy}(\theta)} + \lambda^2\boldsymbol{H}_{Tk}'\boldsymbol{H}_{Tk}\boldsymbol{\underline{\Phi_{0}}}))   
\end{equation}
\begin{equation}\label{psisv}
     \boldsymbol{\tilde{\Psi}} =  \left(\boldsymbol{X'D^{-1}X} + \gamma\boldsymbol{\Gamma_{xx}(\theta)}  +  \lambda^2\boldsymbol{H_{Tk}'H_{Tk}}\right)^{-1}
\end{equation}
\begin{equation}\label{Ssv}
 \tilde{\boldsymbol{S}} = \boldsymbol{Y'D^{-1}Y} + \underline{\underline{\boldsymbol{S}}} + \boldsymbol{\underline{\underline{\Phi}}}'\boldsymbol{\underline{\underline{\Psi}}}^{-1}\boldsymbol{\underline{\underline{\Phi}}} - \boldsymbol{\tilde{\Phi}}'\boldsymbol{\tilde{\Psi}}^{-1}\boldsymbol{{\tilde{\Phi}}}   
\end{equation}
\begin{equation}\label{nusv}
\tilde{\nu} = \underline{\underline{\nu}} + T 
\end{equation}
Finally, in step 3, to draw time varying volatilities stored in $\boldsymbol{D}$, I compute an independence Metropolis step, following \citet{ccm}. Finally, the draws for $\alpha$ and $\sigma^2_{\eta}$ are standard draws respectively from the \textit{Normal} and the \textit{Inverse Gamma} distributions.

\section{Forecasting with the TC-TVP-VAR}\label{oos}

In this section I consider the problem of forecasting the rate of growth of GDP, the inflation rate and the Fed Fund rate using a trivariate TVP-VAR model. In the specific,  I estimate a trivariate TVP-VAR for the US economy using data from 1970 up to 2019  and I compare the forecast accuracy of a standard TVP-VAR model to the forecasts from a TC-TVP-VAR. 

\subsection{Theory from a Small scale New-Keynesian model}
In the TC-TVP-VAR I exploit the New-Keynesian model in \citet{DNS2004} to parametrize the theory coherent prior. The conceptual framework commonly denoted as the 3-equation New-Keynesian model constitutes the nucleus of Michael Woodford's book ``Interest and Prices" \citep{woodford2003interest} and underpins most of modern monetary macroeconomics models.\footnote{See \citet{DNS2004} for the more in depth details on the New-Keynesian Model.} More specifically, the structural model is composed by an IS curve (\ref{IS}), a New-Keynesian Phillips curve (\ref{NKPC}), a monetary policy rule (\ref{MP}) and two equations  that describe the dynamics for the log-deviation from the steady state of technological process (\ref{z_t}) and government spending (\ref{g_t}), namely

%&composed by an IS curve (\ref{IS}), a New-Keynesian Phillips curve (\ref{NKPC}) a monetary policy rule (\ref{MP}) and a dynamics for the growt

\begin{equation}\label{IS}
    \hat{y_t} =  E_t[\hat{y}_{t+1}] - \frac{1}{\tau} \left( \hat{R}_t - E_t[\hat{\pi}_{t+1}]\right) + (1 - \rho_g)\hat{g}_t + \frac{\rho_z }{\tau}\hat{z}_t  \\
    \end{equation}
    \begin{equation}\label{NKPC}
    \hat{\pi}_t =   \frac{\tilde{\gamma}}{r^*}E_{t}[\hat{\pi}_{t+1}]+ \kappa(\hat{y}_t - \hat{g}_t) \\
    \end{equation}
    \begin{equation}\label{MP}
     \hat{R}_t = \rho_R \hat{R}_{t-1} + (1 - \rho_R)(\psi_1\hat{\pi}_t + \psi_2\hat{y}_t) + \varepsilon_{R,t}  \hspace{0.7cm}  \varepsilon_{R,t} \sim \mathcal{N}(0,\sigma_{R}) \\
    \end{equation}
    \begin{equation}\label{z_t}
     \hat{z}_t =   \rho_z\hat{z}_{t-1} + \varepsilon_{z,t}  \hspace{4.2cm}  \varepsilon_{z,t} \sim \mathcal{N}(0,\sigma_{z})\\       
    \end{equation}
    \begin{equation}\label{g_t}
     \hat{g}_t =   \rho_g \hat{g}_{t-1} + \varepsilon_g   \hspace{4.5cm}  \varepsilon_{g,t} \sim \mathcal{N}(0,\sigma_{g})\\    
    \end{equation}
The population moments needed to parametrize the prior are derived from the state-space representation of the New-Keynesian model obtained by solving the system of non-linear rational expectation equations. In particular, the non-linear rational expectation equations are solved using the method based on matrix eigenvalue decomposition by \citet{sims2002} leading to a solution which has the form 
    \begin{equation}\label{sp_states11}
      \boldsymbol{s_t} =  \mathcal{T}(\boldsymbol{\theta})\boldsymbol{s_{t-1}} + \mathcal{R}(\boldsymbol{\theta})\boldsymbol{\epsilon}_t
    \end{equation}
that is complemented with the set of observation equations
\begin{equation} \label{sp_observation1}
     \boldsymbol{y}_{t} = \mathcal{D}+  \mathcal{B}\boldsymbol{s_t}
\end{equation}
which look like:
\begin{equation}\label{obs1}
  YGR_{t} =  ln(\tilde{\gamma}) + \Delta\hat{y_t} + \hat{z}_t
\end{equation}    
\begin{equation}\label{obs2}
INFL_{t} =  ln(\pi^*) + \hat{\pi}_t  
\end{equation}
\begin{equation}\label{obs3}
ln(INT_t) = 4ln(r^*) + 4ln(\pi^*) + 4\hat{R}_t
\end{equation}
%     %where $\boldsymbol{\Omega}(\boldsymbol{\theta})= diag(\sigma_R^,\sigma_g^2,\sigma_z^2 )$  
and relate the unobservable latent states in (\ref{sp_states11}) to the observed time series of output growth, inflation rate and the Fed Fund rate. Equations  (\ref{sp_states11}) and (\ref{sp_observation1}) are used to derive the population moments needed to parametrize the prior. The population moments $\boldsymbol{\Gamma_{xx}}(\boldsymbol{\theta})$, $\boldsymbol{\Gamma_{xy}}(\boldsymbol{\theta})$, $\boldsymbol{\Gamma_{yy}}(\boldsymbol{\theta})$ are derived conditioning on $\boldsymbol{\theta}$, the vector of the deep parameters of the NK model
\begin{equation}
\boldsymbol{\theta} = [ln(\tilde{\gamma}), ln(\pi^*), ln (r^*), \kappa, \tau, \psi_1, \psi_2,\rho_R, \rho_g, \rho_z, \sigma_R^2, \sigma_g^2, \sigma_z^2]    
\end{equation}

\subsection{Forecast comparison}
Table \ref{tab:forecast_accuracy_US} shows the comparison of the forecasts from a standard TVP-VAR model and the TC-TVP-VAR model. The forecasting exercise is designed such that I compute the recursive one quarter, two quarters, and one year ahead forecasts starting from 1985-Q1 up to 2019-Q4.\footnote{Details on the data can be found in the appendix \ref{data_ex1}} To compare relative point forecast accuracy, in the table I report the Root Mean Squared Error (RMSE) while for evaluating density forecast accuracy I report the average Cumulative Ranked Probability Scores (CRPS). In the table I also include the results concerning the forecasts from a constant parameters VAR with flat prior and a constant parameters Bayesian VAR (BVAR) with Minnesota type of prior.\footnote{Appendix \ref{competing} the reports details on the competing forecasting models.} Overall, the TC-TVP-VAR provides the most accurate point and density forecasts for both output growth and inflation rate, outperforming the standard TVP-VAR model at all the horizons considered. For forecasting output growth, the standard TVP-VAR performs poorly relative to the TC-TVP-VAR, but also to the constant parameters BVAR with Minnesota prior, suggesting that the model tends to fit some noise in the time series of output growth. In line with the previous forecasting literature, allowing for time variation of the parameters of the VAR is important for obtaining accurate forecasts of the inflation rate, as the standard TVP-VAR outperforms both the VAR with flat prior and the BVAR with the Minnesota prior. However, economic shrinkage is helpful to obtain more reliable point and density forecasts when modelling the time variation of the coefficients. Indeed the TC-TVP-VAR outperforms the standard TVP-VAR at all the horizons. As a caveat, the standard Minnesota prior centering the autoregressive coefficients on a random walk process outperforms the other competitors, including the TC-TVP-VAR, for forecasting the Fed Fund Rate. This result is consistent with the results of the forecasting exercise in \citet{DNS2004} which use the same small scale NK model to parametrize a prior for a constant parameters VAR. 
Table \ref{tab:sv} presents the same metrics, comparing forecasts from a standard TVP-VAR model with stochastic volatility, as described in Section \ref{sec:tvolatility}, to those from the theory-coherent TVP-VAR model with stochastic volatility. In terms of both point and density forecast accuracy, and across all horizons, the theory-coherent model consistently delivers more accurate predictions of GDP growth and inflation. 
\begin{table}[htp]
\caption{Point and density forecast accuracy 1985Q1-2019Q4}
\scriptsize
\centering
\begin{tabular}{l|ll|ll|ll}
 & \multicolumn{2}{c|}{\textbf{GDP growth}} & \multicolumn{2}{c|}{\textbf{Inflation rate}} & \multicolumn{2}{c}{\textbf{Interest Rate}} \\
\scriptsize
& \textbf{RMSE}  & \textbf{CRPS} & \textbf{RMSE} & \textbf{CRPS} & \textbf{RMSE} & \textbf{CRPS}  \\
\hline\hline
a) \textit{One quarter ahead } \\
\hline 
\textbf{VAR flat prior}              & 0.4417 & 0.3703 & 0.3451 & 0.2877 & $\mathbf{0.0402}$ & 0.0954 \\
\textbf{B-VAR min}        & 0.3422 & 0.3584 & 0.2854 & 0.2637 & 0.0403 & $\mathbf{0.0936}$ \\
\textbf{Standard TVP-VAR} & 0.4283 & 0.3853 & 0.3362 & 0.2682 & 0.0451 & 0.1096 \\
\textbf{TC-TVP-VAR}       & $\mathbf{0.3027}$ & $\mathbf{0.3485}$ & $\mathbf{0.2715 }$& \textbf{0.2461} & 0.0608 & 0.1312 \\
\hline \hline
b) \textit{Two quarters ahead}  \\
\hline
\textbf{VAR flat prior}              & 0.5045 & 0.3885 & 0.4771 & 0.3423 & 0.1384 & 0.1784 \\
\textbf{B-VAR min}        & 0.3990 & 0.3909 & 0.3819 & 0.3110 & $\mathbf{0.1303}$ & $\mathbf{0.1709}$ \\
\textbf{Standard TVP-VAR} & 0.5131 & 0.4151 & 0.3362 & 0.2911 & 0.1548 & 0.2037 \\
\textbf{TC-TVP-VAR}       & $\mathbf{0.3340}$ & $\mathbf{0.3668}$ & $\mathbf{0.3183}$ & $\mathbf{0.2722}$ & 0.1689 & 0.2096 \\
\hline \hline 
c) \textit{One year ahead } \\
\hline
\textbf{VAR flat prior}              & 0.4278 & $\mathbf{0.3728}$ & 0.4965 & 0.3675 & 0.4475 & 0.3385 \\
\textbf{B-VAR min}        & 0.4037 & 0.4065 & 0.4099 & 0.3393 & $\mathbf{0.4142}$ & $\mathbf{0.3229}$ \\
\textbf{Standard TVP-VAR} & 0.6472 & 0.4775 & 0.2855 & 0.3039 & 0.5665 & 0.3854 \\
\textbf{TC-TVP-VAR}       & $\mathbf{0.3617}$ & 0.3887 & $\mathbf{0.2758}$ & $\mathbf{0.2802}$ & 0.5156 & 0.3593 \\
\hline \hline

\end{tabular}
\centering
\vspace{0.25cm}\hspace{2cm}\parbox{1.2\textwidth}{\scriptsize Notes: The Table reports the Root Mean Squared Error (RMSE) and the average  \\ Cumulative Ranked Probability Scores (CRPS). In bold the best model according to \\ each forecast metric. }
\label{tab:forecast_accuracy_US}
\end{table}

\begin{table}[htp]
\caption{Point and density forecast accuracy with stochastic volatility 1985Q1-2019Q4}\label{tab:sv}
\scriptsize
\centering
\begin{tabular}{l|ll|ll|ll}
 & \multicolumn{2}{c|}{\textbf{GDP growth}} & \multicolumn{2}{c|}{\textbf{Inflation rate}} & \multicolumn{2}{c}{\textbf{Interest Rate}} \\
\scriptsize
& \textbf{RMSE}  & \textbf{CRPS} & \textbf{RMSE} & \textbf{CRPS} & \textbf{RMSE} & \textbf{CRPS}  \\
\hline\hline
a) \textit{One quarter ahead } \\
\hline 
\textbf{TVP-VAR-SV}       & 0.4987 & 0.3763 & 0.2997 & 0.2667 & 0.0426 & 0.1033 \\
\textbf{TC-TVP-VAR-SV}    & \textbf{0.4508} & \textbf{0.3714} & \textbf{0.2699} & \textbf{0.2506} & \textbf{0.0421} & \textbf{0.1004} \\
\hline \hline
b) \textit{Two quarters ahead}  \\
\hline
\textbf{TVP-VAR-SV}       & 0.5062 & 0.3830 & 0.3030 & 0.2801 & 0.1551 & 0.1948 \\
\textbf{TC-TVP-VAR-SV}    & \textbf{0.4530} & \textbf{0.3769} & \textbf{0.2745} & \textbf{0.2624} & \textbf{0.1478 }& \textbf{0.1899} \\
\hline \hline 
c) \textit{One year ahead } \\
\hline
\textbf{TVP-VAR-SV}       & 0.5313 & 0.4145 & 0.2760 & 0.2804 & \textbf{0.5511 }& \textbf{0.3674} \\
\textbf{TC-TVP-VAR-SV}    & \textbf{0.4923} & \textbf{0.4113} & \textbf{0.2698} & \textbf{0.2783} & 0.5648 & 0.3744\\
\hline \hline
\end{tabular}
\centering
\vspace{0.25cm}\hspace{2cm}\parbox{1.2\textwidth}{\scriptsize Notes: The Table compares the Root Mean Squared Error (RMSE) and the average \\ Cumulative Ranked Probability Scores (CRPS). In bold the best model according to  \\each forecast metrics.}
\end{table}

\clearpage
\section{Response analysis at the ZLB }\label{sec:application}
\begin{figure}[H]
    \centering
    \includegraphics[scale=0.15]{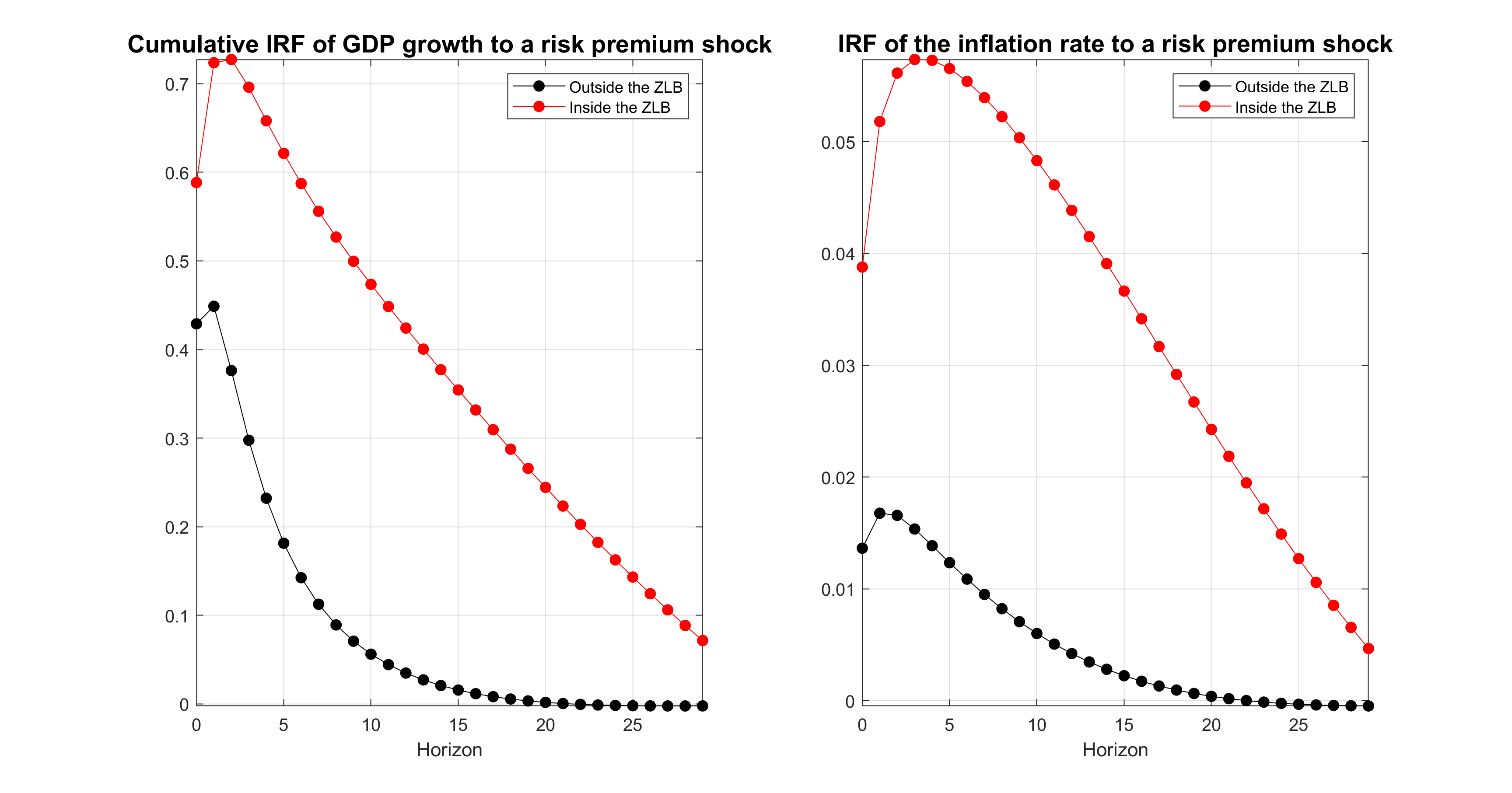}
    \caption{\small{The two plots show the cumulative response of output growth (on the left) and the response of the inflation rate (on the right) to a risk premium shock inside and outside the ZLB period in the New-Keynesian model.}}
    \label{fig:irfs_dsge}
\end{figure}

TVP-VARs are extensively used in applied research not only to make forecasts, but also to infer changes in the response of the economy to macroeconomic shocks. In this section,  I show that the proposed shrinkage prior can be useful also to enhance inference on the impulse response functions estimated from a TVP-VAR. Recent studies in empirical macroeconomics have used TVP-VARs to assess whether the US economy's performance was affected by a binding ZLB constraint \citep{galigambetti,lubikbenati}. As a matter of fact, according to a standard New-Keynesian model, the economy is expected to exhibit different responses when the ZLB constraint is in effect. For example, the model predicts a distinct response of output and inflation following both demand and supply shocks when the conventional stabilizing monetary policy response to aggregate shocks is constrained as a consequence of the Federal Funds Rate  hitting the ZLB. Figure \ref{fig:irfs_dsge} makes this point, by showing the responses to a pure demand shock - the risk premium shock \footnote{The risk premium shock in \citet{SWauters} is an exogenous term which affects the intertemporal margins entering both the consumption and investment Euler equation. In contrast to a discount factor shock, the risk premium shock helps
explaining the co-movement of consumption and investment. More details on the risk premium shocks are presented in the Appendix \ref{sec:rpshocks}.} - in the \citet{SWauters} model version considered in \citet{delnegroschorfheide2005}.\footnote{In the model, the ZLB period is treated as in \citet{delnegroschorfheide2005}, and more details on the model will be explained in the next subsection.} The figure plots the cumulative response of output growth and the response of the inflation rate to an exogenous risk premium shock that reduces households' required return of assets, decreasing firms' cost of capital. As expected, when monetary policy is constrained by the policy rate hitting the zero lower bound, the response of both output and inflation to the risk premium shock is magnified on impact with the effects of the shock taking much more quarters to be reabsorbed by the economy.
Despite the sharp difference in the propagation of the shocks in the economy inside and outside the zero lower bound period predicted by a standard NK model, empirical evidences investigating this issue are mixed and many studies do not find substantial evidence supporting a different response of the US economy during the zero lower bound period. For example, \citet{galigambetti}, support the \textit{irrelevance hypothesis} i.e. the hypothesis that the economy's performance has not been affected by a binding ZLB constraint, embracing the view that unconventional monetary policy have been effective at getting around the zero lower bound (ZLB) constraint. \citet{lubikbenati} recently showed that, given the short length of the zero-lower-bound period, inference based on a standard TVP-VAR where the time varying parameters are regarded as slow moving stochastic processes, doesn't allow to capture the changing relationship among the macroeconomic variables during the ZLB period predicted by the New-Keynesian model. In what follows, based on a simulation study, I also find that a standard TVP-VAR struggles to detect the change in the responses of the economy in the ZLB period generated by a NK model. I show that the TC-TVP-VAR can in principle be used to solve this inferential issue and therefore to recover the distinct response of the economy during the zero lower bound period.  %Afterwards, I estimate the TC-TVP-VAR on true US data, to understand whether the data provide enough evidences supporting the view that the US economy's performance was indeed affected by a binding ZLB constraint as predicted by the standard New-Keynesian model. 

\subsection{Theory from a Medium scale NK model with the ZLB and forward guidance}

The New-Keynesian model is the version of the \citet{SWauters} model considered in \citet{delnegroschorfheide2005}. The model features price and wage stickiness,
investment adjustment costs, habit formation in consumption and 7 shocks being monetary policy shocks, technology shocks, price mark-up shocks, wage mark-up shocks, risk premium shocks, fiscal policy shock and shocks to the marginal efficiency of capital. In the model the monetary policy rule accounts for ZLB period and forward guidance.\footnote{The log-linearized equilibrium conditions of the model can be found in the appendix \ref{sec:eq_cond} The model is labelled ''SW" in \citet{delnegroschorfheide2005} and assumes a constant inflation target.} More specifically, when accounting for the zero lower bound and forward guidance the solution of the model implies a state state space representation which exhibits time varying coefficients. The solution method follows the approach developed by \citet{kulishcagliarini} for linear stochastic rational expectations models in the face of a finite sequence of anticipated structural changes. In particular, it is assumed that at a given period $t$, agents expect the nominal interest rate to be at the ZLB for $\bar{H}$ periods.  That is, the monetary policy rule
\begin{equation}\label{MPrule}
    R_t = \rho_RR_{t-1} + (1 - \rho_R) \left(\psi_1 (\pi_t - \pi_t^*) + \psi_2(y_t - y_t^f) \right) + \psi_3\left((y_t - y_t^f) - (y_{t-1} - y_{t-1}^f)\right) + r_t^m
\end{equation}
becomes 
\begin{equation}
    R_{\tau} = -R_{*} \hspace{1cm}   
\end{equation}
for $\tau = t, \ldots, t + \bar{H}$ while it is determined by (\ref{MPrule}) for $\tau > t + \Bar{H} $. This implies that the equilibrium conditions
\begin{equation}
    \boldsymbol{\Gamma}_{2,\tau}E_{\tau} [\boldsymbol{{s}}_{\tau+1}] +  \boldsymbol{\Gamma}_{0,\tau}\boldsymbol{{s}}_{\tau} = \boldsymbol{\Gamma}_{c,\tau} + \boldsymbol{\Gamma}_{1,\tau}\boldsymbol{{s}}_{\tau-1} + \boldsymbol{\Psi_{\tau}}\boldsymbol{\varepsilon}_{\tau}
\end{equation}
differ over time depending on whether $\tau \leq t + \Bar{H}$, with the row corresponding to the policy rule differing across $\tau$. Following \citet{kulishcagliarini} the solution for the linear rational expectations model under anticipated structural variations takes the form
\begin{equation}\label{sp_states_tau}
      \boldsymbol{s_{\tau}} =  \mathcal{C}_{\tau}^{(\tau, \Bar{H})} + \mathcal{T}_{\tau}^{(\tau, \Bar{H})}\boldsymbol{s_{\tau-1}} + \mathcal{R}_{\tau}^{(\tau, \Bar{H})}\boldsymbol{\epsilon_{\tau}}
\end{equation} 
where  $\mathcal{C}_{\tau}^{(\tau, \Bar{H})}$  $ \mathcal{T}_{\tau}^{(\tau, \Bar{H})}$ and $\mathcal{R}_{\tau}^{(\tau, \Bar{H})}$ are computed by recursion
\begin{equation}
  \mathcal{C}_{\tau}^{(\tau, \Bar{H})} =  ( \boldsymbol{\Gamma}_{2,\tau} \mathcal{T}_{\tau+1}^{(\tau, \Bar{H})} + \boldsymbol{\Gamma}_{0,\tau})^{-1} (\boldsymbol{\Gamma}_{c,\tau} - \boldsymbol{\Gamma}_{2,\tau}\mathcal{C}_{\tau+1}^{(\tau, \Bar{H})}  )
\end{equation}
\begin{equation}
  \mathcal{T}_{\tau}^{(\tau, \Bar{H})} =  ( \boldsymbol{\Gamma}_{2,\tau} \mathcal{T}_{\tau+1}^{(\tau, \Bar{H})} + \boldsymbol{\Gamma}_{0,\tau})^{-1} \boldsymbol{\Gamma}_{1,\tau}
\end{equation}
\begin{equation}
  \mathcal{R}_{\tau}^{(\tau, \Bar{H})} =  ( \boldsymbol{\Gamma}_{2,\tau} \mathcal{T}_{\tau+1}^{(\tau, \Bar{H})} + \boldsymbol{\Gamma}_{0,\tau})^{-1}\boldsymbol{\Psi_{\tau}} 
\end{equation}
starting from $\mathcal{T}_{t+\Bar{H} + 1}^{(\tau, \Bar{H})} = \mathcal{T}$ , $\mathcal{C}_{t+1+\Bar{H}}^{(\tau, \Bar{H})}=0$ and
where the superscript $(t,\Bar{H})$ is used  to indicate that the solution is obtained under the assumption that the announcement of zero interest rates for a duration of $\Bar{H}$ periods was made in period $t$. Following \citet{delnegroschorfheide2005} to measure the number of quarters $\bar{H}$ that the Federal Funds Rate is expected to remain at the ZLB I exploit information based on the overnight index swap (OIS) rates. In particular I identify the ZLB period as the quarters in which the OIS rate is lower then 0.35. This classification leads to the same ZLB period considered in \citet{lubikbenati} and \citet{galigambetti} namely 2009Q1-2015Q3 (28 quarters). Following \citet{chencurdiaferrero} I assume that the number of quarters such that the policy rate is expected to stay fixed, is at most equal to four. According to the model's solution, the matrices $\mathcal{C}_{t}^{(\tau, \Bar{H})},\mathcal{T}_{t}^{(\tau, \Bar{H})},\mathcal{R}_{t}^{(\tau, \Bar{H})}  $ characterize the transition equation of a time varying coefficients state space model
\begin{equation} \label{sp_observation}
     \boldsymbol{y}_{t} = \mathcal{D}+  \mathcal{B}\boldsymbol{s_t} + \boldsymbol{v}_t
\end{equation}
\begin{equation} \label{sp_states}
     \boldsymbol{s}_{t} = \mathcal{C}_t+  \mathcal{T}_t\boldsymbol{s_{t-1}} + \mathcal{R}_t\boldsymbol{\epsilon}_t
\end{equation}
where the constant matrices $\mathcal{D}$, $\mathcal{B}$ and time varying matrices  $\mathcal{C}_{t}$, $\mathcal{T}_{t}$, $\mathcal{R}_{t}$ depend on the structural parameters $\boldsymbol{\theta}$ while $\boldsymbol{v}_t$ is a measurement error. \footnote{Details for the observation equations are in the appendix \ref{sec:eq_cond} together with the list of the deep structural parameters $\boldsymbol{\theta}$ (appendix \ref{secpriors_dsge_medium}).} Therefore, while the space representation of the model's solution features time varying coefficients, the deep structural parameters of the NK model $\boldsymbol{\theta}$ are constant. Defining $\underline{T}^{zlb}$ the first period  inside the ZLB and $\bar{T}^{zlb}$ the last period inside the ZLB,  for $t < \underline{T}^{zlb}$ and $t > \bar{T}^{zlb}$ we have $\mathcal{R}_t = \bar{\mathcal{R}}$ and $\mathcal{T}_t = \bar{\mathcal{T}}$ and $\mathcal{C}_t = \boldsymbol{0}$. Therefore the state space model becomes

\begin{equation}
 \boldsymbol{y}_t =   \mathcal{D} + \mathcal{B}\boldsymbol{s}_t 
\end{equation}
\begin{equation}
 \boldsymbol{s}_t =    \bar{\mathcal{T}}\boldsymbol{s}_{t-1} + \bar{\mathcal{R}}\boldsymbol{\varepsilon}_t  
\end{equation}
To parameterize the prior using the moments from this state space representation we proceed as follows. For $ t = 1, \ldots, \underline{T}^{zlb}-1$ and $  t = \bar{T}^{zlb}+1, \ldots, T$ the moments $\boldsymbol{\Gamma_{xx,t}} = \E [\boldsymbol{x_tx_t'}|\boldsymbol{\theta}]$, $\boldsymbol{\Gamma_{xy,t}} = \E [\boldsymbol{x_ty_t'}|\boldsymbol{\theta}]$ and $\boldsymbol{\Gamma_{yy,t}} \equiv \E [\boldsymbol{y_ty_t'}|\boldsymbol{\theta}]$ are computed assuming $\E[\boldsymbol{s}_t\boldsymbol{s}_t'] = \E[\boldsymbol{s}_{t-1}\boldsymbol{s}_{t-1}']$ and solving the Lyapunov equation
\begin{equation}
    \E[\boldsymbol{s_t}\boldsymbol{s_t}'] = \bar{\mathcal{T}}\E[\boldsymbol{s_{t-1}}\boldsymbol{s_{t-1}}']\bar{\mathcal{T}}' + \bar{\mathcal{R}}\boldsymbol{\Omega}\bar{\mathcal{R}}'
\end{equation}
For the periods inside the ZLB, i.e. for $ \underline{T}^{zlb} \leq t \leq \bar{T}^{zlb}$
,  $\E[\boldsymbol{s}_t\boldsymbol{s}_t']$ and the implied population moments are computed recursively according to the low of motion implied by (\ref{sp_observation}) and (\ref{sp_states}). These moments are then used to compute $\boldsymbol{\Gamma_{xx}(\theta)}$, $\boldsymbol{\Gamma_{xy}(\theta)}$, and $\boldsymbol{\Gamma_{yy}(\theta)}$  as detailed in \ref{popmoments} and parameterize the prior. 

\subsection{Simulation study}\label{simulation_study}
\begin{figure}[H]
    \centering
    \includegraphics[scale=0.2]{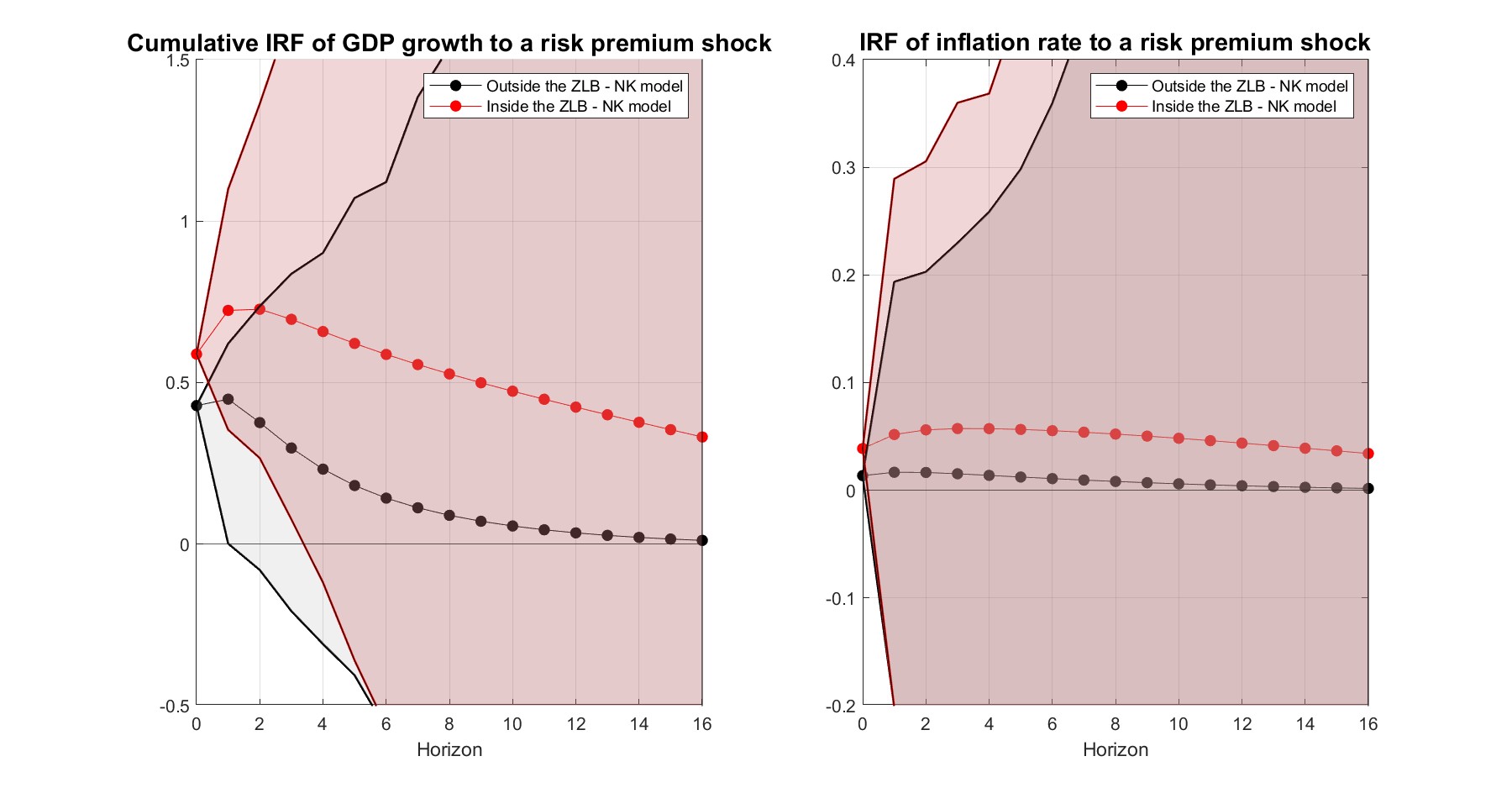}
    \caption{\small{Estimated $90^{th}-10^{th}$ credible sets of the cumulative response of output growth (on the left) and of the response of the inflation rate (on the right) to a risk premium shock inside (shaded red) and outside (shaded grey) the ZLB period by a standard TVP-VAR on data simulated from the NK model.}}
    \label{fig:irf_standardtvp}
\end{figure}

As anticipated above and shown in figure \ref{fig:irfs_dsge}, the NK model predicts a distinct response of the economy to a risk premium shock inside and outside the ZLB period. Conditioning on the vector of deep parameters of the NK model, I simulate data from the state space representation (\ref{sp_observation}) and (\ref{sp_states}).\footnote{Parameters are calibrated according to the posterior mode in \citet{delnegroschorfheide2005}} In the simulation I consider artificial samples with $T=139 $ mimicking quarterly observations for the period 1985Q1-2019Q3. The length of the ZLB period is 28 quarters, covering the period 2009Q1-2015Q3.  I estimate a standard TVP-VAR model \citep{chan2009efficient} on the simulated data to understand whether the model is able to recover the change in the response of the economy during the ZLB period generated by the NK model.\footnote{The standard TVP-VAR model is the model in \citet{chan2009efficient}. To estimate the model I use the MATLAB codes kindly made available by Joshua Chan on his personal website.} Figure \ref{fig:irf_standardtvp} shows the estimated responses of output and inflation to a risk premium shock obtained from a standard TVP-VAR.  The figure plots the responses of output growth (cumulative) and the inflation rate to a one standard deviation risk premium shock in two reference dates, one outside and the other one inside the ZLB period. In order to identify the shocks in the structural TVP-VAR I exploit the true impact matrix of the NK model given by
    \begin{equation}\label{Adsge}
    \frac{\partial \boldsymbol{y_t}}{\partial\boldsymbol{\epsilon}'_t} = \underbrace{\mathcal{B}(\boldsymbol{\theta})\boldsymbol{\mathcal{R}_t}(\boldsymbol{\theta})\boldsymbol{\Omega}(\boldsymbol{\theta})}_{\text{$\boldsymbol{A_{0,t}}(\boldsymbol{\theta})^{NK}$}}
    \end{equation}
where $\boldsymbol{\Omega}(\boldsymbol{\theta})= diag(\sigma_g^2, \sigma_b^2, \sigma_{\mu}^2, \sigma_z^2, \sigma_{\lambda_f}^2, \sigma_{\lambda_w}^2, \sigma_{r}^2)$  is the diagonal matrix containing the volatility of the structural shocks. The idea is that, by conditioning on the correct identification of risk premium shocks, we aim to determine whether the standard TVP-VAR model can reveal a distinct propagation of these shocks both inside and outside the ZLB. As figure  \ref{fig:irf_standardtvp} shows, inference based on a standard TVP-VAR struggles to provide convincing evidences supporting a distinct response of both output growth and the inflation rate to a risk premium shock inside and outside the ZLB period. As a matter of fact, the estimates of the impulse responses are so imprecise that the model doesn't allow to detect any change in the response of the economy inside and outside the ZLB. \footnote{As shown in the appendix in figure \ref{multiple_sim}, this result is not driven by this specific simulated sample of artificial observations.}

\begin{figure}[htp!]
\centering
% First row of two figures
\begin{subfigure}{.5\linewidth}
  \includegraphics[width=\linewidth]{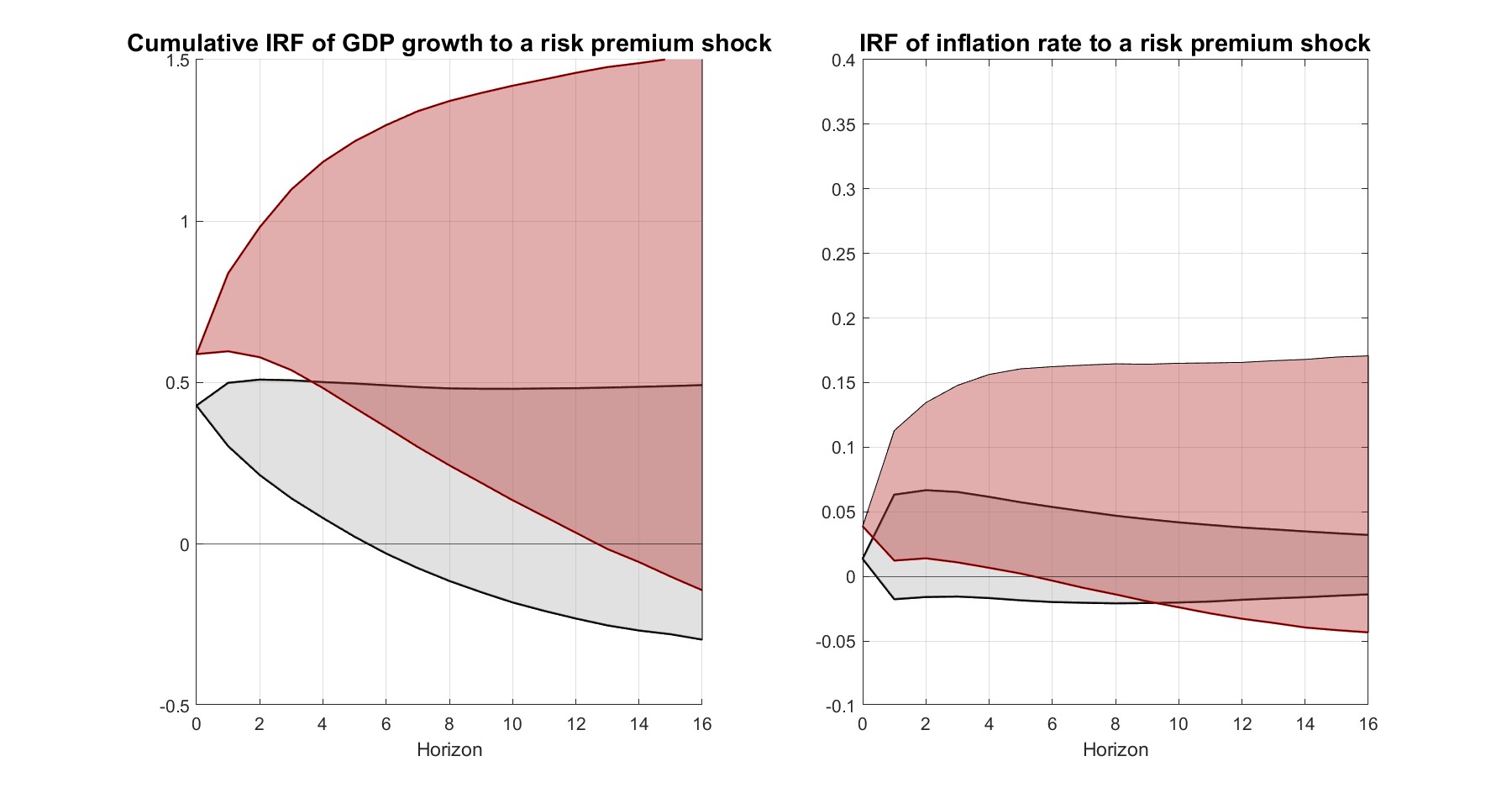}
  \caption{$\gamma = 10$}
  \label{panel_a}
\end{subfigure}\hfill % Fills the space between the subfigures in the same row
\begin{subfigure}{.5\linewidth}
  \includegraphics[width=\linewidth]{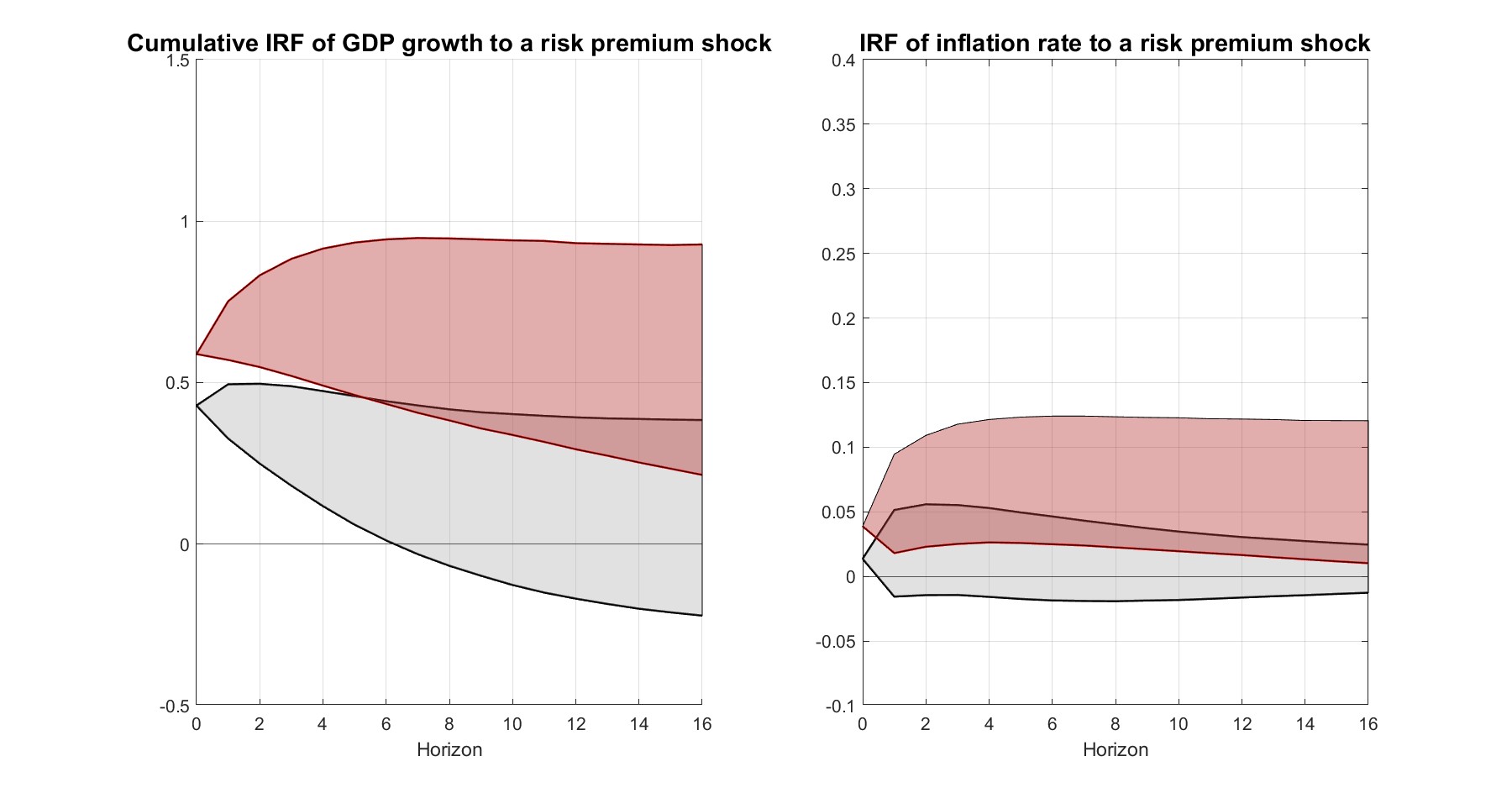}
  \caption{$\gamma = 100$}
  \label{panel_b}
\end{subfigure}

\medskip % Creates vertical separation between the two rows
% Second row of two figures
\begin{subfigure}{.5\linewidth}
  \includegraphics[width=\linewidth]{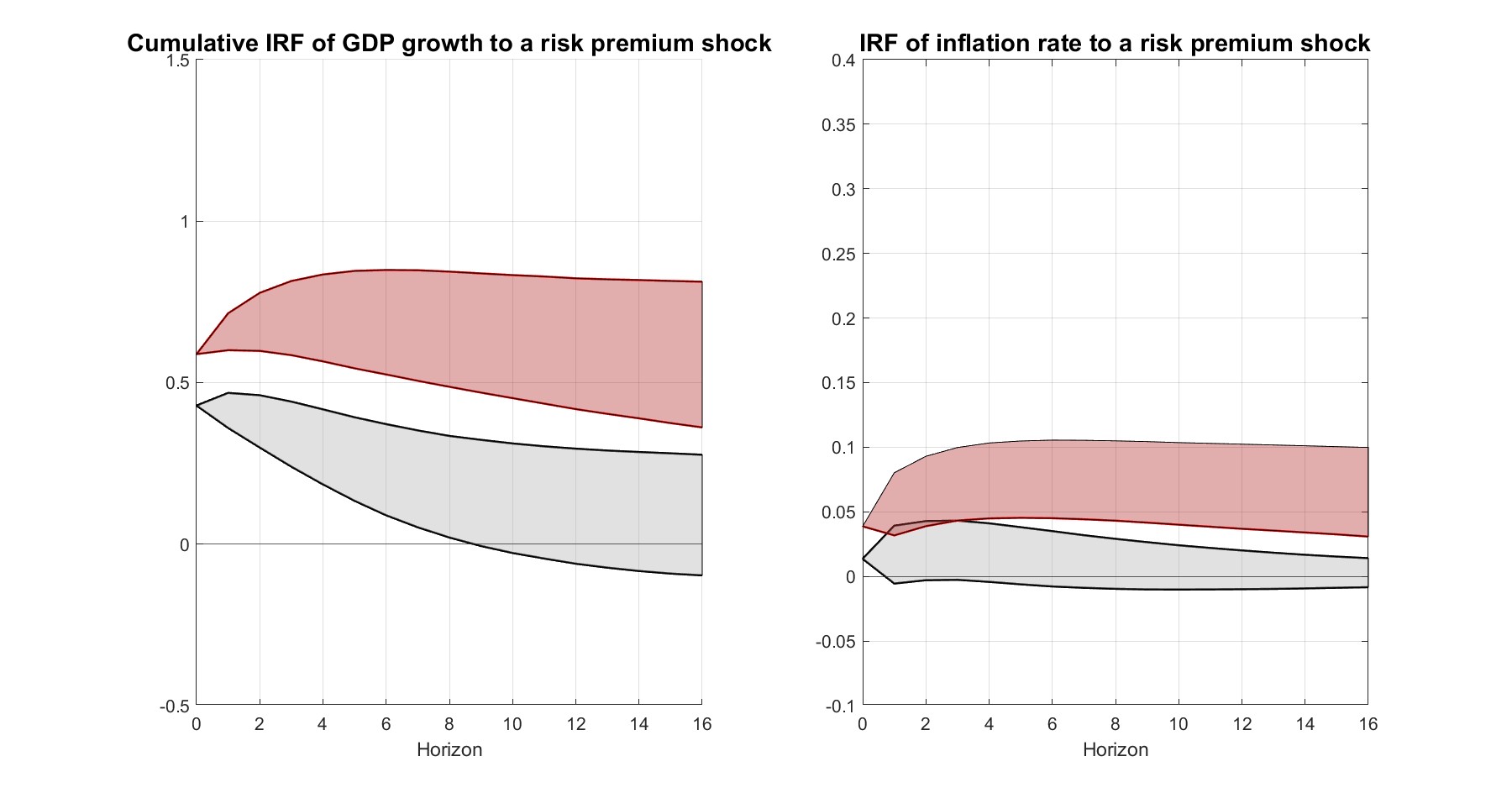}
  \caption{$\gamma = 300$}
  \label{panel_c}
\end{subfigure}\hfill
\begin{subfigure}{.5\linewidth}
  \includegraphics[width=\linewidth]{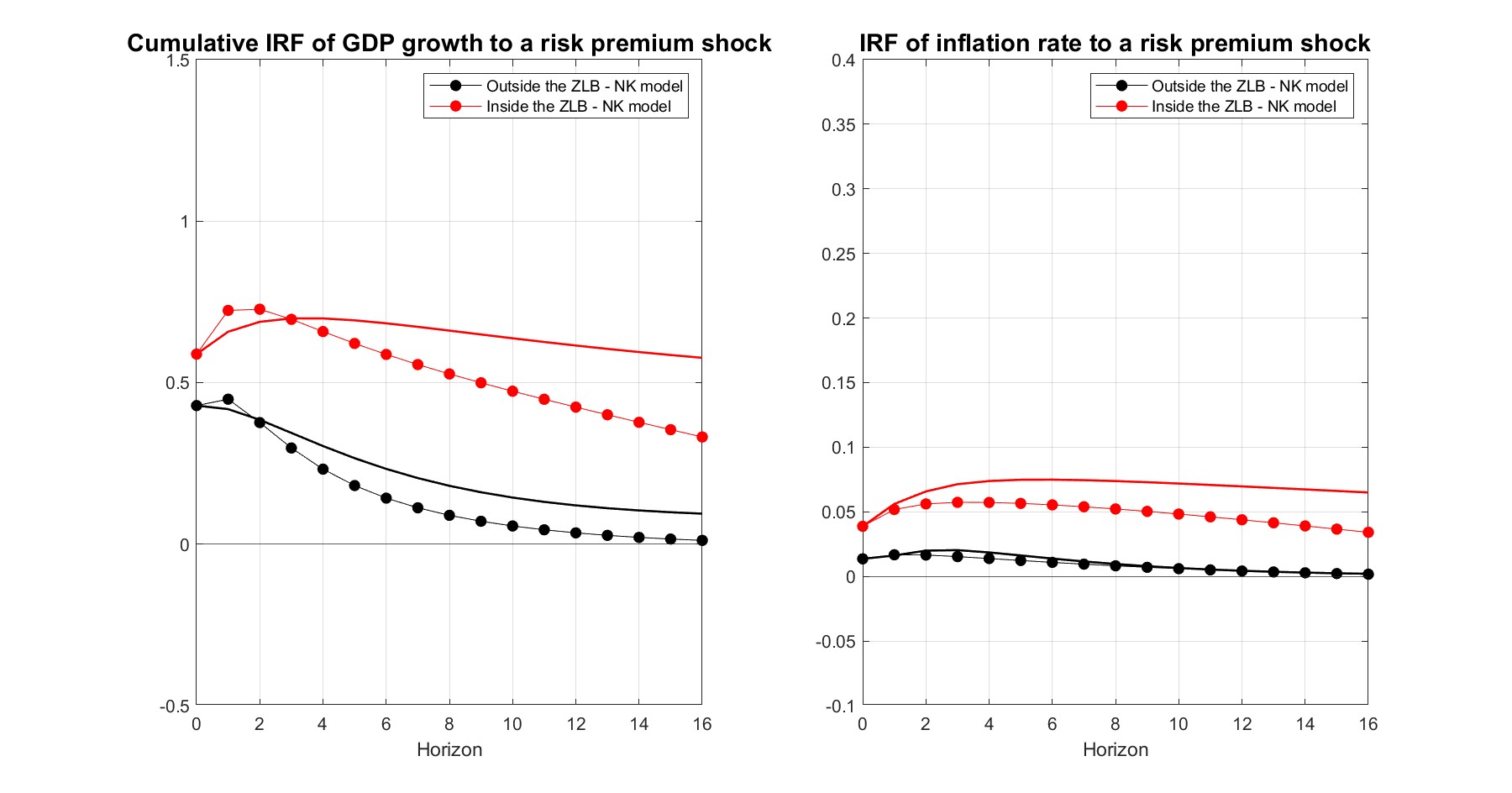}
  \caption{$\gamma = \infty$}
  \label{panel_d}
\end{subfigure}

\caption{\small{Estimated $10^{th}-90^{th}$ credible sets of the cumulative response of output growth (left) and the response of the inflation rate (right) to a risk premium shock inside (shaded red) and outside (shaded grey) the ZLB period for different values of $\gamma$. Data are simulated from the NK model. Horizons are in quarters. a) $\gamma = 10$  b) $\gamma = 100$ c) $\gamma = 300$   d) $\gamma = \infty$}. For $\gamma = \infty$ the figure also reports the true responses from the NK model.}
\label{fig:estimates_tc_tvpvar}
\end{figure}

Figure \ref{fig:estimates_tc_tvpvar}, instead, shows the responses obtained by estimating a TC-TVP-VAR that exploits the NK model as a prior for the time varying coefficients. The moment matrices $\boldsymbol{\Gamma_{xx}}$, $\boldsymbol{\Gamma_{xy}}$, $\boldsymbol{\Gamma_{yy}}$ used to parametrize the \textit{Normal-Inverse-Wishart} prior are derived from the equations of the state representation  (\ref{sp_observation}) and (\ref{sp_states}). Clearly, due to the time variation of the coefficients in the state space representation, the moments of the structural model are time varying with $\E[\boldsymbol{x_tx_t'}|\boldsymbol{\theta}]$, $\E[\boldsymbol{x_ty_t'}|\boldsymbol{\theta}]$ and $\E[\boldsymbol{y_ty_t'}|\boldsymbol{\theta}]$ changing over time. This, in turns, implies time varying restriction functions for the coefficients of the TVP-VAR encoded in the prior, resulting from $\boldsymbol{\Phi(\theta)^*} = \boldsymbol{\Gamma_{xx}(\theta)^{-1}\Gamma_{xy}(\theta)}$. Since the time varying restrictions incorporate the change in the response of the economy foreseen by the NK model, the responses of output growth and the inflation rate to the risk premium shock are more precisely estimated as the time varying coefficients are shrinked towards these restrictions. This is shown in figure \ref{fig:estimates_tc_tvpvar}, which plots the responses of output growth and of the inflation rate to a risk premium shock for different fixed values of $\gamma$, that is for a different amount of shrinkage towards the restrictions implied by the NK model.\footnote{Also in this case, in the TC-TVP-VAR we condition on the correct identification scheme for the structural shocks.}  As the figure shows, a sufficiently high $\gamma$ is needed to detect a precise and distinct response of the economy inside and outside the ZLB period as predicted by the NK model. As $\gamma$ increases and gets big enough, we get more and more precise estimates of the time varying coefficients. This allows to detect the different propagation of the shocks inside and outside the ZLB period, as predicted by the New-Keynesian model (panels \ref{panel_a}, \ref{panel_b}, \ref{panel_c}). Letting $\gamma \rightarrow \infty$ the estimated model is a restricted TVP-VAR in which the time varying coefficients exactly satisfy $\boldsymbol{\Phi(\theta)^*} = \boldsymbol{\Gamma_{xx}(\theta)^{-1}\Gamma_{xy}(\theta)}$. As expected, the  estimated responses from this model almost exactly resemble the responses from the NK model (panel \ref{panel_d}).

%\footnote{Note that the responses from the TC-TVP-VAR with $ \gamma \rightarrow \infty $ slightly differ from the responses of the NK model, due to the fact that the state space representation of the NK model has a TVP-VAR$(\infty)$ representation, while the estimated model is a TVP-VAR(2). } 

\subsection{Estimating the TC-TVP-VAR on US data}\label{zlb_usdata}

\begin{figure}[H]
\centering
\includegraphics[width=1\linewidth]{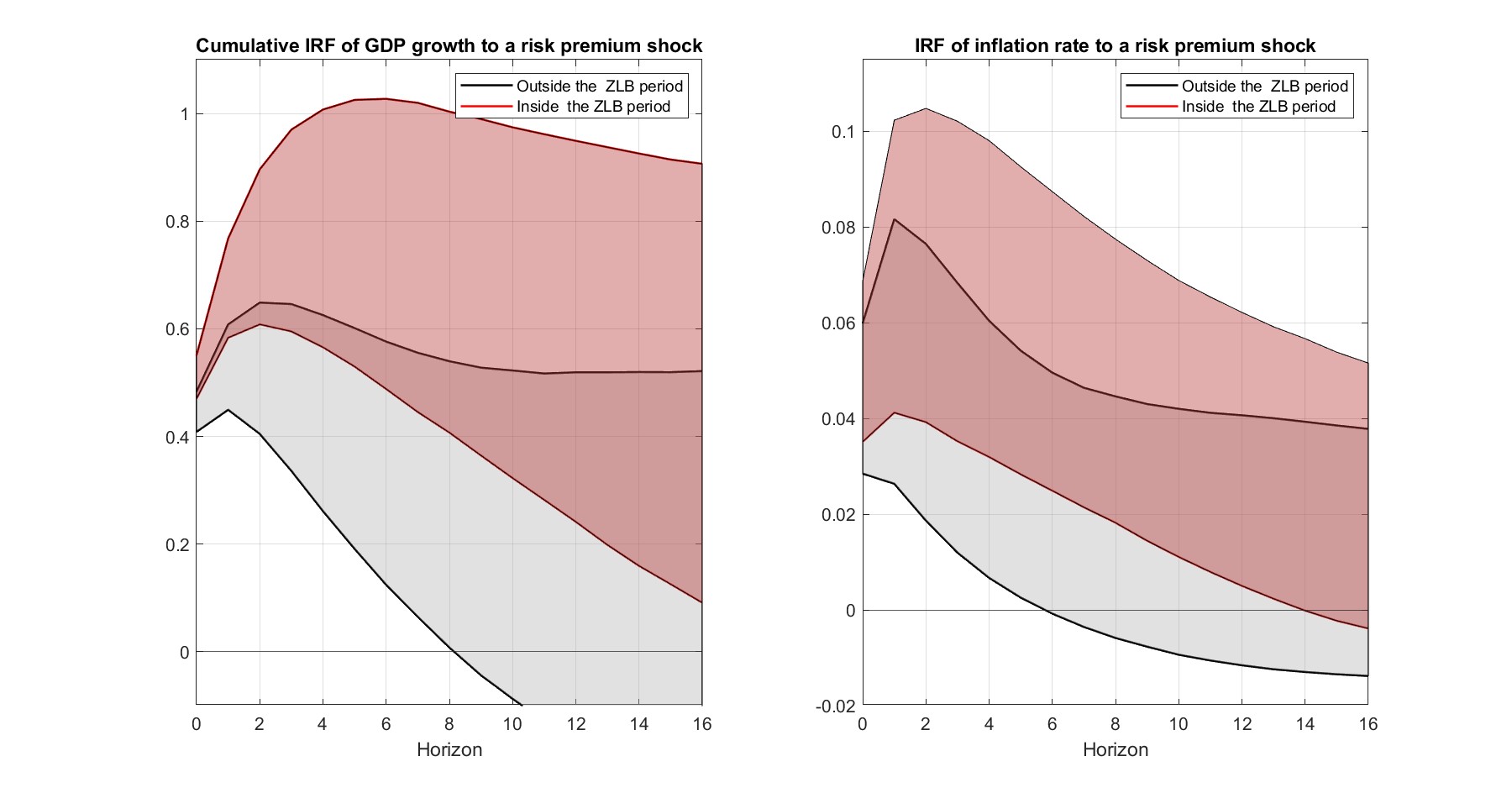}
\caption{ \small{Estimated $90^{th}-10^{th}$ credible sets of the cumulative response of output growth (on the left) and of the response of the inflation rate (on the right) to a risk premium shock inside (shaded red) and outside (shaded grey) the ZLB period from the TC-TVP-VAR on US data.} }
\label{tctvp_usdata}
\end{figure}

When analyzing real data, it is reasonable to think of the NK model just as an approximate (most likely misspecified) tightly parameterized representation of the true data generating process. In other words, we do not expect the restrictions implied by the NK model to hold exactly. However, when encoded into a prior, these restrictions might prove to be useful to get more precise estimates of the time varying coefficients and of nonlinear functions of these coefficients, such as the impulse response functions. In particular, in our context, shrinkage might turn out to be particularly useful to detect the change in the response of the economy during the ZLB period  predicted by the NK model. In order to analyze to what extent the data support the changing behavior during the ZLB, my approach consists in exploiting the New-Keynesian model as a prior for the TC-TVP-VAR. The extent to which the predictions from the New-Keynesian model will be supported by the data will depend on the estimated posterior distribution of $\gamma$, the hyper-parameter governing the degree of shrinkage of the parameters towards the restriction implied by the New-Keynesian model. If the posterior distribution of $\gamma$ is concentrated relatively far from zero, the coefficients of the TVP-VAR will be shrinked towards the restrictions implied by the NK model and the estimated responses from the TC-TVP-VAR will resemble the predictions of the NK model. This would happen in practice if the restrictions from the NK model find enough support on the data. Conversely, if the restrictions from the NK are deemed implausible by the data, the posterior distribution of $\gamma$ will be concentrated around zero and the estimates of the time varying coefficients will not reflect the restrictions coming from the prior. The estimated model is a 7-variable TC-TVP-VAR for the US economy including output growth, consumption growth, investment growth, real wage growth, hours worked, inflation and the Fed Fund rate and it is estimated over the sample 1985Q1-2019Q3.\footnote{Details on the variables and their transformation are available in the appendix \ref{sec:eq_cond}.} In order to identify the shocks in the structural TVP-VAR, I exploit the impact matrix of the NK model. The deep parameters of the NK model are treated as unknown and estimated along with the other parameters of the model, which implies that on impact there is uncertainty on the effect of the shocks on the variables of the system.  %\footnote{Table \ref{tabposterior_dsge} in appendix \ref{secposterior_dsge} reports the posterior estimates of these parameters in the TC-TVP-VAR where $\gamma$ is estimated along with the other parameters.}
Figure \ref{tctvp_usdata} shows the estimated responses from the TC-TVP-VAR where the hyper-parameter $\gamma$ is estimated along with the other parameters of the model. The posterior distribution of $\gamma$ is estimated to be concentrated far from zero, meaning that the time varying restriction functions from the NK model find sufficient support on the data. This, in turns, is reflected on the estimates of the $10^{th}-90^{th}$ credible sets which provide some evidences supporting a distinct responses of the economy inside and outside the ZLB period. As predicted by the NK model, the estimates suggest that when monetary policy is constrained by the nominal rate hitting the ZLB,  the risk premium shock is reabsorbed at a much slower pace by the economy. Following the positive demand shock both output growth and the inflation rate increase, with the effect on output growth being more precisely estimated at shorter horizons. The effect on output peaks after almost two quarters outside the ZLB period, while  about after five quarters inside the ZLB period. As for inflation, the effect peaks after one quarter outside the ZLB period and after two quarters inside the ZLB period. After the peak, the effect of the risk premium shock on both output and inflation dies at a much slower pace inside the ZLB period, as foreseen by the NK model. Clearly,  given the symmetric nature of the impulse response functions in our econometric model, figure \ref{tctvp_usdata} implies that in the face of an increase in the risk premium, the economy would experience more persistent decreases of both inflation and output inside the ZLB period. Hence, risk premium shocks become more important when the ZLB binds as found in \citet{gourio2020risk}.
More in general, this finding is compatible with the findings in \citet{scorfZLB}, which building on \citet{mavroeidis2021identification} develop a structural VAR in which an occasionally-binding constraint generates censoring of one of the dependent variables. They find that the presence of the ZLB is empirically relevant for the propagation of macroeconomic shocks. Differently from both \citet{mavroeidis2021identification} and \citet{scorfZLB}, but similarly to \citet{galigambetti} and \citet{lubikbenati} our approach  assumes the ZLB period to be completely observable. As well, while both \citet{mavroeidis2021identification} and \citet{scorfZLB} leverage censoring to find identifying information on the propagation of macroeconomic shocks we directly resort to economic theory encoded in the NK model to identify the macroeconomic shocks. One key difference is that while in \citet{scorfZLB} the structural coefficients switch across unobserved regimes, in our TVP-SVAR the autoregressive time varying parameters are allowed to slowly drift inside, outside and in the transition to and from the ZLB period. \footnote{As for the impact matrix $A_{0,t}(\boldsymbol{\theta})$, in our framework deterministically changes over time  (conditioning on the NK parameters $\boldsymbol{\theta}$) as a function of $\bar{H}_t$, the number of periods agents think the ZLB remains binding.}

%Despite, ,  

%\begin{figure}[H]
%\centering
%\begin{subfigure}[b]{0.85\textwidth}
% \includegraphics[width=1\linewidth]{modelversion_usdata_01.jpg}
%   \caption{}
%   \label{fig:Ng1} 
%\end{subfigure}
%\begin{subfigure}[b]{0.85\textwidth}
%\includegraphics[width=1\linewidth]{modelversion_usdata_est.jpg}
%\caption{}
%\label{fig:Ng2}
%\end{subfigure}
%\caption{ \small{Estimated $90^{th}-10^{th}$ credible sets of the cumulative response of output growth (on the left) and of the response of the inflation rate (on the right) to a risk premium shock inside (shaded grey) and outside (shaded red) the ZLB period from the TC-TVP-VAR on US data. Panel (a) shows the estimated responses fixing the hyperparameter $\gamma = 0.01$. Panel (b) shows the estimated responses estimating the hyperparameter along with the other parameters of the model.} }
%\label{tctvp_usdata}
%\end{figure}
%I use the TC-TVP-VAR to understand whether the US economy during the ZLB period reacted consistently to the predictions of a standard New-Keynesian model. 

%The shortcomes of the analysys are the following:
%    

%\subsection{Forecasting exercise (in progress)}
%\subsection{Out of sample forecasts (\textcolor{red}{in progress}) }

%In this application, this identification strategy turns out to be particularly successful since identification based on short run restrictions a la Cholesky delivers the price puzzle, with inflation increasing in the face of a restrictive monetary policy shock and identification through sign restrictions is computationally demanding given the medium size of the TVP-VAR considered.   
%\subsection{Comparing competing economic theories}
\section{Conclusion}\label{conclòus}
%Econometricians often encounter the dilemma of constructing models that strike a balance between being comprehensive enough to analyze complex problems and restrictive enough to provide precise inferences based on finite samples. One approach to tackle this challenge is by employing an over-parametrized model, that has more parameters than can be feasibly estimated using the available data and then by seeking plausible restrictions that enable reliable inference. TVP-VARs are a prominent example. On the top of modelling the persistence of the time varying coefficients through the usual set of linear (\textit{fuzzy}) restrictions that model a priori the distance of the time varying coefficients in two consecutive time periods, this paper propose to use economic theory to seek another set of plausible nonlinear (\textit{fuzzy}) restrictions among the time varying coefficients. It does so by introducing a shrinkage prior that centers the time varying coefficients on the cross equation restrictions implied by an economic theory about the variables in the model. The paper shows that  ``economic shrinkage'' can successfully be exploited
%On the top of modellingthe persistence of the time varying coefficients through the usual set of linear (fuzzy) restrictionsthat govern a priori the distance of the time varying coefficients in two consecutive time periods,this paper propose to use economic theory to seek another set of plausible nonlinear (fuzzy)restrictions among the time varying coefficients

TVP-VARs are flexible statistical models used both for prediction and semi-structural analysis in macroeconomics.   Despite their flexibility allows to capture changes in the dynamic relationship among the macroeconomic variables, inference on the time varying parameters is often very imprecise when using standard priors. This translates into poor forecasting performances  and imprecise inference on typical objects of interests such as the impulse responses.  On the other side, models from the economic theory typically provide a more tightly parameterized representation of the macroeconomy and therefore have the opposite tendency of fitting the data rather poorly.  This paper exploits economic theory to formulate a prior for the parameters of TVP-VARs. More specifically, the paper introduces a novel shrinkage prior that centers the time varying coefficients on a path implied by an underlying economic theory about the variables in the system.  The paper shows that ``economic shrinkage'' can be  successfully used to obtain more accurate forecasts and more precise estimates of the impulse response functions. In an application, I exploit a medium scale New-Keynesian model that accounts for forward guidance and the Zero Lower bound period, to formulate a prior for a medium scale TVP-VAR. Using this theory coherent shrinkage prior allows to estimate more precisely the response of the economy to macroeconomic shocks inside and outside the ZLB period, helping to solve the inferential
problems faced by the standard TVP-VAR. On US data, the paper finds indeed a distinct propagation of a pure demand shock - the risk premium shock - inside and outside the ZLB period, with the effects of the shock reabsorbing at a much slower pace inside the ZLB period. This finding has important implications for the conduct of both fiscal and macroprudential policy at the ZLB.   \\

%On the top of modelling the persistence of the time varying coefficients through the usual set of linear (\textit{fuzzy}) restrictions that govern a priori the distance of the time varying coefficients in two consecutive time periods, this paper propose to use economic theory to seek another set of plausible nonlinear (\textit{fuzzy}) restrictions among the time varying coefficients. 

\textbf{Future research} \hspace{0.2cm} For future research the prior proposed in this paper could be modified to accommodate for multiple competing theories à la \citet{LORIA2022105}. As well the prior could be adapted to allow for different forms of heteroskedasticity but this extension is non-trivial since in TVP-VAR-SV a la \citet{primiceri2005} the specification of the model breaks the Kronecker structure of the likelihood. More in general, restrictions from economic theory could successfully be used to reduce overfitting and sharpen inference in non-parametric VARs and Gaussian Procesess VARs \citep{hauzenberger2022gaussian}. 
% In an application I exploit a small scale New-Keynesian model to form a prior for the parameters of a trivariate TVP-VAR for GDP-growth inflation and interest rate on US data. 
%\textbf{Future research} 
\printbibliography

@article{COGLEY2005262,
title = {Drifts and volatilities: monetary policies and outcomes in the post WWII US},
journal = {Review of Economic Dynamics},
volume = {8},
number = {2},
pages = {262-302},
year = {2005},
note = {Monetary Policy and Learning},
issn = {1094-2025},
doi = {https://doi.org/10.1016/j.red.2004.10.009},
url = {https://www.sciencedirect.com/science/article/pii/S1094202505000049},
author = {Timothy Cogley and Thomas J. Sargent}
}

@article{primiceri2005,
  title={Time varying structural vector autoregressions and monetary policy},
  author={Primiceri, Giorgio E},
  journal={The Review of Economic Studies},
  volume={72},
  number={3},
  pages={821--852},
  year={2005},
  publisher={Wiley-Blackwell}
}

@article{SW2012,
Author = {Stock, James H. and Watson, Mark W.},
Title = {Vector Autoregressions},
Journal = {Journal of Economic Perspectives},
Volume = {15},
Number = {4},
Year = {2001},
Month = {12},
Pages = {101-115},
DOI = {10.1257/jep.15.4.101},
URL = {https://www.aeaweb.org/articles?id=10.1257/jep.15.4.101}}

@article{DNS2004,
author = {Del Negro, Marco and Schorfheide, Frank},
title = {Priors from General Equilibrium Models for VARS*},
journal = {International Economic Review},
volume = {45},
number = {2},
pages = {643-673},
doi = {https://doi.org/10.1111/j.1468-2354.2004.00139.x},
url = {https://onlinelibrary.wiley.com/doi/abs/10.1111/j.1468-2354.2004.00139.x},
eprint = {https://onlinelibrary.wiley.com/doi/pdf/10.1111/j.1468-2354.2004.00139.x},
year = {2004}
}

@article{BITTO201975,
title = {Achieving shrinkage in a time-varying parameter model framework},
journal = {Journal of Econometrics},
volume = {210},
number = {1},
pages = {75-97},
year = {2019},
note = {Annals Issue in Honor of John Geweke “Complexity and Big Data in Economics and Finance: Recent Developments from a Bayesian Perspective”},
issn = {0304-4076},
doi = {https://doi.org/10.1016/j.jeconom.2018.11.006},
url = {https://www.sciencedirect.com/science/article/pii/S0304407618302070},
author = {Angela Bitto and Sylvia Frühwirth-Schnatter}
}

@article{belmontekoopkorobilis,
author = {Belmonte, Miguel A.G. and Koop, Gary and Korobilis, Dimitris},
title = {Hierarchical Shrinkage in Time-Varying Parameter Models},
journal = {Journal of Forecasting},
volume = {33},
number = {1},
pages = {80-94},
doi = {https://doi.org/10.1002/for.2276},
url = {https://onlinelibrary.wiley.com/doi/abs/10.1002/for.2276},
eprint = {https://onlinelibrary.wiley.com/doi/pdf/10.1002/for.2276},
year = {2014}
}

@article{FRUHWIRTHSCHNATTER201085,
title = {Stochastic model specification search for Gaussian and partial non-Gaussian state space models},
journal = {Journal of Econometrics},
volume = {154},
number = {1},
pages = {85-100},
year = {2010},
issn = {0304-4076},
doi = {https://doi.org/10.1016/j.jeconom.2009.07.003},
url = {https://www.sciencedirect.com/science/article/pii/S0304407609001614},
author = {Sylvia Frühwirth-Schnatter and Helga Wagner}
}

@misc{coulombe2021timevarying,
      title={Time-Varying Parameters as Ridge Regressions}, 
      author={Philippe Goulet Coulombe},
      year={2021},
      eprint={2009.00401},
      archivePrefix={arXiv},
      primaryClass={econ.EM}
}

@TechReport{gambettidewind,
  author={Joris de Wind and Luca Gambetti},
  title={{Reduced-rank time-varying vector autoregressions}},
  year=2014,
  month=Mar,
  institution={CPB Netherlands Bureau for Economic Policy Analysis},
  type={CPB Discussion Paper},
  url={https://ideas.repec.org/p/cpb/discus/270.html},
  number={270},
  doi={},
}

@Article{stevanovic,
  author={Dalibor Stevanovic},
  title={{Common time variation of parameters in reduced-form macroeconomic models}},
  journal={Studies in Nonlinear Dynamics \& Econometrics},
  year=2016,
  volume={20},
  number={2},
  pages={159-183},
  month={4},
  doi={10.1515/snde-2014-0064},
  url={https://ideas.repec.org/a/bpj/sndecm/v20y2016i2p159-183n3.html}
}

@article{CHAN2020105,
title = {Reducing the state space dimension in a large TVP-VAR},
journal = {Journal of Econometrics},
volume = {218},
number = {1},
pages = {105-118},
year = {2020},
issn = {0304-4076},
doi = {https://doi.org/10.1016/j.jeconom.2019.11.006},
url = {https://www.sciencedirect.com/science/article/pii/S0304407620300348},
author = {Joshua C.C. Chan and Eric Eisenstat and Rodney W. Strachan}
}

@article{kalli,
title = {Time-varying sparsity in dynamic regression models},
journal = {Journal of Econometrics},
volume = {178},
number = {2},
pages = {779-793},
year = {2014},
issn = {0304-4076},
doi = {https://doi.org/10.1016/j.jeconom.2013.10.012},
url = {https://www.sciencedirect.com/science/article/pii/S0304407613002273},
author = {Maria Kalli and Jim E. Griffin}
}

@article{chan2009efficient,
  title={Efficient simulation and integrated likelihood estimation in state space models},
  author={Chan, Joshua CC and Jeliazkov, Ivan},
  journal={International Journal of Mathematical Modelling and Numerical Optimisation},
  volume={1},
  number={1-2},
  pages={101--120},
  year={2009},
  publisher={Inderscience Publishers}
}

@article{giannonelenzaprimicieri,
    author = {Giannone, Domenico and Lenza, Michele and Primiceri, Giorgio E.},
    title = "{Prior Selection for Vector Autoregressions}",
    journal = {The Review of Economics and Statistics},
    volume = {97},
    number = {2},
    pages = {436-451},
    year = {2015},
    month = {05},
    issn = {0034-6535},
    doi = {10.1162/REST_a_00483},
    url = {https://doi.org/10.1162/REST\_a\_00483},
    eprint = {https://direct.mit.edu/rest/article-pdf/97/2/436/1917922/rest\_a\_00483.pdf},
}

@Article{kasyfessler2019,
  author={Pirmin Fessler and Maximilian Kasy},
  title={{How to Use Economic Theory to Improve Estimators: Shrinking Toward Theoretical Restrictions}},
  journal={The Review of Economics and Statistics},
  year=2019,
  volume={101},
  number={4},
  pages={681-698},
  month= {10},
  doi={},
  url={https://ideas.repec.org/a/tpr/restat/v101y2019i4p681-698.html}
}

@article{GIACOMINI2014145,
title = {Theory-coherent forecasting},
journal = {Journal of Econometrics},
volume = {182},
number = {1},
pages = {145-155},
year = {2014},
note = {Causality, Prediction, and Specification Analysis: Recent Advances and Future Directions},
issn = {0304-4076},
doi = {https://doi.org/10.1016/j.jeconom.2014.04.014},
url = {https://www.sciencedirect.com/science/article/pii/S0304407614000736},
author = {Raffaella Giacomini and Giuseppe Ragusa}
}

@article{LORIA2022105,
title = {Economic theories and macroeconomic reality},
journal = {Journal of Monetary Economics},
volume = {126},
pages = {105-117},
year = {2022},
issn = {0304-3932},
doi = {https://doi.org/10.1016/j.jmoneco.2021.12.001},
url = {https://www.sciencedirect.com/science/article/pii/S0304393221001434},
author = {Francesca Loria and Christian Matthes and Mu-Chun Wang}
}

@Article{WHITEMAN,
  author={Ingram, Beth F. and Whiteman, Charles H.},
  title={{Supplanting the 'Minnesota' prior: Forecasting macroeconomic time series using real business cycle model priors}},
  journal={Journal of Monetary Economics},
  year=1994,
  volume={34},
  number={3},
  pages={497-510},
  month={12},
  doi={},
  url={https://ideas.repec.org/a/eee/moneco/v34y1994i3p497-510.html}
}

@ARTICLE{sims2002,
title = {Solving Linear Rational Expectations Models},
author = {Sims, Christopher},
year = {2002},
journal = {Computational Economics},
volume = {20},
number = {1-2},
pages = {1-20},
url = {https://EconPapers.repec.org/RePEc:kap:compec:v:20:y:2002:i:1-2:p:1-20}
}

@article{sims1980,
 ISSN = {00129682, 14680262},
 URL = {http://www.jstor.org/stable/1912017},
 author = {Christopher A. Sims},
 journal = {Econometrica},
 number = {1},
 pages = {1--48},
 publisher = {[Wiley, Econometric Society]},
 title = {Macroeconomics and Reality},
 urldate = {2023-10-21},
 volume = {48},
 year = {1980}
}

@article{ccm,
author = {Carriero, Andrea and Clark, Todd E. and Marcellino, Massimiliano},
title = {No-arbitrage priors, drifting volatilities, and the term structure of interest rates},
journal = {Journal of Applied Econometrics},
volume = {36},
number = {5},
pages = {495-516},
doi = {https://doi.org/10.1002/jae.2828},
url = {https://onlinelibrary.wiley.com/doi/abs/10.1002/jae.2828},
eprint = {https://onlinelibrary.wiley.com/doi/pdf/10.1002/jae.2828},
year = {2021}
}

@article{smith1993,
author = {Smith Jr., A. A.},
title = {Estimating nonlinear time-series models using simulated vector autoregressions},
journal = {Journal of Applied Econometrics},
volume = {8},
number = {S1},
pages = {S63-S84},
doi = {https://doi.org/10.1002/jae.3950080506},
url = {https://onlinelibrary.wiley.com/doi/abs/10.1002/jae.3950080506},
eprint = {https://onlinelibrary.wiley.com/doi/pdf/10.1002/jae.3950080506},
year = {1993}
}

@article{delnegroschorfheide2005,
Author = {Del Negro, Marco and Giannoni, Marc P. and Schorfheide, Frank},
Title = {Inflation in the Great Recession and New Keynesian Models},
Journal = {American Economic Journal: Macroeconomics},
Volume = {7},
Number = {1},
Year = {2015},
Month = {1},
Pages = {168-96},
DOI = {10.1257/mac.20140097},
URL = {https://www.aeaweb.org/articles?id=10.1257/mac.20140097}}

@article{SWauters,
Author = {Smets, Frank and Wouters, Rafael},
Title = {Shocks and Frictions in US Business Cycles: A Bayesian DSGE Approach},
Journal = {American Economic Review},
Volume = {97},
Number = {3},
Year = {2007},
Month = {6},
Pages = {586-606},
DOI = {10.1257/aer.97.3.586},
URL = {https://www.aeaweb.org/articles?id=10.1257/aer.97.3.586}}

@article{kulishcagliarini,
    author = {Cagliarini, Adam and Kulish, Mariano},
    title = "{Solving Linear Rational Expectations Models with Predictable Structural Changes}",
    journal = {The Review of Economics and Statistics},
    volume = {95},
    number = {1},
    pages = {328-336},
    year = {2013},
    month = {03},
    issn = {0034-6535},
    doi = {10.1162/REST_a_00240},
    url = {https://doi.org/10.1162/REST\_a\_00240},
    eprint = {https://direct.mit.edu/rest/article-pdf/95/1/328/1917161/rest\_a\_00240.pdf},
}

@article{FARMER20091849,
title = {Understanding Markov-switching rational expectations models},
journal = {Journal of Economic Theory},
volume = {144},
number = {5},
pages = {1849-1867},
year = {2009},
issn = {0022-0531},
doi = {https://doi.org/10.1016/j.jet.2009.05.004},
url = {https://www.sciencedirect.com/science/article/pii/S0022053109000787},
author = {Roger E.A. Farmer and Daniel F. Waggoner and Tao Zha}
}

@article{MILANI20072065,
title = {Expectations, learning and macroeconomic persistence},
journal = {Journal of Monetary Economics},
volume = {54},
number = {7},
pages = {2065-2082},
year = {2007},
issn = {0304-3932},
doi = {https://doi.org/10.1016/j.jmoneco.2006.11.007},
url = {https://www.sciencedirect.com/science/article/pii/S0304393206002406},
author = {Fabio Milani}
}

@book{woodford2003interest,
  title={Interest and Prices: Foundations of a Theory of Monetary Policy},
  author={Woodford, Michael},
  year={2003},
  publisher={Princeton University Press}
}

@TechReport{lubikbenati,
  author={Luca Benati and Thomas A. Lubik},
  title={{Impulse Response Analysis at the Zero Lower Bound}},
  year=2023,
  month=Jun,
  institution={Universitaet Bern, Departement Volkswirtschaft},
  type={Diskussionsschriften},
  url={https://ideas.repec.org/p/ube/dpvwib/dp2306.html},
  number={dp2306},
  doi={},
}

@InCollection{galigambetti,
  author={Davide Debortoli and Jordi Galí and Luca Gambetti},
  title={{On the Empirical (Ir)relevance of the Zero Lower Bound Constraint}},
  booktitle={{NBER Macroeconomics Annual 2019, volume 34}},
  publisher={National Bureau of Economic Research, Inc},
  year=2019,
  month={10},
  volume={},
  number={},
  series={NBER Chapters},
  edition={},
  chapter={},
  pages={141-170},
  doi={},
  url={https://ideas.repec.org/h/nbr/nberch/14241.html}
}

@INCOLLECTION{CS2002,
title = {Evolving Post-World War II US Inflation Dynamics},
author = {Cogley, Timothy and Sargent, Thomas},
year = {2002},
pages = {331-388},
booktitle = {NBER Macroeconomics Annual 2001, Volume 16},
publisher = {National Bureau of Economic Research, Inc},
url = {https://EconPapers.repec.org/RePEc:nbr:nberch:11068}
}

@article{cogleysbordone,
Author = {Cogley, Timothy and Sbordone, Argia M.},
Title = {Trend Inflation, Indexation, and Inflation Persistence in the New Keynesian Phillips Curve},
Journal = {American Economic Review},
Volume = {98},
Number = {5},
Year = {2008},
Month = {12},
Pages = {2101-26},
DOI = {10.1257/aer.98.5.2101},
URL = {https://www.aeaweb.org/articles?id=10.1257/aer.98.5.2101}}

@article{sbordoneascari,
 ISSN = {00220515},
 URL = {http://www.jstor.org/stable/24434108},
 author = {Guido Ascari and Argia M. Sbordone},
 journal = {Journal of Economic Literature},
 number = {3},
 pages = {679--739},
 publisher = {American Economic Association},
 title = {The Macroeconomics of Trend Inflation},
 urldate = {2023-10-07},
 volume = {52},
 year = {2014}
}

@article{ascari2023long,
  title={The long-run phillips curve is... a curve},
  author={Ascari, Guido and Bonomolo, Paolo and Haque, Qazi},
  year={2023},
  publisher={CAMA Working Paper 37/2023}
}

@article{kc1994,
 ISSN = {00063444},
 URL = {http://www.jstor.org/stable/2337125},
 author = {C. K. Carter and R. Kohn},
 journal = {Biometrika},
 number = {3},
 pages = {541--553},
 publisher = {[Oxford University Press, Biometrika Trust]},
 title = {On Gibbs Sampling for State Space Models},
 urldate = {2023-10-09},
 volume = {81},
 year = {1994}
}

@misc{hauzenberger2022gaussian,
      title={Gaussian Process Vector Autoregressions and Macroeconomic Uncertainty}, 
      author={Niko Hauzenberger and Florian Huber and Massimiliano Marcellino and Nico Petz},
      year={2022},
      eprint={2112.01995},
      archivePrefix={arXiv},
      primaryClass={econ.EM}
}

@article{chencurdiaferrero,
author = {Chen, Han and Cúrdia, Vasco and Ferrero, Andrea},
title = {The Macroeconomic Effects of Large-scale Asset Purchase Programmes*},
journal = {The Economic Journal},
volume = {122},
number = {564},
pages = {F289-F315},
doi = {https://doi.org/10.1111/j.1468-0297.2012.02549.x},
url = {https://onlinelibrary.wiley.com/doi/abs/10.1111/j.1468-0297.2012.02549.x},
eprint = {https://onlinelibrary.wiley.com/doi/pdf/10.1111/j.1468-0297.2012.02549.x},
year = {2012}
}

@article{giannonegambetti,
 ISSN = {08837252, 10991255},
 URL = {http://www.jstor.org/stable/23355996},
 author = {Antonello D'Agostino and Luca Gambetti and Domenico Giannone},
 journal = {Journal of Applied Econometrics},
 number = {1},
 pages = {82--101},
 publisher = {Wiley},
 title = {Macroeconomic forecasting and structural change},
 urldate = {2023-10-31},
 volume = {28},
 year = {2013}
}

@article{gallant,
 ISSN = {01621459},
 URL = {http://www.jstor.org/stable/40591904},
 author = {A. Ronald Gallant and Robert E. McCulloch},
 journal = {Journal of the American Statistical Association},
 number = {485},
 pages = {117--131},
 publisher = {[American Statistical Association, Taylor & Francis, Ltd.]},
 title = {On the Determination of General Scientific Models With Application to Asset Pricing},
 urldate = {2023-11-07},
 volume = {104},
 year = {2009}
}

@Article{onorante,
  author={Florian Huber and Gary Koop and Luca Onorante},
  title={{Inducing Sparsity and Shrinkage in Time-Varying Parameter Models}},
  journal={Journal of Business \& Economic Statistics},
  year=2021,
  volume={39},
  number={3},
  pages={669-683},
  month={7},
  doi={10.1080/07350015.2020.171},
  url={https://ideas.repec.org/a/taf/jnlbes/v39y2021i3p669-683.html}
}

@article{gourio2020risk,
  title={Risk premia at the ZLB: a macroeconomic interpretation},
  author={Gourio, Fran{\c{c}}ois and Ngo, Phuong},
  year={2020},
  publisher={FRB of Chicago Working Paper No. WP-2020-01}
}

@article{gz2012,
Author = {Gilchrist, Simon and Zakrajšek, Egon},
Title = {Credit Spreads and Business Cycle Fluctuations},
Journal = {American Economic Review},
Volume = {102},
Number = {4},
Year = {2012},
Month = {June},
Pages = {1692-1720},
DOI = {10.1257/aer.102.4.1692},
URL = {https://www.aeaweb.org/articles?id=10.1257/aer.102.4.1692}}

@article{scorfZLB,
title = {SVARs with occasionally-binding constraints},
journal = {Journal of Econometrics},
volume = {231},
number = {2},
pages = {477-499},
year = {2022},
note = {Special Issue: The Econometrics of Macroeconomic and Financial Data},
issn = {0304-4076},
doi = {https://doi.org/10.1016/j.jeconom.2021.07.013},
url = {https://www.sciencedirect.com/science/article/pii/S0304407621002487},
author = {S. Borağan Aruoba and Marko Mlikota and Frank Schorfheide and Sergio Villalvazo}
}

@article{mavroeidis2021identification,
  title={Identification at the zero lower bound},
  author={Mavroeidis, Sophocles},
  journal={Econometrica},
  volume={89},
  number={6},
  pages={2855--2885},
  year={2021},
  publisher={Wiley Online Library}
}

@article{jcc_kron,
author = {Joshua C. C. Chan},
title = {Large Bayesian VARs: A Flexible Kronecker Error Covariance Structure},
journal = {Journal of Business \& Economic Statistics},
volume = {38},
number = {1},
pages = {68--79},
year = {2020},
publisher = {Taylor \& Francis},
doi = {10.1080/07350015.2018.1451336},


URL = { 
    
        https://doi.org/10.1080/07350015.2018.1451336
    
    

},
eprint = { 
    
        https://doi.org/10.1080/07350015.2018.1451336
    
    

}

}

@article{ccm2016,
  title={Common drifting volatility in large Bayesian VARs},
  author={Carriero, Andrea and Clark, Todd E and Marcellino, Massimiliano},
  journal={Journal of Business \& Economic Statistics},
  volume={34},
  number={3},
  pages={375--390},
  year={2016},
  publisher={Taylor \& Francis}
}

@article{leon,
    author = {Leiva-León, Danilo and Uzeda, Luis},
    title = "{Endogenous Time Variation in Vector Autoregressions}",
    journal = {The Review of Economics and Statistics},
    volume = {105},
    number = {1},
    pages = {125-142},
    year = {2023},
    month = {01},
    issn = {0034-6535},
    doi = {10.1162/rest_a_01038},
    url = {https://doi.org/10.1162/rest\_a\_01038},
    eprint = {https://direct.mit.edu/rest/article-pdf/105/1/125/2066625/rest\_a\_01038.pdf},
}
\clearpage
\appendix
\section{Appendix}
\subsection{Theory Coherent TVP-VAR}
\subsubsection{Time Varying Parameters by dummy observations}\label{sec:dummy_tvp}
Starting from:
\begin{equation*}    \underbrace{\boldsymbol{y}'_t}_{\text{$1 \times N$}} = \underbrace{\boldsymbol{x}'_t}_{\text{$1 \times k$}} \underbrace{\boldsymbol{\Phi}_t}_{\text{$k \times N$}} + \underbrace{\boldsymbol{u_t}}_{\text{$1 \times N$}} \hspace{3cm}  \boldsymbol{u_t} \sim \mathcal{N}(\boldsymbol{0}_{1 \times N}, \boldsymbol{\Sigma_u})
\end{equation*}
we can write the TVP-VAR in static compact form as: 

\begin{equation*}    \underbrace{\boldsymbol{Y}}_{\text{$T \times N$}} = \underbrace{\boldsymbol{X}}_{\text{$T \times Tk$}} \underbrace{\boldsymbol{\Phi}}_{\text{$Tk \times N$}} + \underbrace{\boldsymbol{U}}_{\text{$T \times N$}} \hspace{3cm} \boldsymbol{U} \sim MVN(\boldsymbol{0},\boldsymbol{\Sigma_u} ,\boldsymbol{I_T}) 
\end{equation*}
%
%Log-Likelihood: 
%
%\begin{equation}
%    L(\boldsymbol{y_{1:T}}|\boldsymbol{\Sigma_{1:T}, \Phi_{1:T}}) = -\frac{NT}{2}log(2\pi) -\frac{1}{2} \sum_{t=1}^T(log(|\boldsymbol{\Sigma_t}|) - \frac{1}{2}\sum_{t=1}^T\left[ (\boldsymbol{y_t} - \boldsymbol{x_t\Phi_t})\boldsymbol{\Sigma_t^{-1}}(\boldsymbol{y_t} - \boldsymbol{x_t\Phi_t})' \right] 
%   \hspace{1cm} 
%\end{equation}
%
Suppose we want to specify independent RW stochastic processes for all the coefficients in $\boldsymbol{\Phi}$ as: 
\begin{equation}\label{rw}
\phi_{jt}^{(i)} = \phi_{jt-1}^{(i)} + \xi_{jt}^{(i)} \hspace{2cm} \xi_{jt}^{(i)} \sim \mathcal{N}\left(0, \frac{\Sigma_{ii}}{\lambda_j^2}\right)     
\end{equation}
for $t = 1, \ldots, T$ $i = 1, \ldots, N$ and $j = 1, \ldots, k$. Defining $\boldsymbol{\Lambda_k} = diag(\lambda_1, \ldots, \lambda_k)$ we can use a set of dummy observations to model the time variation of the coefficients. The dummy observations  imply the linear  \textit{fuzzy}  restrictions in (\ref{rw}) on the coefficients in $\boldsymbol{\Phi}$, in the specific we can write: 

\begin{equation}
\underbrace{
\begin{bmatrix}
\boldsymbol{\Lambda}_k\boldsymbol{\Phi_0} \\
    \boldsymbol{0} \\ 
    \boldsymbol{0} \\
    \vdots         \\
    \boldsymbol{0} \\
\end{bmatrix}}_{\text{$\boldsymbol{Y^*}$}}
=
\underbrace{
\begin{bmatrix}
 \boldsymbol{\Lambda_k} & \boldsymbol{0}  &\ldots& \boldsymbol{0}  \\
0 & \boldsymbol{\Lambda_k}& \ldots & \boldsymbol{0}  \\
\boldsymbol{0} & 0& \ldots & \boldsymbol{0}  \\
\vdots & \vdots&\vdots & \vdots\\
\boldsymbol{0}  & \boldsymbol{0} & 0& \boldsymbol{\Lambda_k} \\
\end{bmatrix}
\begin{bmatrix}
 \boldsymbol{I}_k & \boldsymbol{0}  &\ldots& \boldsymbol{0}  \\
-\boldsymbol{I}_k & \boldsymbol{I}_k& \ldots & \boldsymbol{0}  \\
\boldsymbol{0} & -\boldsymbol{I}_k& \ldots & \boldsymbol{0}  \\
\vdots & \vdots&\vdots & \vdots\\
\boldsymbol{0}  & \boldsymbol{0} & -\boldsymbol{I}_k & \boldsymbol{I}_k \\
\end{bmatrix}}_{\text{$\boldsymbol{X}^*$}} 
\begin{bmatrix}
    \boldsymbol{\Phi_1} \\
    \boldsymbol{\Phi_2} \\ 
    \boldsymbol{\Phi_3}  \\
    \vdots              \\
    \boldsymbol{\Phi_T} \\
\end{bmatrix} +
\underbrace{\begin{bmatrix}
    \boldsymbol{U}^*_1 \\
    \boldsymbol{U}^*_2  \\ 
    \boldsymbol{U}^*_3  \\
    \vdots         \\
    \boldsymbol{U}^*_T  \\
\end{bmatrix}}_{\text{$\boldsymbol{U}^*$}} 
\end{equation}

\begin{equation}\label{dummyo}
   \boldsymbol{Y}^* = \boldsymbol{X}^*\boldsymbol{\Phi}  + \boldsymbol{U}^*
\end{equation}
This is just another way of writing: 
\begin{equation}
 \underbrace{
\begin{bmatrix}
 \boldsymbol{I}_k & \boldsymbol{0}  &\ldots& \boldsymbol{0}  \\
 -\boldsymbol{I}_k & \boldsymbol{I}_k& \ldots & \boldsymbol{0}  \\
\boldsymbol{0} & -\boldsymbol{I}_k& \ldots & \boldsymbol{0}  \\
\vdots & \vdots&\vdots & \vdots\\
\boldsymbol{0}  & \boldsymbol{0} & -\boldsymbol{I}_k & \boldsymbol{I}_k \\
\end{bmatrix}}_{\text{$\boldsymbol{H_{Tk}}$}}
\underbrace{
\begin{bmatrix}
    \boldsymbol{\Phi_1} \\
    \boldsymbol{\Phi_2} \\ 
    \boldsymbol{\Phi_3}  \\
    \vdots              \\
    \boldsymbol{\Phi_T} \\
\end{bmatrix}}_{\text{$\boldsymbol{\Phi}$}} = 
\underbrace{
\begin{bmatrix}
    \boldsymbol{\Phi_0} \\
    \boldsymbol{0} \\ 
    \boldsymbol{0} \\
    \vdots         \\
    \boldsymbol{0} \\
\end{bmatrix}}_{\text{$\boldsymbol{\Phi_{00}}$}}  - 
\underbrace{
\begin{bmatrix}
 \boldsymbol{\Lambda_k^{-1}} & \boldsymbol{0}  &\ldots& \boldsymbol{0}  \\
0 & \boldsymbol{\Lambda_k^{-1}}& \ldots & \boldsymbol{0}  \\
\boldsymbol{0} & 0& \ldots & \boldsymbol{0}  \\
\vdots & \vdots&\vdots & \vdots\\
\boldsymbol{0}  & \boldsymbol{0} & 0& \boldsymbol{\Lambda_k^{-1}} \\
\end{bmatrix}}_{\text{$\boldsymbol{\Lambda}^{-1}$}}
\begin{bmatrix}
    \boldsymbol{U}^*_1 \\
    \boldsymbol{U}^*_2  \\ 
    \boldsymbol{U}^*_3  \\
    \vdots         \\
    \boldsymbol{U}^*_T  \\
\end{bmatrix}    
\end{equation}
Since the generic $t^{th}$  block of dimension $k \times N$  reads as follows : 
\begin{equation}
\boldsymbol{\Phi_t} = \boldsymbol{\Phi}_{t-1} + \boldsymbol{\Lambda_k}^{-1}\boldsymbol{U_t}^* \hspace{2cm} vec(\boldsymbol{U}_t^*) \sim \mathcal{N}(0, \boldsymbol{\Sigma_u} \otimes \boldsymbol{I}_k)    
\end{equation}
%%%%%%%%%%%%%%%%%%%%%%%%%%%%%%%%%%%%%%%%%%%%%%%%%%%%%%%%%%%%%%%%%%%%%%%%%%
\iffalse
In other words the innovations in the state equations for a generic time $t$ are the $t^{th}$ $k \times N$ block of  $ \boldsymbol{\Lambda}^{-1}\boldsymbol{U^*}$ that reads as follows: 
\[
\boldsymbol{N_t} = \boldsymbol{\Lambda^{-1}}\boldsymbol{u}^*_t 
\]
Notice that $vec(\boldsymbol{N}_t) = \boldsymbol{\eta_t}$, therefore taking vec we get: 
\[
\boldsymbol{\eta_t} = (\boldsymbol{I_N \otimes\Lambda^{-1}})vec(\boldsymbol{u}^*_t)
\]
\fi
%%%%%%%%%%%%%%%%%%%%%%%%%%%%%%%%%%%%%%%%%%%%%%%%%%%%%%%%%%%%%%%%%%%%%%%%%%
defining $\boldsymbol{\Xi_t} = \boldsymbol{\Lambda_k}^{-1}\boldsymbol{U_t}^*$ we obtain: 
\begin{equation}
vec(\boldsymbol{\Xi_t}) \sim  \mathcal{N}(0, \boldsymbol{\Sigma_u}\otimes (\boldsymbol{\Lambda_k'\Lambda_k)}^{-1})     
\end{equation}
that is equivalent to the dynamic linear model : 
\begin{equation}    \underbrace{\boldsymbol{y}'_t}_{\text{$1 \times N$}} = \underbrace{\boldsymbol{x}'_t}_{\text{$1 \times k$}} \underbrace{\boldsymbol{\Phi}_t}_{\text{$k \times N$}} + \underbrace{\boldsymbol{u_t}'}_{\text{$1\times N$}} \hspace{3cm}  \boldsymbol{u_t} \sim \mathcal{N}(\boldsymbol{0}_{N \times 1}, \boldsymbol{\Sigma_u})
\end{equation}
\begin{equation}  
vec(\boldsymbol{\Phi_t}) = vec(\boldsymbol{\Phi_{t-1}}) + vec(\boldsymbol{\Xi_t})\hspace{2cm} vec(\boldsymbol{\Xi_t}) \sim \mathcal{N}(0, \boldsymbol{\Sigma} \otimes (\boldsymbol{\Lambda_k}'\boldsymbol{\Lambda_k})^{-1}) 
\end{equation}
where $\boldsymbol{\Omega} = (\boldsymbol{\Lambda_k'\Lambda_k)}^{-1}$. Therefore, in this model, modeling the time variation of the coefficients can be achieved simply by adding the dummy observations, defined in (\ref{dummyo}) to the system. 

\subsubsection{Details on the moment matrices}\label{popmoments}
The likelihood of the observations simulated from the model of the economic theory $\boldsymbol{Y(\theta)}$ is:
\begin{equation}\scriptsize
\begin{aligned}
p(\boldsymbol{Y(\theta)}| \boldsymbol{\Phi}, \boldsymbol{\Sigma_u}) & =  (2\pi)^{-\frac{T N}{2}}|\boldsymbol{ \Sigma_u}|^{-\frac{T}{2}} exp \left[-\frac{1}{2}tr\left(( \boldsymbol{\Sigma_u}^{-1})(\boldsymbol{Y(\theta)'Y(\theta)} - \boldsymbol{\Phi'X(\theta)'Y(\theta)} - \boldsymbol{Y(\theta)'X(\theta)\Phi} + \boldsymbol{\Phi'X(\theta)'X(\theta)\Phi} )\right)  \right]  \\
\end{aligned}
\end{equation}
Considering $\gamma$ replications of $Y(\boldsymbol{\theta})$ we get 
\begin{equation}\scriptsize
\begin{aligned}
p(\boldsymbol{Y(\theta)}| \boldsymbol{\Phi}, \boldsymbol{\Sigma_u}) & =  (2\pi)^{-\frac{\gamma T N}{2}}|\boldsymbol{ \Sigma_u}|^{-\frac{\gamma T}{2}} exp \left[-\frac{1}{2}tr\left(( \boldsymbol{\Sigma_u}^{-1})(\boldsymbol{\gamma Y(\theta)'Y(\theta)} -\gamma \boldsymbol{\Phi'X(\theta)'Y(\theta)} - \gamma \boldsymbol{Y(\theta)'X(\theta)\Phi} + \gamma \boldsymbol{\Phi'X(\theta)'X(\theta)\Phi} )\right)  \right]  \\
\end{aligned}
\end{equation}
Next, following the approach of \citet{DNS2004}  the sample moments $\boldsymbol{X(\theta)'X(\theta)}$, $\boldsymbol{
X(\theta)'Y(\theta)}$, $\boldsymbol{
Y(\theta)'Y(\theta)}$ are replaced by their expected values. Taking expectations conditionally on $\boldsymbol{\theta}$ we define:

\begin{equation}
 \boldsymbol{\Gamma_{xx}(\theta)} \equiv \E[\boldsymbol{X'X}|\boldsymbol{\theta}]  = 
\begin{bmatrix}
   \boldsymbol{\Gamma_{xx,1}(\theta)} & \boldsymbol{0}_k & \ldots & \boldsymbol{0}_k \\
    \boldsymbol{0}_k & \boldsymbol{\Gamma_{xx,2}(\theta)} & \ldots & \boldsymbol{0}_k \\
    \vdots & \boldsymbol{0}_k & \ddots & \boldsymbol{0}_k                        \\
    \boldsymbol{0}_k & \ldots & \boldsymbol{0}_k & \boldsymbol{\Gamma_{xx,T}(\theta)} \\
\end{bmatrix}    
\end{equation}
\begin{equation}
 \boldsymbol{\Gamma_{xy}(\theta)} \equiv \E [\boldsymbol{X'X}|\boldsymbol{\theta}]  = 
\begin{bmatrix}
    \boldsymbol{\Gamma_{xy,1}(\theta)} \\
    \boldsymbol{\Gamma_{xy,2}(\theta)} \\
    \ldots                             \\
    \boldsymbol{\Gamma_{xy,T}(\theta)}
\end{bmatrix}   
\end{equation}
\begin{equation}
 \boldsymbol{\Gamma_{yy}(\theta)} \equiv \E [\boldsymbol{Y'Y}|\boldsymbol{\theta}] 
 = \sum_{t=1}^T \boldsymbol{\Gamma_{yy,t}(\theta)}    
\end{equation}
where $\boldsymbol{\Gamma_{xx,t}} = \E [\boldsymbol{x_tx_t'}|\boldsymbol{\theta}]$, $\boldsymbol{\Gamma_{xy,t}} = \E [\boldsymbol{x_ty_t'}|\boldsymbol{\theta}]$ and $\boldsymbol{\Gamma_{yy,t}} \equiv \E [\boldsymbol{y_ty_t'}|\boldsymbol{\theta}]$ for $t = 1, \ldots, T$. Substituting in the likelihood we get \begin{equation}\scriptsize
\begin{aligned}\label{dsgeobs}
p(\boldsymbol{Y(\theta)}^*| \gamma, \boldsymbol{\Phi}, \boldsymbol{\Sigma_u}) & =  (2\pi)^{-\frac{\gamma T N}{2}}|\boldsymbol{ \Sigma_u}|^{-\frac{\gamma T}{2}} exp \left[-\frac{1}{2}tr\left(( \boldsymbol{\Sigma_u}^{-1})(\gamma\boldsymbol{\Gamma_{yy}(\theta)} - \gamma\boldsymbol{\Phi'\Gamma_{xy}(\theta))} - \gamma\boldsymbol{\Gamma_{yx}(\theta)\Phi} + \gamma\boldsymbol{\Phi'\Gamma_{xx}(\theta)\Phi} )\right)  \right]  \\
\end{aligned}
\end{equation}
%%%%%%%%%%%%%%%%%%%%%%%%%%%%%%%%%%%%%%%%%%%%%%%%%%%%%%%%%%%%%%%%%%%%%
\iffalse
Define $\tilde{\boldsymbol{s}}_t = \boldsymbol{s}_t - \mathcal{C}_t $
Rewrite the state space model as

\begin{equation}
 \boldsymbol{y}_t =   \mathcal{D} + \mathcal{B}[\tilde{\boldsymbol{s}}_t +  \mathcal{C}_t]
\end{equation}
\begin{equation}
 \tilde{\boldsymbol{s}_t} =   \mathcal{T}_t\boldsymbol{s}_{t-1} + \mathcal{R}_t\boldsymbol{\varepsilon}_t  
\end{equation}

\begin{equation}
\begin{aligned}
     \E[\boldsymbol{y}_t\boldsymbol{y}_t '] & = \E[(\mathcal{D} + \mathcal{B}\boldsymbol{s}_t) ( \mathcal{D}' + \mathcal{B}'\boldsymbol{s}_t')]    \\ & = \mathcal{D}\mathcal{D}' + \mathcal{B}\boldsymbol{s}_t) (  + \mathcal{B}'\boldsymbol{s}_t')] 
\end{aligned}
\end{equation}
\fi
%%%%%%%%%%%%%%%%%%%%%%%%%%%%%%%%%%%%%%%%%%%%%%%%%%%%%%%%%%%%%%%%%%%%%%%
\subsubsection{Integrating constant of the theory coherent prior}\label{ic_prior}
The integrating constant of the \textit{Normal-Inverse-Wishart} prior 

\begin{equation}
 c(\lambda,\boldsymbol{\theta},\gamma) = \int_{-\infty}^{\infty}  p(\boldsymbol{\Phi, \Sigma_u}|\lambda)p(\boldsymbol{Y(\theta)}| \gamma, \boldsymbol{\Phi}, \boldsymbol{\Sigma_u}) d\boldsymbol{\Phi} d\boldsymbol{\Sigma_u}  
\end{equation}
is given by: 

\label{intconst}
\begin{equation}
c(\boldsymbol{\lambda,\theta},\gamma) = \left( \pi\right)^{-\frac{\gamma N}{2}}\frac{\Gamma_N\left(\frac{\underline{\underline{\nu}} }{2} \right) |\underline{\underline{S}}|^{\frac{-\underline{\underline{\nu}}}{2}} |\underline{\underline{\boldsymbol{\Psi}}}|^{\frac{N}{2}}}{\Gamma_N\left(\frac{\underline{\nu}}{2} \right) |\underline{S}|^{-\frac{\underline{\nu}}{2}}|\boldsymbol{\underline{\Psi}}|^{\frac{N}{2}}}  
\end{equation}
    where the definitions for $\underline{\underline{\boldsymbol{S}}}$, $ \boldsymbol{\underline{\underline{\Psi}}}$, $\boldsymbol{\underline{\underline{\Phi}}}$,$ \boldsymbol{\underline{\Psi}}$, $ \underline{\boldsymbol{S}}$, $\boldsymbol{\underline{\Psi}}$, are given above in the text while  $\boldsymbol{\Gamma}_{N}(.)$ is the Gamma function.

\subsubsection{Marginal likelihood and fit-complexity trade off}\label{mdd_anal}
The marginal likelihood is given by: 
\begin{equation}
p(\boldsymbol{Y}|\lambda,\boldsymbol{\theta},\gamma) = \left(\pi\right)^{-\frac{T N}{2}}\frac{\Gamma_N\left(\frac{\tilde{{\nu}} }{2} \right) |\tilde{{\boldsymbol{S}}}|^{-\frac{\tilde{\nu}}{2}} |\tilde{{\boldsymbol{\Psi}}}|^{\frac{N}{2}}}{\Gamma_N\left(\frac{\underline{\nu}}{2} \right) |\underline{\underline{\boldsymbol{S}}}|^{-\frac{\underline{\underline{\nu}}}{2}}|\boldsymbol{\underline{\underline{\Psi}}}|^{\frac{N}{2}}} 
\end{equation}
Following the same steps as in \citep{giannonelenzaprimicieri} it can be re-written as : 
\begin{equation}  
p(\boldsymbol{Y}|\lambda,\boldsymbol{\theta},\gamma)  = const | \boldsymbol{(V_{\varepsilon}^{post})^{-1}V^{prior}_{\varepsilon}}|^{\frac{T + \underline{\underline{\nu}}}{2}} \prod_{t=1}^{T}|\boldsymbol{V_{t|t-1}}|^{-\frac{1}{2}} 
\end{equation}
where 
%\[ 
% \boldsymbol{\tilde{\phi}} = vec(\boldsymbol{\hat{\Phi}}) = vec\left((\boldsymbol{X'X}  + \boldsymbol{\underline{\underline{\Psi}}^{-1}})^{-1}(\boldsymbol{X'Y + \underline{\underline{\Psi}}^{-1}\underline{\underline{\Phi}}})\right)
%\]

\begin{equation}
vec(\boldsymbol{\hat{\varepsilon}_t}) = vec( \boldsymbol{Y - X\tilde{\Phi} })     
\end{equation} 

\begin{equation}
 \boldsymbol{V}_{\varepsilon}^{prior} = E [\boldsymbol{\Sigma}] = \frac{\boldsymbol{\underline{\underline{S}}}}{\underline{\underline{\nu}} - N -1}   
\end{equation}
\begin{equation}
\boldsymbol{V}_{\varepsilon}^{post} = E [\boldsymbol{\Sigma}|\boldsymbol{Y}] =\frac{\boldsymbol{\underline{\underline{S}}} + \boldsymbol{\hat{\varepsilon}'\hat{\varepsilon}} + (\boldsymbol{\tilde{\Phi}} - \boldsymbol{\underline{\underline{\Phi}}})'\boldsymbol{\underline{\underline{\Psi}}}^{-1} (\boldsymbol{\tilde{\Phi}} - \boldsymbol{\underline{\underline{\Phi}}})}{T + \underline{\underline{\nu}} -N - 1}    
\end{equation}
\begin{equation}
\boldsymbol{V}_{t|t-1} = \frac{\boldsymbol{{\underline{\underline{S}}}}}{\underline{\underline{\nu}} - N - 1 } \boldsymbol{\otimes} \left( 1 + \boldsymbol{X_t'(X_{t-1}'X_{t-1} + \underline{\underline{\Psi}}^{-1})^{-1}X_{t}}\right)
\end{equation}
%\boldsymbol{V_{t|t-1}} = \frac{\boldsymbol{\underline{S}}}{\underline{\underline{\nu}} - N - 1 } \otimes \left( 1 + \boldsymbol{X^t'(X^{t-1}'X^{t-1} + \underline{\underline{\Psi^{-1}}})^{-1}X^t}\right)   
\begin{equation}
    const = \left( \frac{1}{\pi} \right)^{\frac{NT}{2}} \frac{\Gamma_N\left(\frac{T + \underline{\underline{\nu}}}{2} \right)}{\Gamma_N\left( \frac{\underline{\underline{\nu}}}{2}\right)}\frac{ (T + \underline{\underline{\nu}} - N - 1)^{\underline{\underline{\nu}}/2}}{(\underline{\underline{\nu}} - N - 1)^{T +{\underline{\underline{\nu}}}/2}}
\end{equation}

\subsubsection{Formulas with distinct \texorpdfstring{$\lambda_j$}{TEXT} for \texorpdfstring{$j=1, \ldots, K$}{TEXT}}\label{sec:distinctlambda}

In the case we consider regressor specific hyper-parameters $\lambda_j$ with $j=1, \ldots, K$, defining $\boldsymbol{\Lambda_k} = diag(\lambda_1, \ldots,\lambda_k )$ the theory coherent prior becomes: 

\begin{equation}
 p(vec(\boldsymbol{\Phi})|\boldsymbol{\Sigma}_u, \boldsymbol{\Lambda},\boldsymbol{\theta},\gamma) \sim \mathcal{N}(vec(\boldsymbol{\underline{\underline{\Phi}}}),\boldsymbol{\Sigma}_u \otimes \boldsymbol{\underline{\underline{\Psi}}})    
\end{equation}
\begin{equation}
 vec(\boldsymbol{\underline{\underline{\Phi}}})  = vec( \left(\gamma\boldsymbol{\Gamma_{xx}(\theta)} + \boldsymbol{H_{Tk}'H_{Tk}(I_T \otimes \Lambda_k'\Lambda_k})\right)^{-1}( \gamma\boldsymbol{\Gamma_{xy}(\theta)} + \boldsymbol{H}_{Tk}'\boldsymbol{H}_{Tk}(\boldsymbol{I_T \otimes \Lambda_k'\Lambda_k})\boldsymbol{\underline{\Phi_{0}}} ))   
\end{equation}
\begin{equation}
 \boldsymbol{\underline{\underline{\Psi}}} = \left( (\gamma\boldsymbol{\Gamma_{xx}(\theta)}  +  \boldsymbol{H_{Tk}'H_{Tk}(I_T \otimes \Lambda_k'\Lambda_k})\right)^{-1}   
\end{equation}
  
\begin{equation}
p(\boldsymbol{\Sigma}_u|\boldsymbol{\Lambda},\boldsymbol{\theta},\gamma) \sim \mathcal{IW} \left({\underline{\underline{\boldsymbol{S}}}}, \underline{\underline{\nu}}\right)    
\end{equation}

\begin{equation}
\underline{\underline{\boldsymbol{S}}} = \underline{S} + \gamma \boldsymbol{\Gamma_{yy}(\theta)} +   \boldsymbol{\underline{\Phi_{0}}}'(\boldsymbol{H_{Tk}}' \boldsymbol{H_{Tk}})(\boldsymbol{I_T} \otimes \boldsymbol{\Lambda'\Lambda})\boldsymbol{\underline{\Phi_{0}}} - \boldsymbol{\underline{\underline{\Phi}}}'\boldsymbol{\underline{\underline{\Psi}}}^{-1}\boldsymbol{\underline{\underline{\Phi}}}    
\end{equation}
\begin{equation}
    \underline{\underline{\nu}} = \underline{\nu} + \gamma
\end{equation}
The formula of the conditional posterior of $\boldsymbol{\Phi}$ and $\boldsymbol{\Sigma_u}$ becomes

\begin{equation}
\begin{aligned}
  p(vec(\boldsymbol{\Phi})|\boldsymbol{\Sigma}_u, \boldsymbol{\Lambda},\boldsymbol{\theta},\gamma, \boldsymbol{Y}) & \sim \mathcal{N}(vec(\boldsymbol{\tilde{\Phi}}),\boldsymbol{\Sigma}_u \otimes \boldsymbol{\tilde{\Psi}})  \\
  p(\boldsymbol{\Sigma}_u| \boldsymbol{\Lambda},\boldsymbol{\theta},\gamma, \boldsymbol{Y}) & \sim \mathcal{IW} \left({\boldsymbol{\tilde{S}}}, \tilde{\nu}\right)
\end{aligned}
\end{equation}
\begin{equation}\small
  vec(\boldsymbol{\tilde{\Phi}}) = vec(\left( \boldsymbol{X'X} 
+ \gamma\boldsymbol{\Gamma_{xx}(\theta)} + \boldsymbol{H_{Tk}'H_{Tk}(I_T \otimes \Lambda_k'\Lambda_k})\right)^{-1}( \boldsymbol{X'Y} +  \gamma\boldsymbol{\Gamma_{xy}(\theta)} + \boldsymbol{H_{Tk}'H_{Tk}(I_T \otimes \Lambda_k'\Lambda_k})\boldsymbol{\underline{\Phi_{0}}}))   
\end{equation}
\begin{equation}
     \boldsymbol{\tilde{\Psi}} =  \left(\boldsymbol{X'X} + \gamma\boldsymbol{\Gamma_{xx}(\theta)}  +  \boldsymbol{H_{Tk}'H_{Tk}(I_T \otimes \Lambda_k'\Lambda_k})\right)^{-1}
\end{equation}
\begin{equation}
 \tilde{\boldsymbol{S}} = \boldsymbol{Y'Y} + \underline{\underline{\boldsymbol{S}}} + \boldsymbol{\underline{\underline{\Phi}}}'\boldsymbol{\underline{\underline{\Psi}}}^{-1}\boldsymbol{\underline{\underline{\Phi}}} - \boldsymbol{\tilde{\Phi}}'\boldsymbol{\tilde{\Psi}}^{-1}\boldsymbol{{\tilde{\Phi}}}   
\end{equation}
\begin{equation}
\tilde{\nu} = \underline{\underline{\nu}} + T 
\end{equation}
The marginal likelihood is just (\ref{ML}), updated with the new definitions of $ \underline{\underline{\nu}}$, $\underline{\underline{\boldsymbol{S}}}$, $ \boldsymbol{\underline{\underline{\Psi}}}$, $\boldsymbol{\underline{\underline{\Phi}}}$,$ \boldsymbol{\tilde{\Psi}}$, $ \tilde{\boldsymbol{S}}$, $\boldsymbol{\tilde{\Psi}}$.

%%%%%%%%%%%%%%%%%%%%%%%%%%%%%%%%%%%%%%%%%%%%%%%%%%%%%%%%%%%%%%%%%%%%%%%%%%%%%%%%%%%%%%%%%%%%%%%%%%%%%%%%%%%%%%%%%%%%%%%%%%%%%%%%%%%%%%%%%%%%%%%%%%%%%%%%%%%%%%%%%%%%%%%%%%%%%%%%%%%%%%%%%%%%%%%%%%%%%%%%%%%%%%%%%%%%%%%%%%%%%%%%%%%%%%%%%%%%%%%%%%%%%%%%%%%%%%%%%%%%%%%%%%%%%%

%

\subsubsection{TVP-VAR with time varying error covariance matrix}\label{sec:tctvp_var_sv}

In order to estimate the TVP-VAR with time varying error covariance matrix discussed in section \ref{sec:tvolatility} the MCMC sampler iterates the following steps:

    \begin{enumerate}
    \item Draw from $p(\lambda, \boldsymbol{\theta}, \gamma|\boldsymbol{Y}, \boldsymbol{D}) \propto p(\boldsymbol{Y}|\boldsymbol{D},\lambda,\boldsymbol{\theta},\gamma)p(\lambda, \boldsymbol{\theta}, \gamma)$. 
    \item Draw from $p(\boldsymbol{\Phi, \Sigma_u} | \lambda,\boldsymbol{\theta}, \gamma,\boldsymbol{D},\boldsymbol{Y})$ %\sim \mathcal{NIW} $
    \item Draw from $p(\boldsymbol{D},\rho, \sigma^2_{\eta}| \boldsymbol{\Phi}, \boldsymbol{\Sigma}_u,\boldsymbol{Y})$ 
    \end{enumerate}
In the random walk metropolis in step 1, $p(\boldsymbol{Y}|\boldsymbol{D},\lambda,\boldsymbol{\theta},\gamma)$ is given by:

\begin{equation}
p(\boldsymbol{Y}|\boldsymbol{D},\lambda,\boldsymbol{\theta},\gamma) = \left(\pi\right)^{-\frac{T N}{2}}\frac{\Gamma_N\left(\frac{\tilde{{\nu}} }{2} \right) |\tilde{{\boldsymbol{S}}}|^{-\frac{\tilde{\nu}}{2}} |\tilde{{\boldsymbol{\Psi}}}|^{\frac{N}{2}}}{\Gamma_N\left(\frac{\underline{\nu}}{2} \right) |\underline{\underline{\boldsymbol{S}}}|^{-\frac{\underline{\underline{\nu}}}{2}}|\boldsymbol{\underline{\underline{\Psi}}}|^{\frac{N}{2}}} 
\end{equation}
where $ \underline{\underline{\nu}}$, $\underline{\underline{\boldsymbol{S}}}$, $ \boldsymbol{\underline{\underline{\Psi}}}$ are defined in (\ref{prior_nu}), (\ref{prior_S}), (\ref{prior_PSI}) while $\tilde{{\boldsymbol{S}}}$ $\tilde{{\boldsymbol{\Psi}}}$ and ${\tilde{\nu}}$ are defined in equations (\ref{Ssv}) (\ref{psisv}) and (\ref{nusv}).

\clearpage
\subsection{Forecasting exercise and small scale NK model}\label{SMALLMODEL}
\subsubsection{Data}\label{data_ex1}
The data for the out of sample forecasting exercise in Section \ref{oos} are taken from the \href{https://research.stlouisfed.org/wp/more/2020-005}{FRED-QD Dataset} from the Federal bank of St. Louis. The series IDs of the time series are \texttt{GDPC1}, \texttt{CPIAUCSL}, \texttt{FEDFUNDS}. Quarterly GDP growth and quarterly inflation rate are obtained transforming the series according to  $\Delta \% y_t = 100\left(\frac{y_t- y_{t-1}}{y_{t-1}}\right)$. As for the Fed Funds interest rate it is transformed by taking the logarithm.

\subsubsection{Competing models in the forecasting exercise:}\label{competing}
The competing models in the out of sample forecasting exercise in Section \ref{oos} are 
\begin{itemize}
    \item A constant parameters VAR with flat prior.
    \item A constant parameters VAR with \textit{Normal Inverse-Wishart} prior.  
    \item A standard TVP-VAR model, with the prior specification considered in \citet{chan2009efficient}. 
\end{itemize}
The VAR with \textit{Normal Inverse-Wishart} prior is given by:

\begin{equation}
    \boldsymbol{Y} = \boldsymbol{X\Pi} + \boldsymbol{U}  \hspace{2cm}  \boldsymbol{U} \sim MVN (\boldsymbol{0}, \boldsymbol{\Sigma}_u,\boldsymbol{I_T})
\end{equation}
where $\boldsymbol{Y}$ is $T \times N$, $\boldsymbol{X}$ is $T \times k$ with $k = Np + 1$, $\boldsymbol{\Pi}$ is $k \times n$ , $\boldsymbol{U}$ is $T \times N$ and $MVN$ stands for the matricvariate normal. The prior for the autoregressive coefficients and the variance covariance matrix is: 

\begin{equation}
         vec(\boldsymbol{\Pi}) \sim \mathcal{N}(vec(\boldsymbol{\uline{\mu_{\Pi}}}), \boldsymbol{\Sigma}_u \otimes \boldsymbol{\uline{\Omega_{\Pi}}})
\end{equation}
\begin{equation}
    \boldsymbol{\Sigma} \sim \mathcal{IW}(\boldsymbol{S_0},v_0)
\end{equation}
where $vec(\boldsymbol{\Pi})$ centers the non-stationary series on a random walk process and the stationary ones on a white noise process and  $\boldsymbol{\uline{\Omega_{\Pi}}}$ is 
a diagonal matrix with first element element equal to 
\begin{equation}
  \omega_{1} =  100 \\   
\end{equation}
and 
\begin{equation}
  \omega_{s} =  \frac{\theta_1}{\sigma_i^2l^{2}} \\   
\end{equation}
for the other elements $s = 2, \ldots,  Np$ with $\theta_1 = 0.1$ and $\sigma_i^2$ is set equal to $\hat{\sigma}_i^2$ being the estimated variance of the $i^{th}$ variable in a VAR model using a pre-sample of observations. $l$ is the lag order of the variable associated to that  variable. $\boldsymbol{S_0}$ is set equal to $\boldsymbol{\hat{S}_0}$ the variance covariance matrix estimated variance from a VAR model on a pre-sample of observations  and $v_0 = N + 2$. The standard TVP-VAR model and the corresponding choice of the prior distributions follows \citet{chan2009efficient}. The model is given by:
\begin{equation}  
\underbrace{\boldsymbol{y}'_t}_{\text{$1 \times N$}} = \underbrace{\boldsymbol{x'}_t}_{\text{$1 \times k$}} \underbrace{\boldsymbol{\Phi}_t}_{\text{$k \times N$}} + \underbrace{\boldsymbol{u_t}'}_{\text{$1 \times N$}} \hspace{3cm}  \boldsymbol{u_t} \sim \mathcal{N}(\boldsymbol{0}_{N \times 1}, \boldsymbol{\Sigma_u})
\end{equation}
\begin{equation}
    vec(\boldsymbol{\Phi_t}) = vec(\boldsymbol{\Phi_{t-1}}) + \boldsymbol{\eta}_t \hspace{2cm} \boldsymbol{\eta}_t \sim \mathcal{N}(\boldsymbol{0}, \boldsymbol{\Omega} )
\end{equation} 
where $\boldsymbol{\Omega} = diag(\omega_1, \ldots, \omega_{Nk})$  %\footnote{Note that in this model the variances in the state equation of the time varying coefficients are not constrained to be proportional across equations (the structure of $\boldsymbol{\Omega}$  is not constrained to be a Kronecker product $\boldsymbol{\Omega} = \boldsymbol{\Sigma}_u \otimes \boldsymbol{ \Lambda_k}$).}
\begin{equation}
    \omega_i \sim \mathcal{IG} 
\left( \nu , s^2\right) 
\end{equation}
where $\nu = 3$ $s^2 = 0.005$ and the prior for the variance covariance matrix is set as in the constant parameter BVAR with \textit{Normal-Inverse-Wishart} prior. 
\clearpage
\subsubsection{Prior for deep parameters of the Small Scale DSGE model}\label{secpriors_dsge}
\begin{table}[ht!] 
\centering
\caption{Prior distribution for the parameters of the New-Keynesian model}\label{priordsge}
\begin{tabular}{c|c|c|c}
Parameter    & Prior distribution & Mean & Standard Deviation  \\
\hline \hline 
$ln(\tilde{\gamma}s)$ &      Normal              &   0.500   &          0.250   \\
$ln(\pi^*)$  &      Normal             &    1.000  &          0.500    \\
$ln(r^*)$    &      Gamma              &   0.500   &   0.250      \\
$\kappa$     &      Gamma              &   0.300    &   0.150       \\
$\tau$       &      Gamma              &   2.000   &         0.500             \\
$\psi_1$     &      Gamma               &   1.500   &        0.250             \\
$\psi_2$     &      Gamma               &   0.500   &     0.200     \\
$\rho_R$     &      Beta               &   0.500   &        0.250             \\
$\rho_g$     &     Beta               &  0.800    &                    0.100  \\
$\rho_z $    &    Beta                & 0.300     &      0.100 \\
$\sigma_R$   &    Inverse Gamma                &   0.251   &                 0.139  \\
$\sigma_g$   &    Inverse Gamma                 &   0.630   &                 0.323  \\
$\sigma_z$   &    Inverse Gamma                &   0.875 
     &      0.430          \\
\hline \hline
\end{tabular}
\label{fig:priors_dsge}
\vspace{.25cm}\hspace{.85cm}\parbox{1.2\textwidth}{\scriptsize{The table reports the details on the prior distribution of the parameters of the New-Keynesian model.}}
\end{table}
\clearpage

\subsection{Medium scale New-Keynesian model}\label{sec:eq_cond}
The model is taken from \citet{delnegroschorfheide2005} and it is a version of the popular medium scale New-Keynesian model in \citet{SWauters}.
The observation equations are:
\begin{equation}
    \text{Output growth} = \gamma + 100(y_t - y_{t-1} + z_t)
\end{equation}
\begin{equation}
    \text{Consumption growth} = \gamma + 100(c_t - c_{t-1} + z_t)    
\end{equation}
\begin{equation}
    \text{Investment growth} = \gamma + 100(i_t - i_{t-1} + z_t)        
\end{equation}
\begin{equation}
    \text{Real wage growth} = \gamma + 100(w_t - w_{t-1} + z_t)        
\end{equation}
\begin{equation}
    \text{Hours worked} = \bar{l} + 100 l_t       
\end{equation}
\begin{equation}
    \text{Inflation} = \pi_{*} + 100\pi_t       
\end{equation}
\begin{equation}\label{meas_ffr}
    \text{FFR} = R_{*} + 100(R_t)        
\end{equation}
Data sources and transformations are as in \citet{delnegroschorfheide2005} to which I refer for further details. During the ZLB period, the observation equation for the interest rate includes a measurement error, namely: 
\begin{equation}
    \text{FFR} = R_{*} - R_{*}  + v_{r, t}    
\end{equation}
 with $\E [v_{r, t}] = 0$ and $var(v_{r, t}) = 0.001$. This is to account for the fact that the Fed Fund Rate was not exactly equal to zero, remaining slightly above zero in the ZLB.% As well, this condition ensures invertibility of $\boldsymbol{\Gamma_{xx,t}}$ during the ZLB period. 

%%%%%%%%%%%%%%%%%%%%%%%%%%%%%%%%%%%%%%%%%%%%%%%%%%%%%%%%%%%%%%%%%%%%%%%%%%%%%%%%%%%%%%%%%%%%%%%%%%%%%

\begin{landscape}
\centering
\begin{figure}
\subfloat{\includegraphics[width = 2.4in]{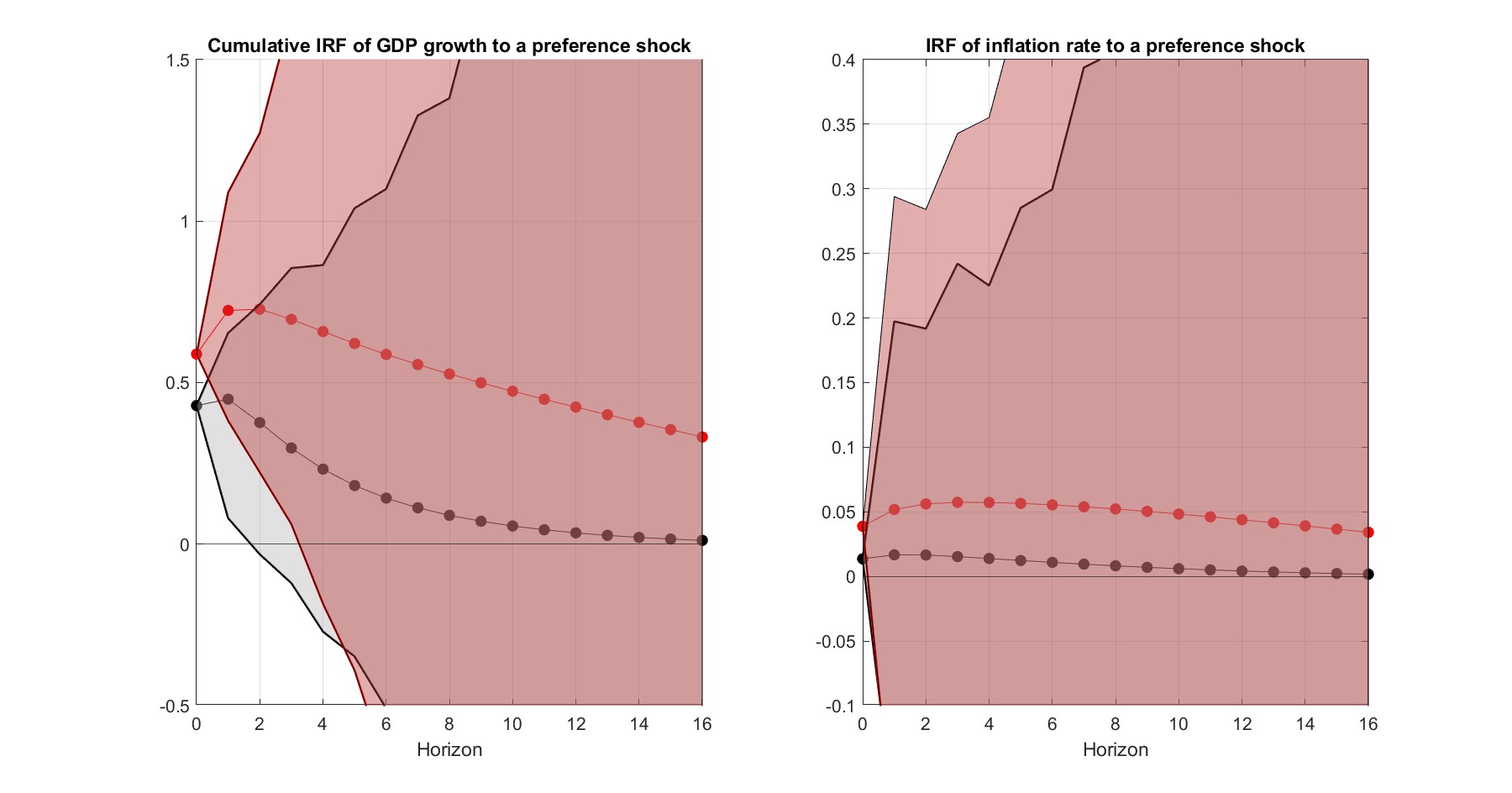}} 
\subfloat{\includegraphics[width = 2.4in]{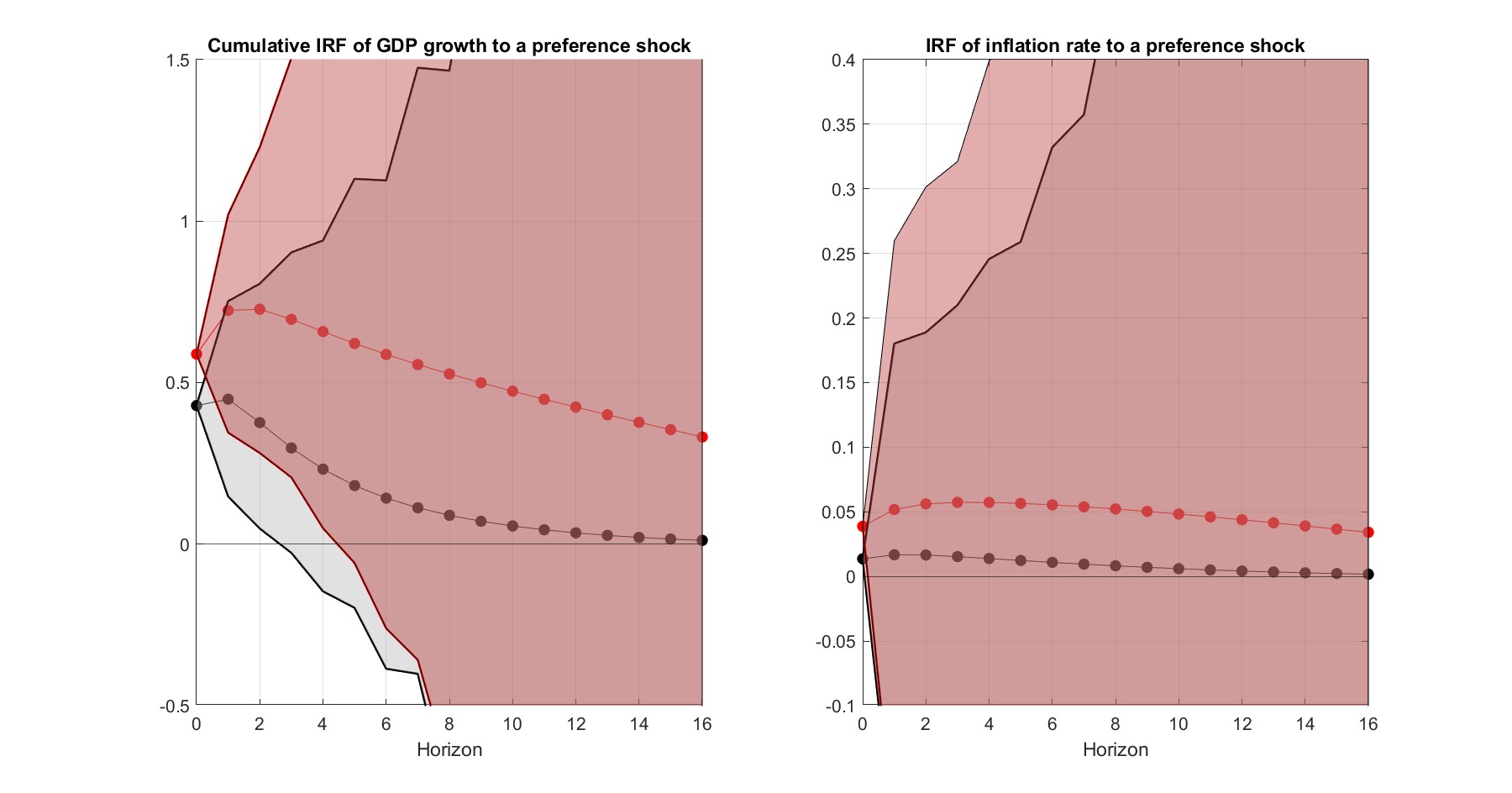}}
\subfloat{\includegraphics[width = 2.4in]{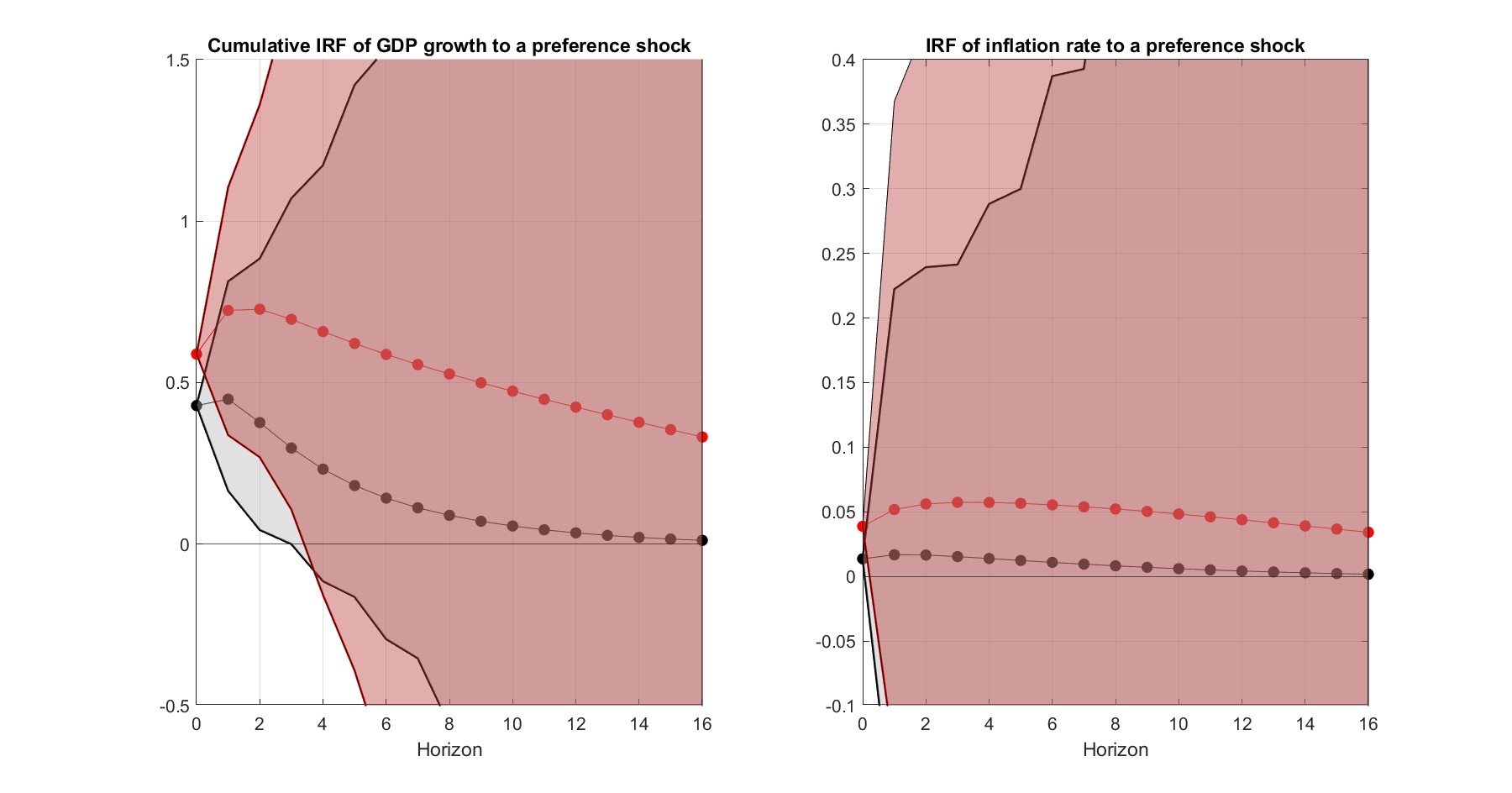}}
\subfloat{\includegraphics[width = 2.4in]{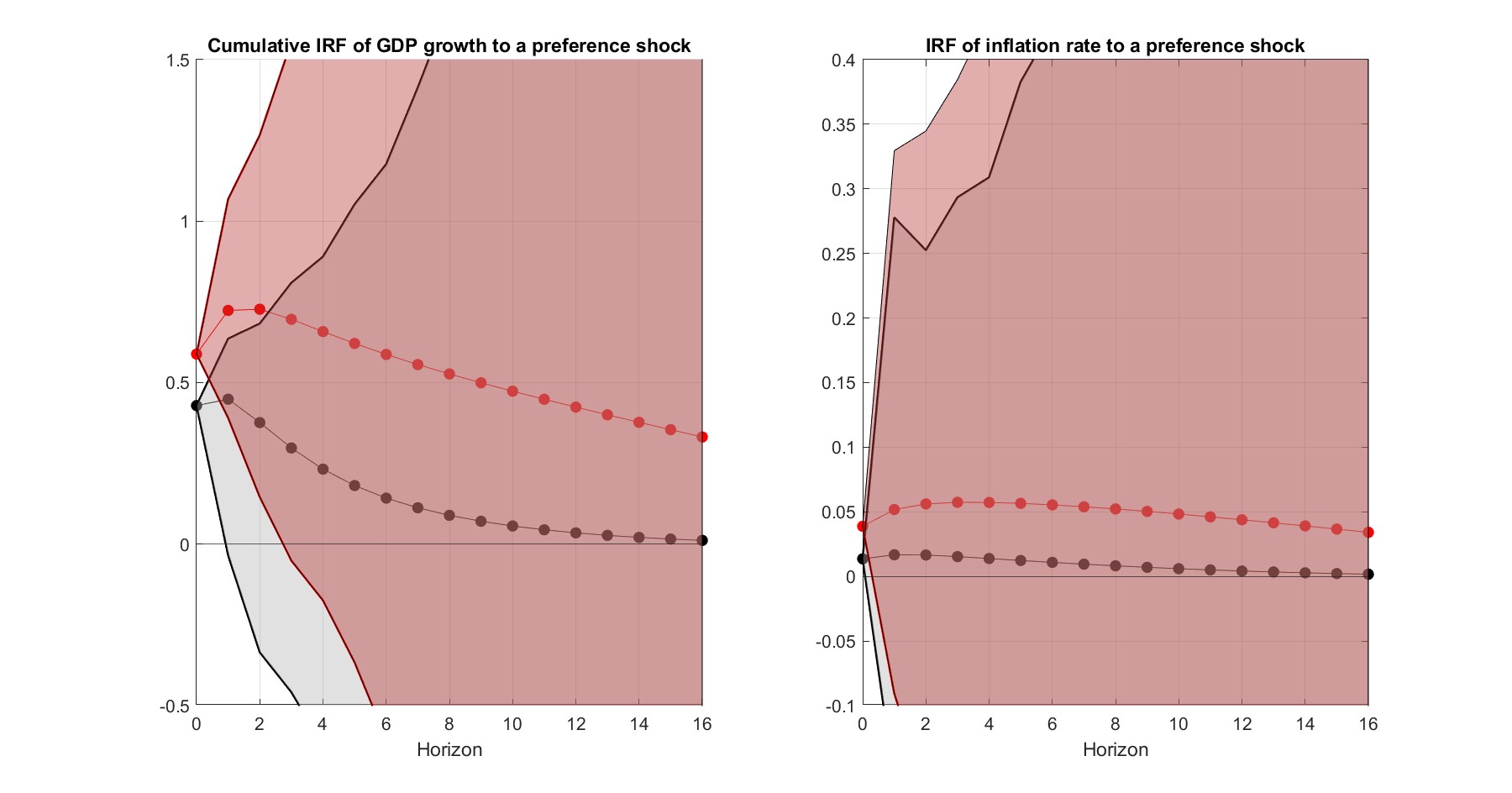}}\\
\subfloat{\includegraphics[width = 2.4in]{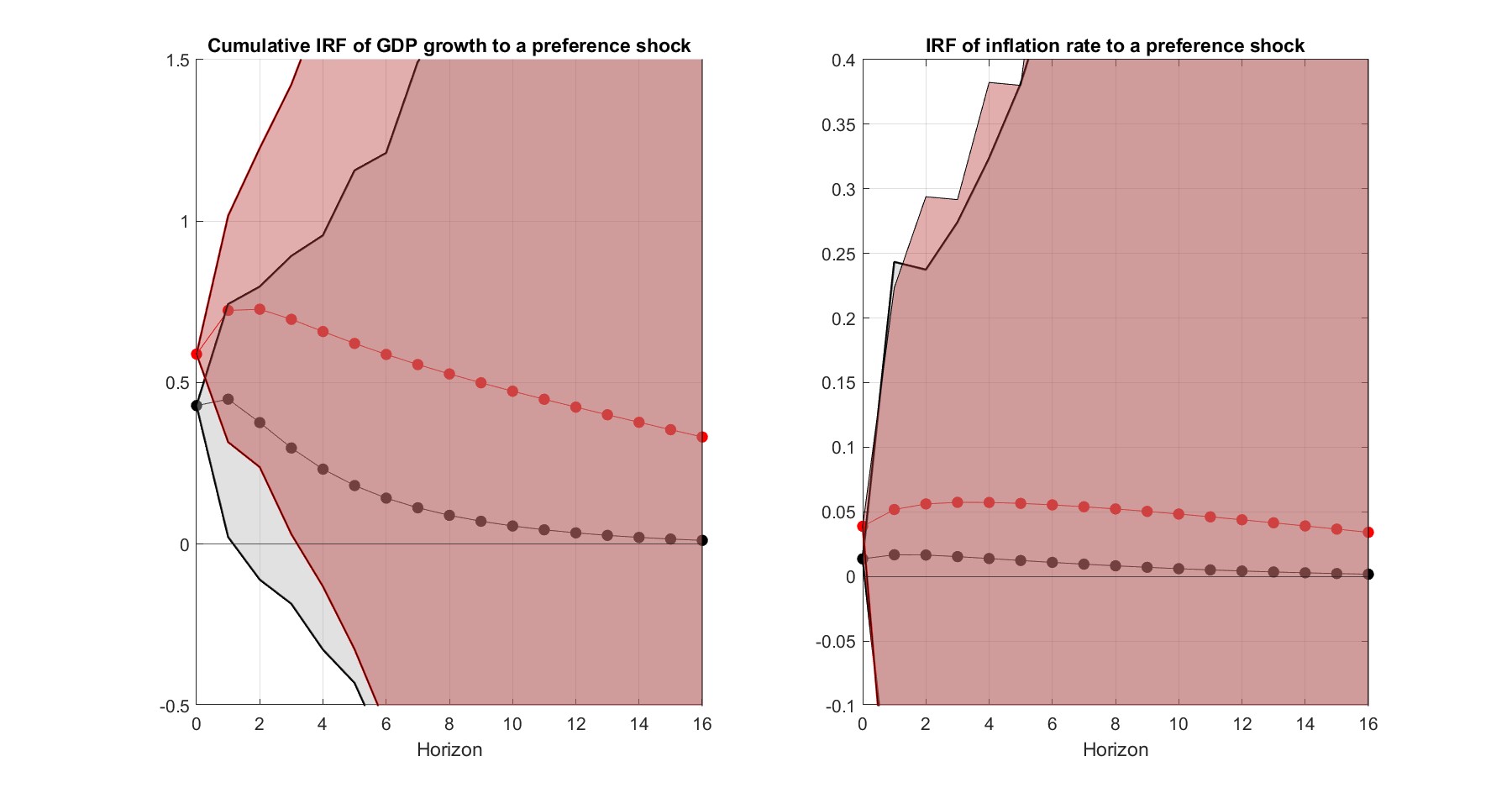}} 
\subfloat{\includegraphics[width = 2.4in]{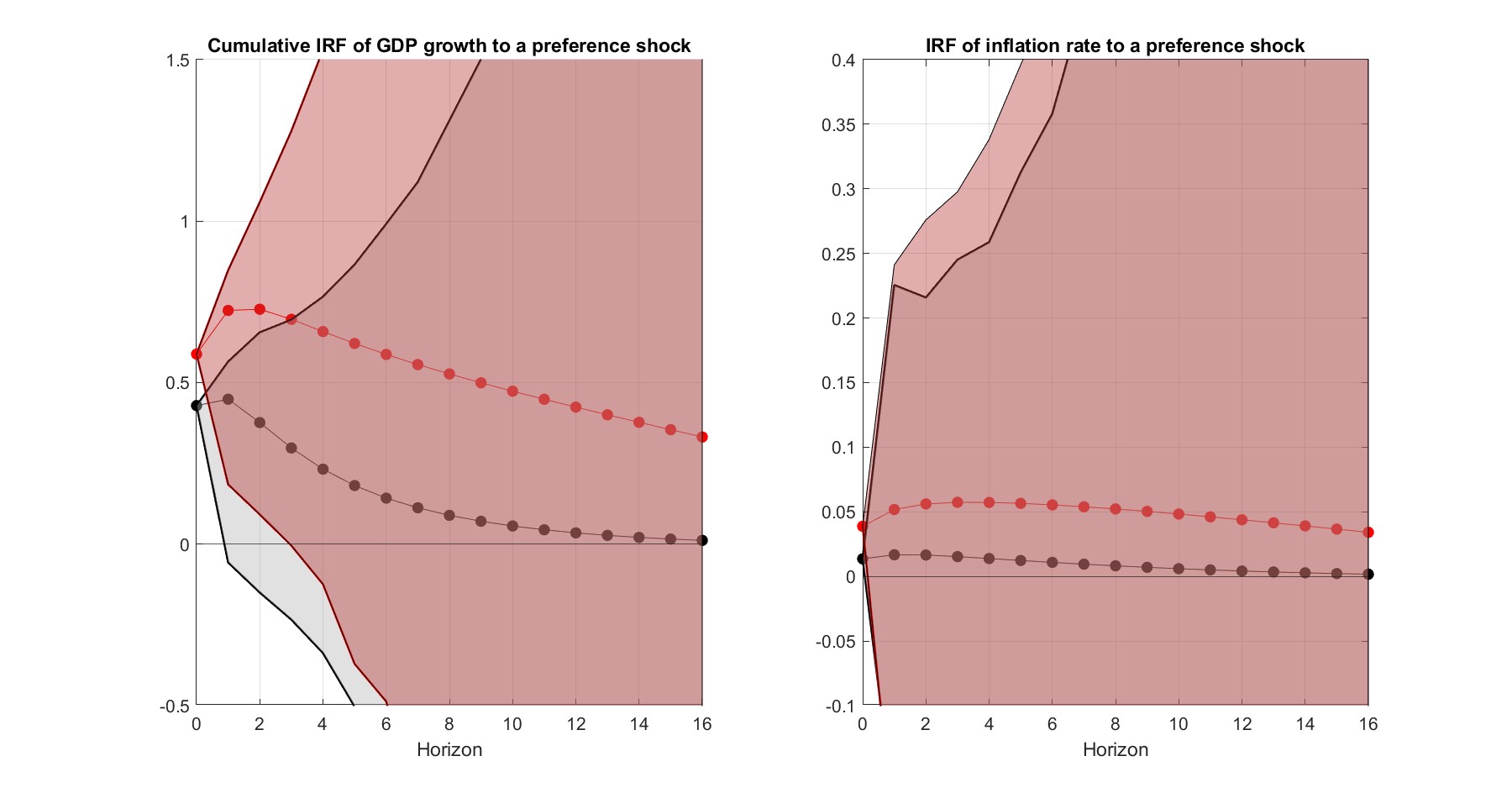}}
\subfloat{\includegraphics[width = 2.4in]{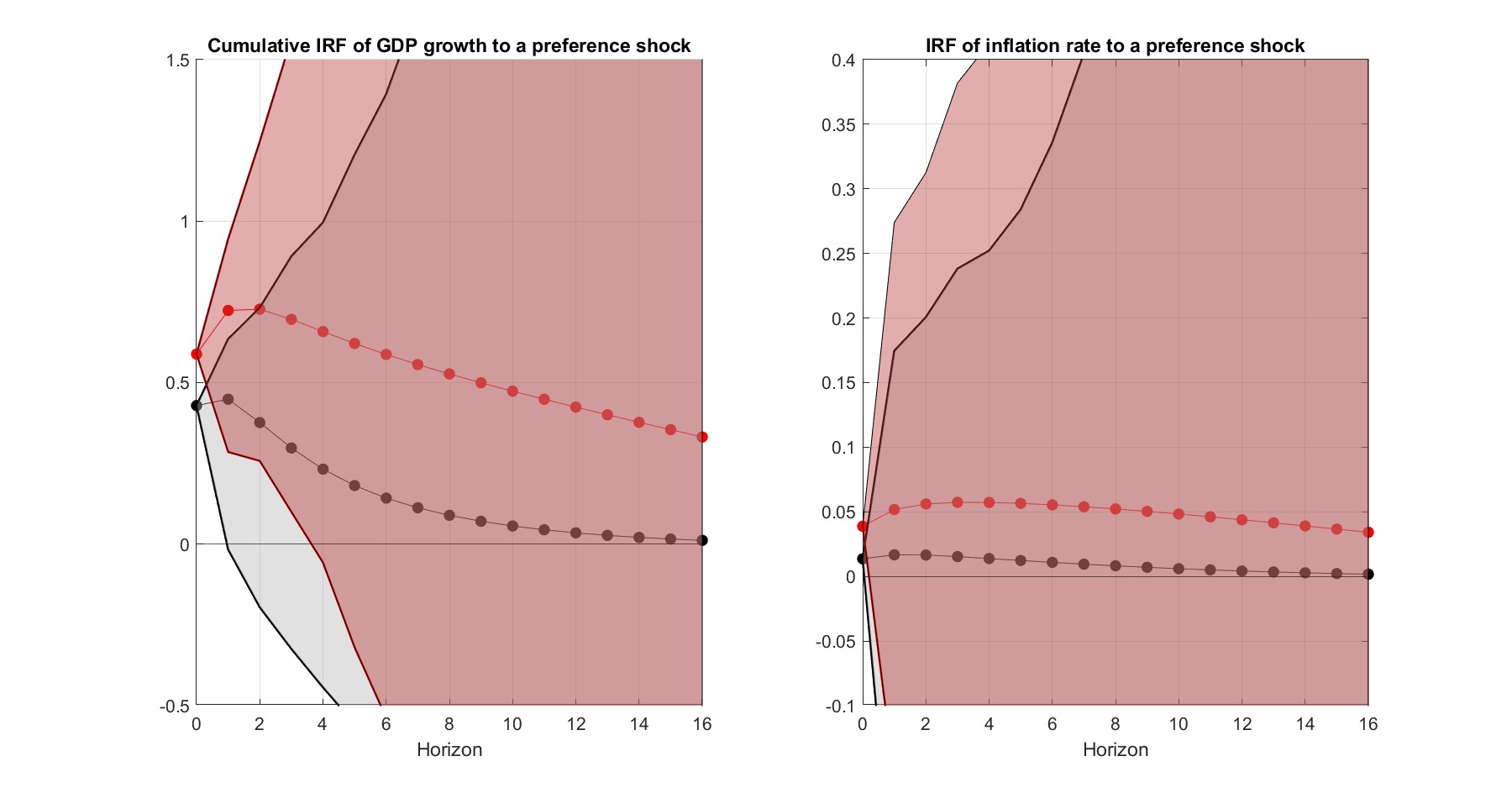}}
\subfloat{\includegraphics[width = 2.4in]{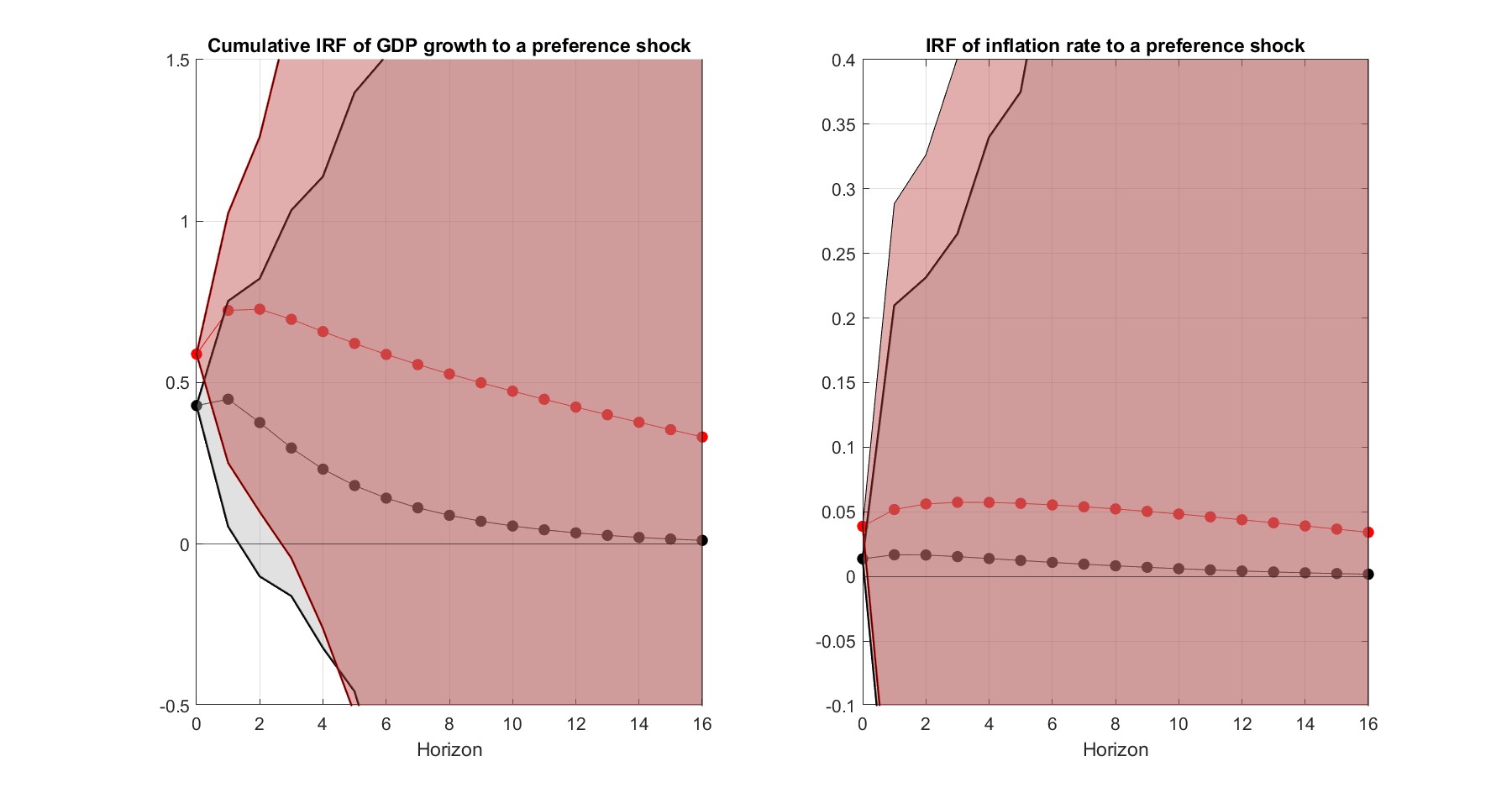}}\\
\subfloat{\includegraphics[width = 2.4in]{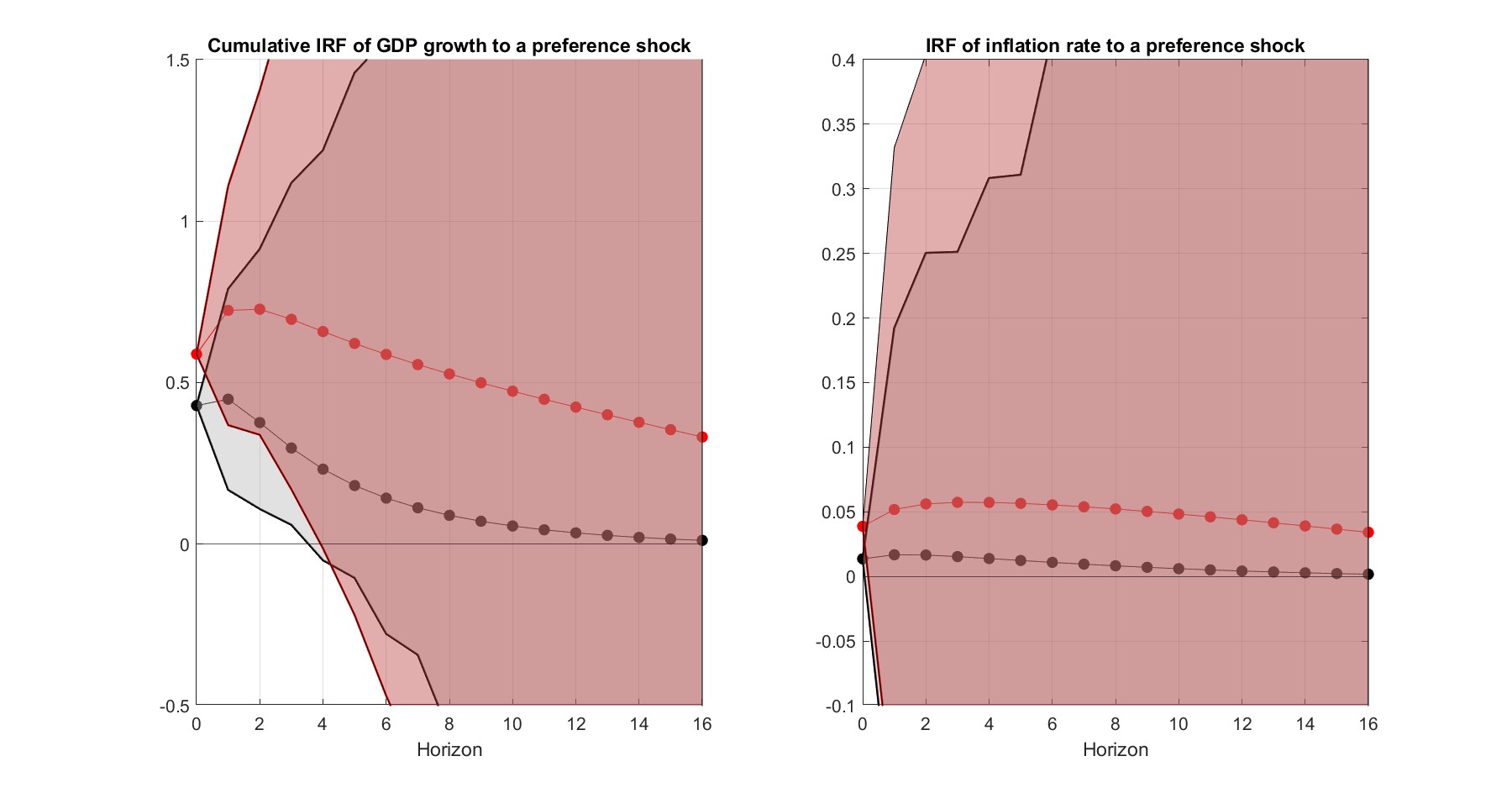}} 
\subfloat{\includegraphics[width = 2.4in]{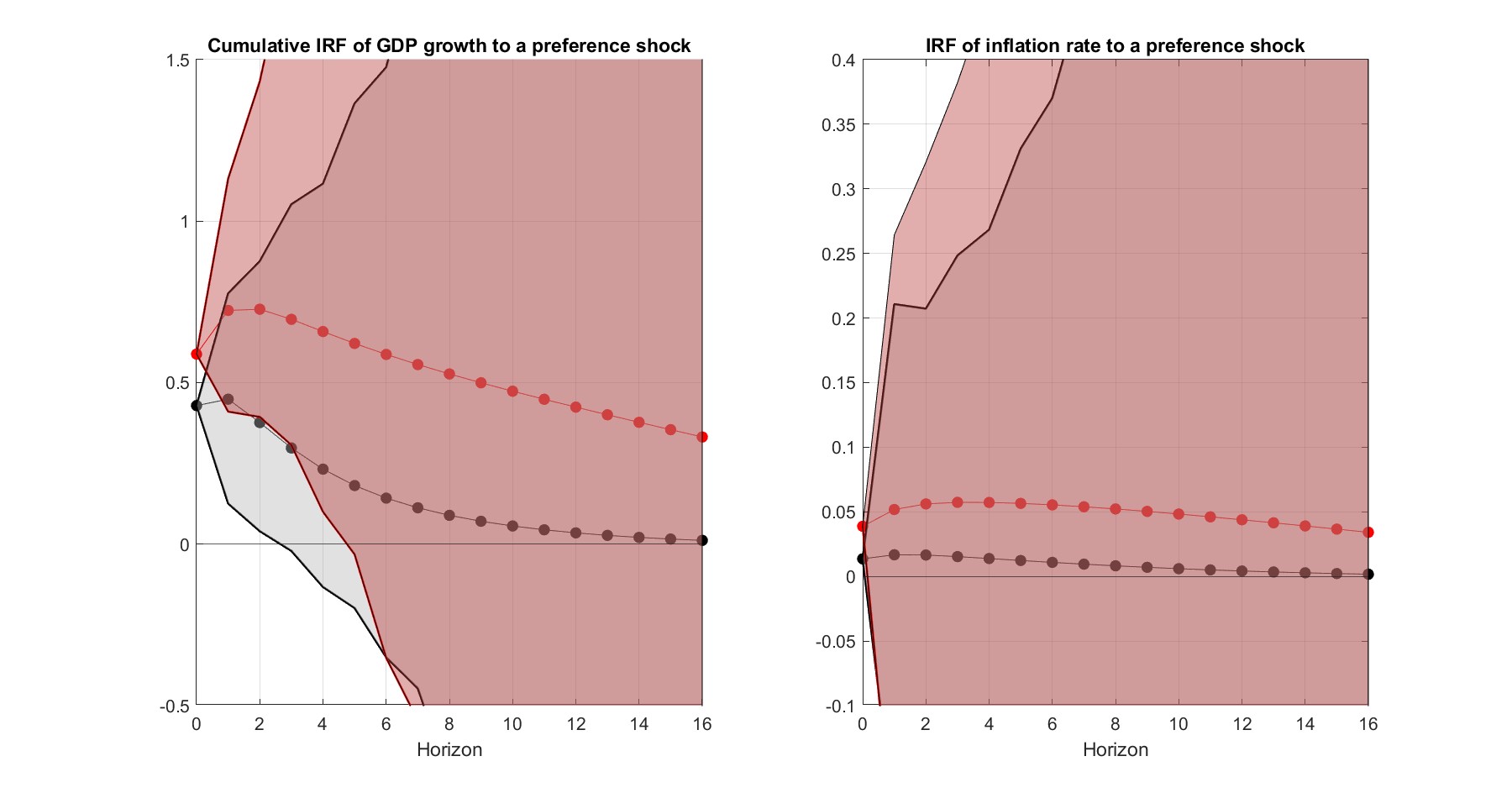}} 
\subfloat{\includegraphics[width = 2.4in]{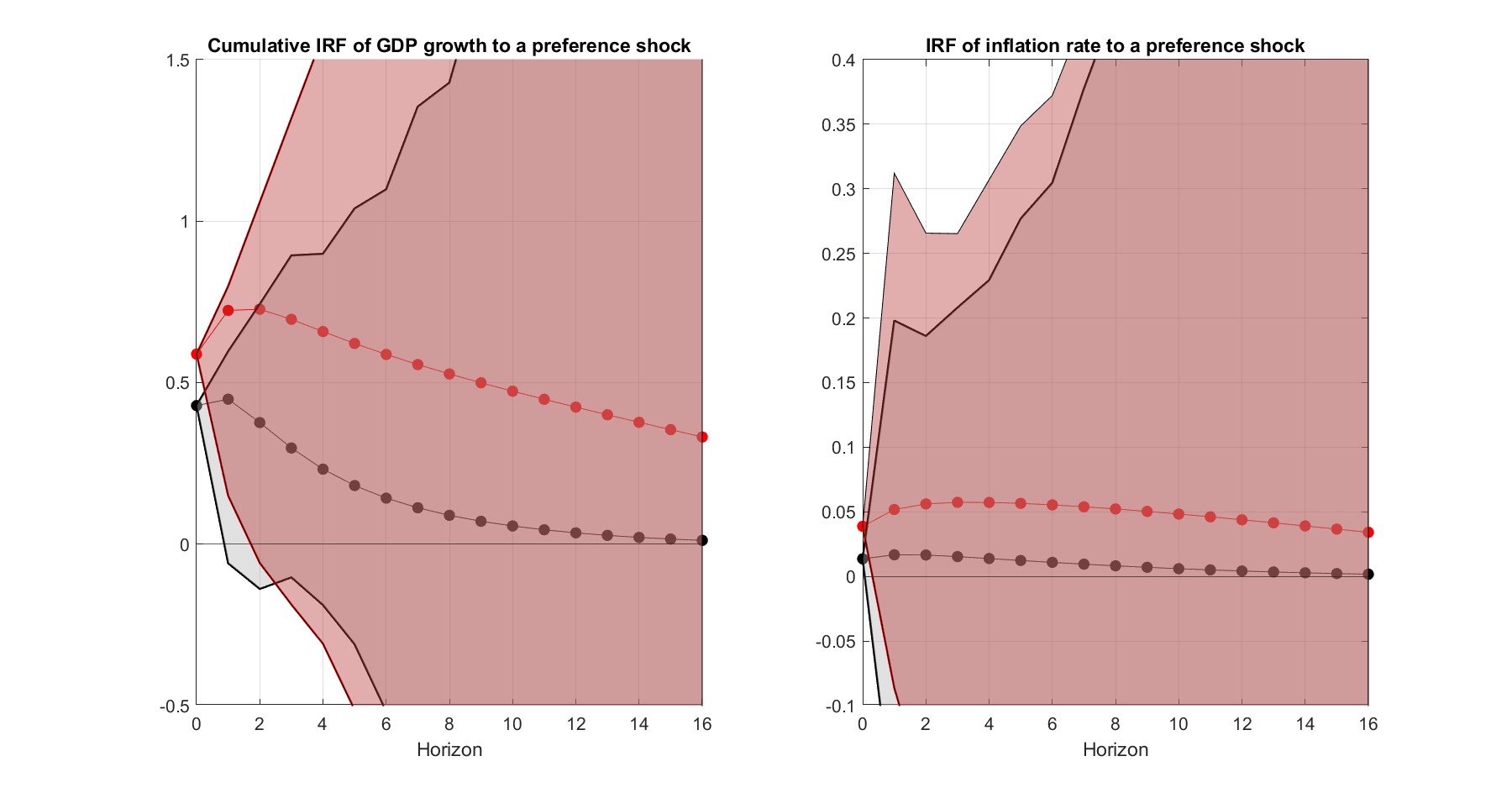}} 
\subfloat{\includegraphics[width = 2.4in]{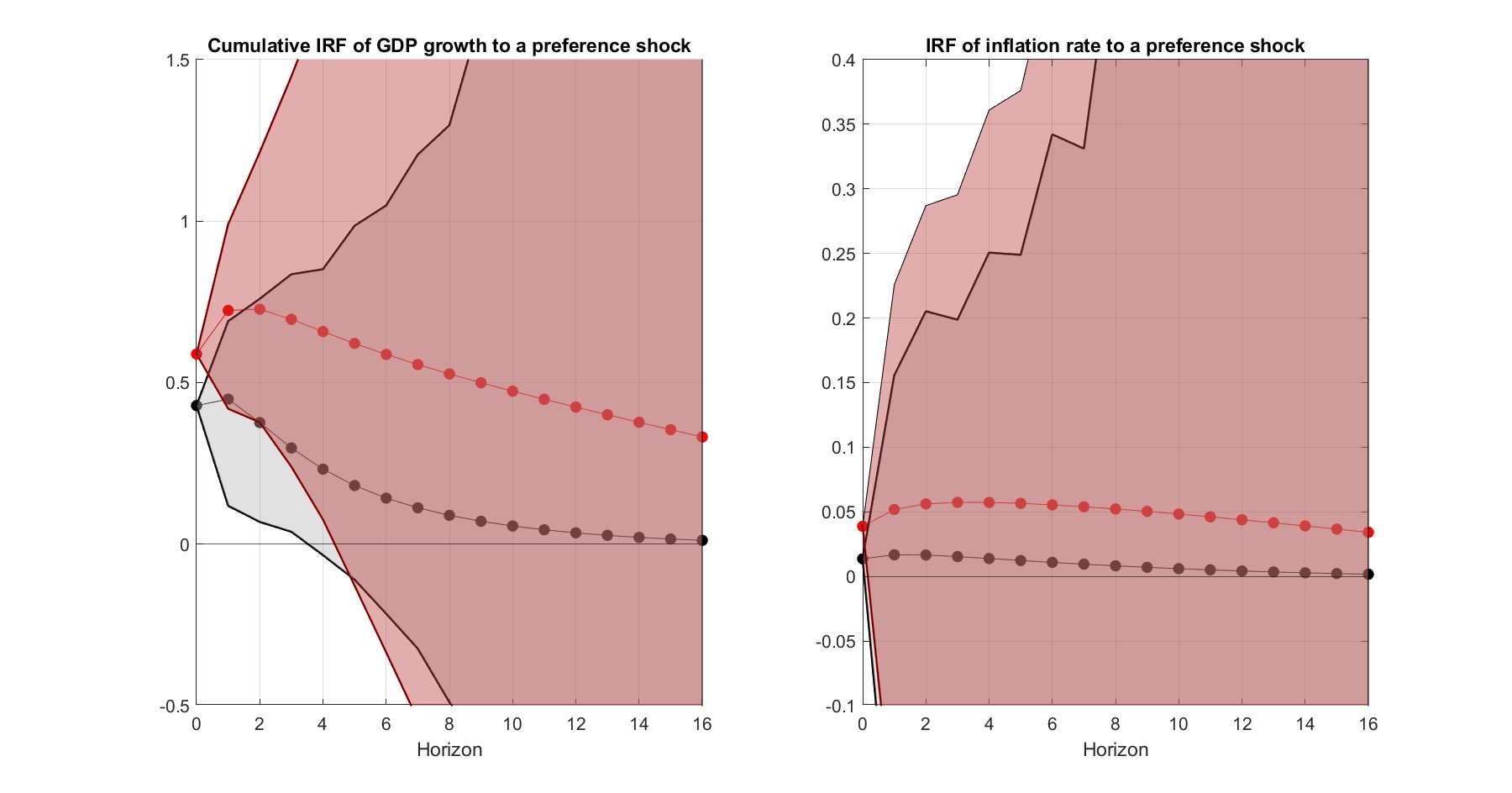}}\\
\subfloat{\includegraphics[width = 2.4in]{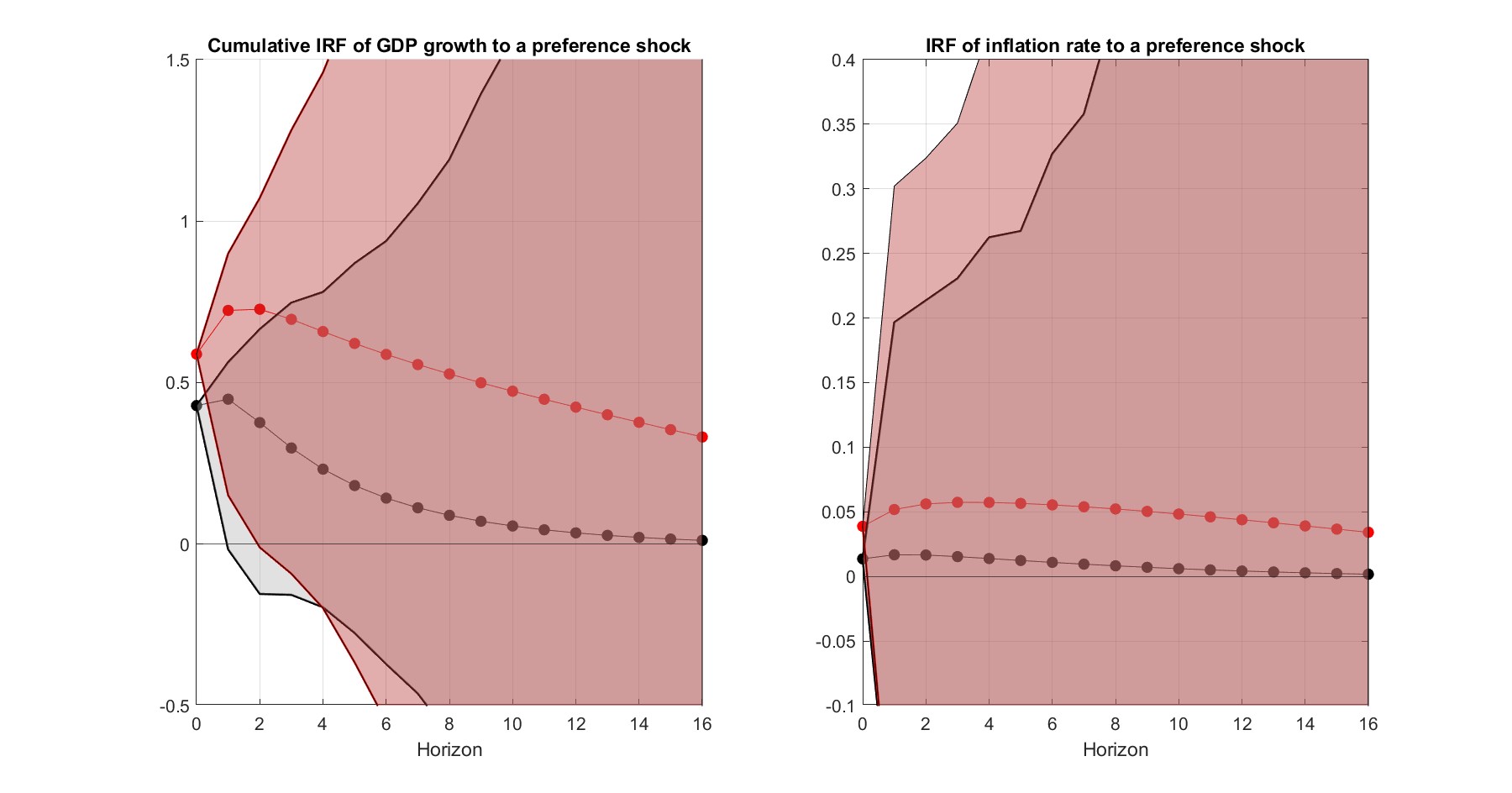}} 
\subfloat{\includegraphics[width = 2.4in]{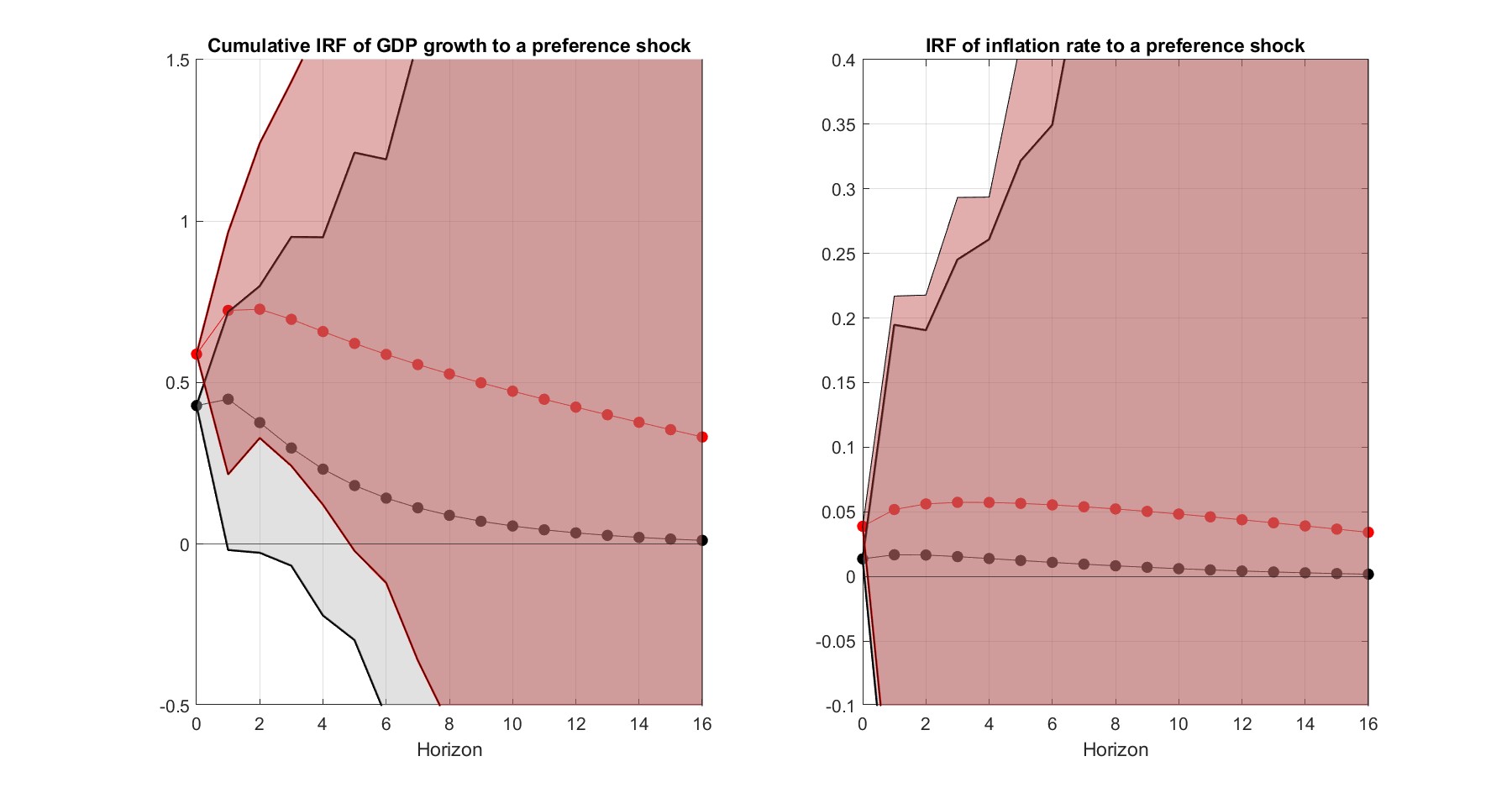}} 
\subfloat{\includegraphics[width = 2.4in]{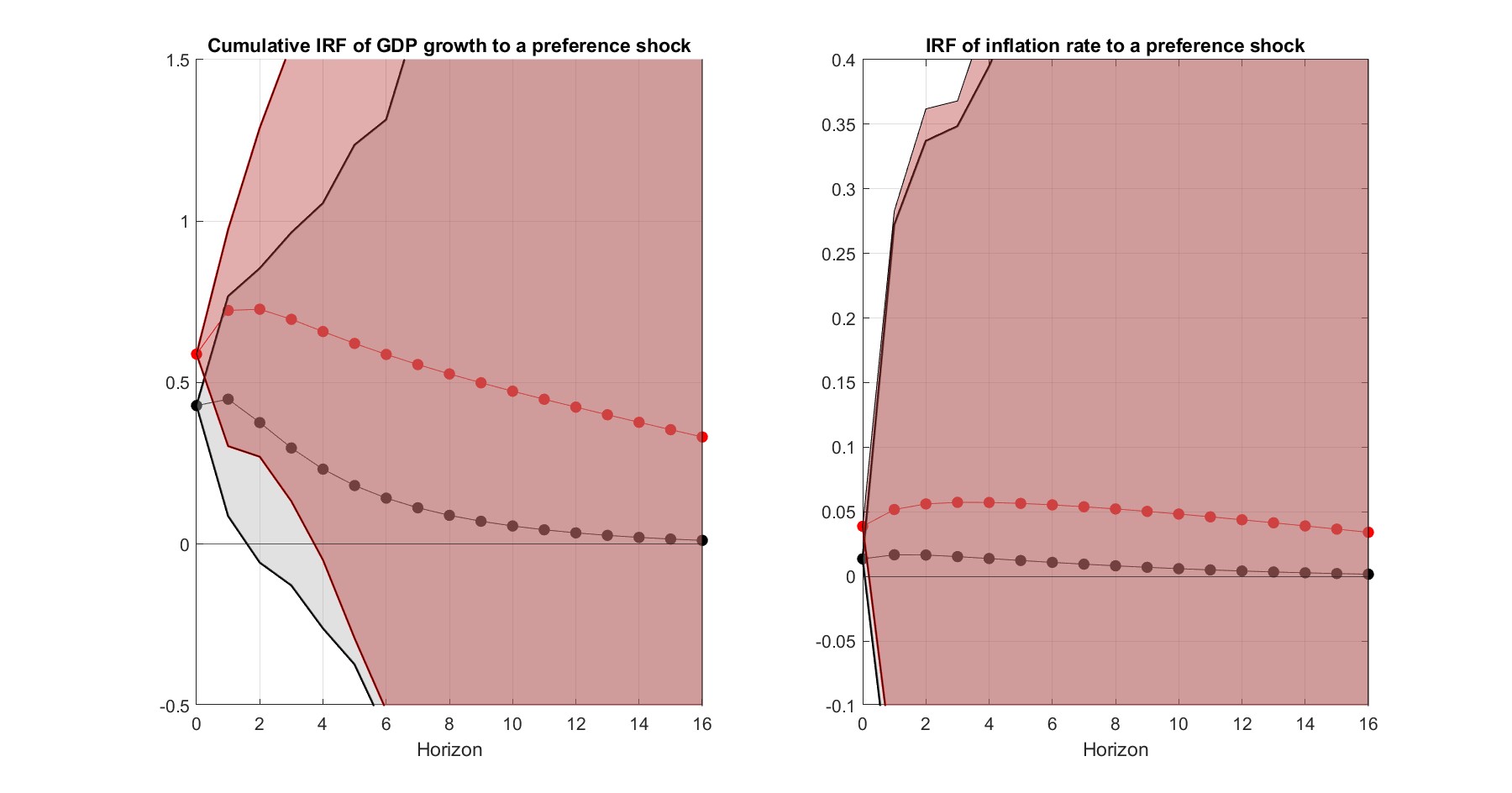}} 
\subfloat{\includegraphics[width = 2.4in]{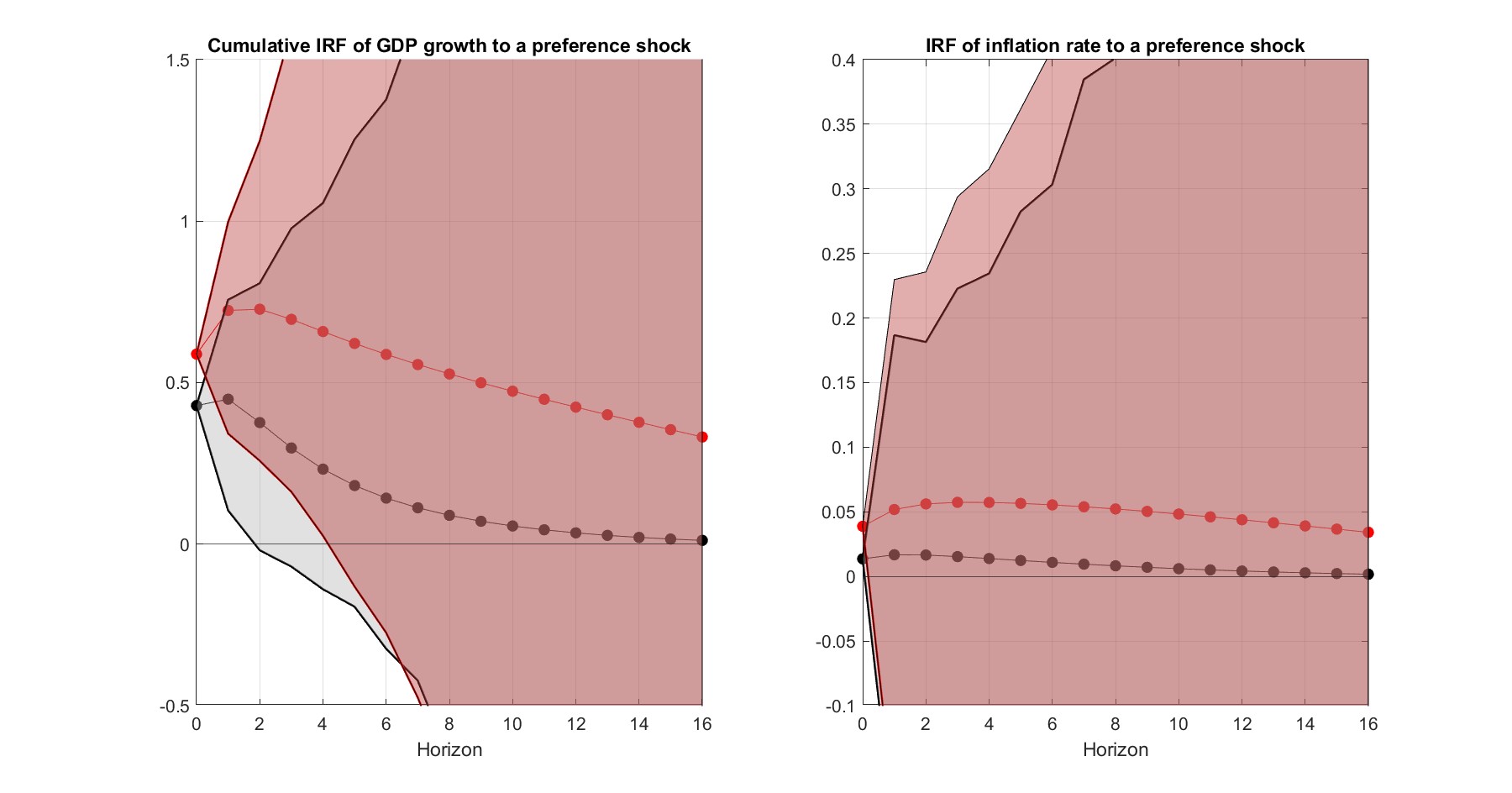}}\\
\caption{\small{Estimated $10^{th}$ and $90^{th}$ credible intervals of responses of output growth (cumulative) and inflation rate inside (in shaded red) and outside (in shaded grey) the ZLB period from a standard TVP-VAR on distinct simulated samples from the NK model. The dotted line plots are the responses inside (in red) and outside (in black) the ZLB period in the NK model.}}
\label{multiple_sim}
\end{figure}

\end{landscape}

\afterpage{
\begin{landscape}  
\subsubsection{Prior for the structural parameters}\label{secpriors_dsge_medium}
\begin{table}[htp!]
\centering
\caption{Prior distribution for the parameters of the New-Keynesian model}
\label{priordsge_smallscale}
\begin{tabular}{c|c|c|c|c|c|c|c}
Parameter    & Prior distribution & Mean & Standard Deviation & Parameter    & Prior distribution & Mean & Standard Deviation \\
\hline \hline 
$\psi_1$     & Gamma & 1.500 & 0.250 & $\rho_R$     & Beta & 0.750 & 0.100 \\
$\psi_2$     & Gamma & 0.120 & 0.050 & $\psi_2$     & Beta & 0.500 & 0.200 \\
$\psi_3$     & Gamma & 1.12  & 0.050 & $\psi_3$     & Inverse-Gamma & 0.10  & 2.000 \\
$\zeta_p$  & Beta  & 0.500 & 0.100 & $\zeta_w$  & Beta  & 0.500 & 0.100 \\
$\alpha$     & Normal & 0.300 & 0.050 & $\pi^*$     & Gamma & 0.750 & 0.400 \\
$\Phi$       & Normal & 1.250 & 0.120 & $\gamma$       & Normal & 0.400 & 0.100 \\
$h$          & Beta   & 0.700 & 0.100 & $S''$          & Normal   & 4.000 & 1.500 \\
$\nu_l$      & Normal & 2.000 & 0.750 & $\sigma_c$      & Normal & 1.500 & 0.370 \\
$\iota_p$    & Beta   & 0.500 & 0.500 & $\iota_w$    & Beta   & 0.500 & 0.150 \\
$r_*$        & Gamma  & 1.500 & 0.250 & $\psi$        & Beta  & 0.500 & 0.150 \\
$\rho_z$     & Beta   & 0.500 & 0.200 & $\sigma_z$     & Inverse-Gamma   & 0.100 & 2.000 \\
$\rho_b$     & Beta   & 0.500 & 0.250 & $\sigma_b$     & Inverse-Gamma   & 0.100 & 2.000 \\
$\rho_{\lambda_f}$ & Beta & 0.500 & 0.200 & $\sigma_{\lambda_f}$ & Inverse-Gamma & 0.100 & 2.000 \\
$\rho_{\lambda_w}$ & Beta & 0.500 & 0.200 & $\sigma_{\lambda_w}$ & Inverse-Gamma& 0.100 & 2.000 \\
$\rho_{\mu}$ & Beta & 0.500 & 0.200 & $\sigma_{\mu}$ & Inverse-Gamma & 0.100 & 2.000 \\
$\rho_g$     & Beta   & 0.500 & 0.200 & $\sigma_g$     & Inverse-Gamma   & 0.100 & 2.000 \\
$\eta_{\lambda_f}$ & Beta & 0.500 & 0.200 & $\eta_{\lambda_f}$ & Beta & 0.500 & 0.200 \\
$\eta_{gz}$   & Beta   & 0.500 & 0.200 &   &    &  & \\
\hline \hline
\end{tabular}
\label{fig:priors_dsge_medium}
\vspace{.25cm}\hspace{.85cm}\parbox{1.2\textwidth}{\scriptsize{The table reports the details on the prior distribution of the parameters of the medium scale New-Keynesian model which accounts for the ZLB period and forward guidance. The choice of the prior follows \citet{delnegroschorfheide2005}.}}
\end{table}
\end{landscape}}

\clearpage
\subsubsection{Risk premium shocks}\label{sec:rpshocks}
\begin{figure}[htp!]
    \centering
\includegraphics[scale=0.15]{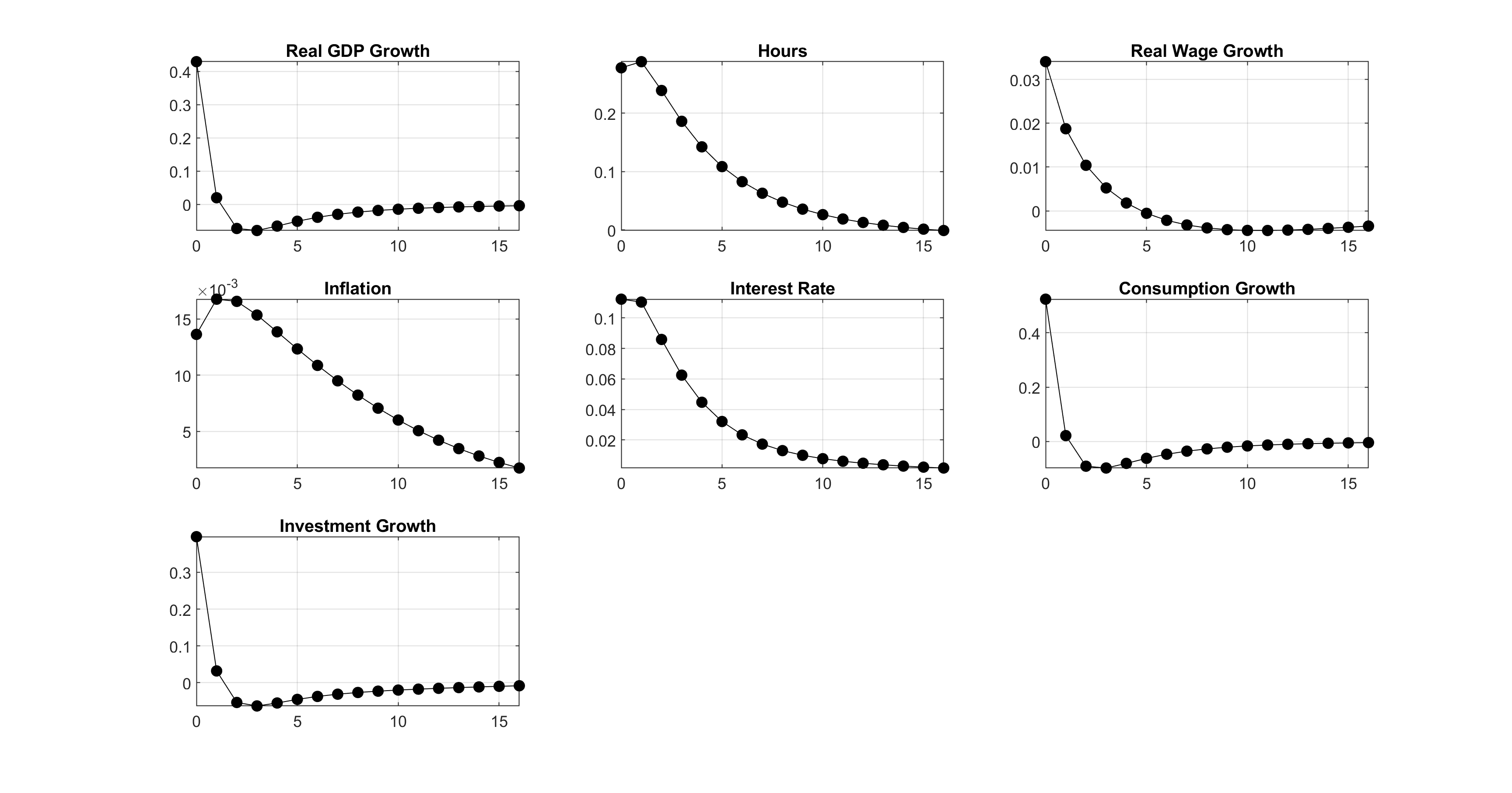}
    \caption{The figure shows the impulse response functions to one standard deviation risk premium shock in the \citet{SWauters} model.}
    \label{fig:irf_rp_sw}
\end{figure}

In the empirical analysis we exploit the impact matrix of the NK model to identify the risk premium shocks. Risk premium shocks in the \citet{SWauters} are financial shocks that drive a wedge between the risk-free interest rate and the actual interest rates faced by borrowers. Figure \ref{fig:irf_rp_sw} shows the impulse response functions to a negative risk premium shock in the \citet{SWauters} model. The shock propagates through the economy like a pure demand shock, by raising both real activity  and inflation.  

\begin{figure}[htp!]
    \centering
    \includegraphics[scale=0.18]{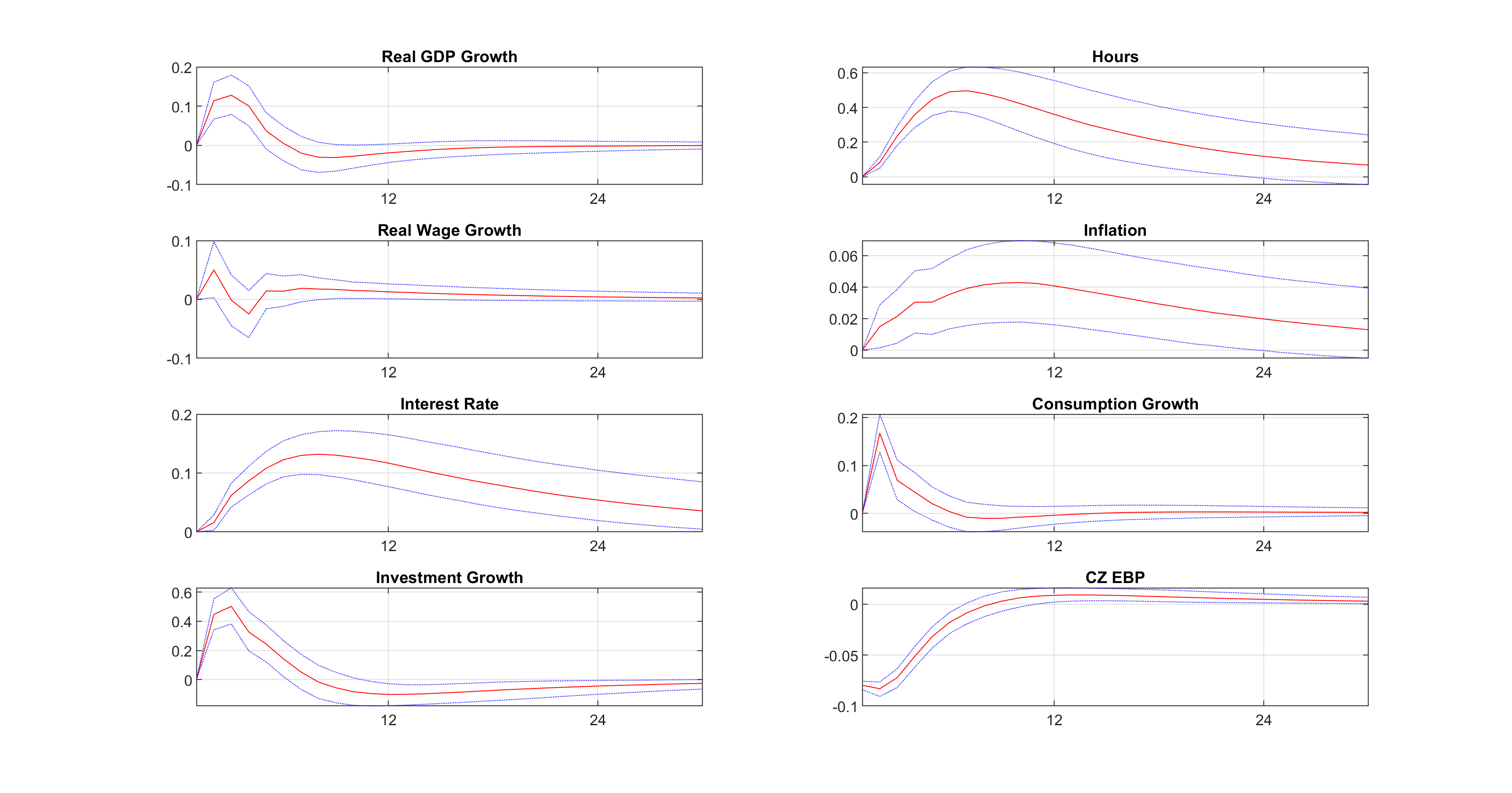}
    \caption{The figure reports the impulse response functions to a negative one standard deviation shock to the excess bond premium of \citet{gz2012}. In red the posterior median, while in blue dotted line the $84^{th}-16^{th}$ credible intervals.}
    \label{fig:EPB}. 
\end{figure}

Considering
a standard VAR model with the seven variables of the \citet{SWauters} model augmented with the Excess Bond Premium (EBP) series by \citet{gz2012},
we think of the orthogonal component of the EPB as akin to the risk premium shocks in the Smets and Wouters model. Figure \ref{fig:EPB}, shows the impulse response function to a one standard deviation shock to the EBP \citet{gz2012}. The shocks are identified through recursive identification by ordering the EBP series last in the VAR, that is allowing the excess bond premium series to contemporaneously respond to the other shocks in the VAR. 

\end{document}